\documentclass[pdflatex,sn-mathphys-num]{sn-jnl}

\usepackage{graphicx}%
\usepackage{multirow}%
\usepackage{amsmath,amssymb,amsfonts}%
\usepackage{amsthm}%
\usepackage{mathrsfs}%
\usepackage[title]{appendix}%
\usepackage{xcolor}%
\usepackage{textcomp}%
\usepackage{manyfoot}
\usepackage{booktabs}%
\usepackage{algorithm}%
\usepackage{algorithmicx}%
\usepackage{algpseudocode}%
\usepackage{listings}%
\usepackage{indentfirst}
\usepackage{upgreek}
\usepackage{slashed}
\usepackage{tensor}
\usepackage{qmech}

\theoremstyle{thmstyletwo}%

\theoremstyle{thmstylethree}%

\newcommand{\ee}{\mathrm{e}}
\newcommand{\B}{\mathrm{B}}
\newcommand{\V}{\mathrm{V}}
\newcommand{\LH}{\mathrm{L}}
\newcommand{\RH}{\mathrm{R}}

\renewcommand{\op}[1]{\hat{#1}}

\begin{document}

\title[Article Title]{Dileptons in heavy-ion collisions}

\author[1,2]{\fnm{Hendrik} \sur{van Hees}}\email{hees@itp.uni-frankfurt.de}

\affil[1]{\orgdiv{Institute for Theoretical Physics}, \orgname{Goethe
    University Frankfurt}, \orgaddress{\street{Max-von-Laue-Stra{\ss}e 1},
    \city{Frankfurt am Main}, \postcode{D-60438}, \country{Germany}}}

\affil[2]{\orgdiv{Helmholtz Research Academy Hessen for FAIR
(HFHF)}, \orgname{GSI Helmholtz Center,
Campus Frankfurt}, \orgaddress{\street{Max-von-Laue-Stra{\ss}e 12}, \city{Frankfurt am
Main}, \postcode{D-60438}, \country{Germany}}}

\abstract{In this review we give an overview about dileptons as probes
  for the properties of the strongly interacting hot and dense matter
  created in heavy-ion collisions at various beam energies. As
  penetrating probes they leave the fireball unaffected from final-state
  interactions and thus provide a space-time-evolution weighted average
  of the in-medium properties of hot and dense QCD matter.}

\maketitle

\section{Introduction}\label{sec:intro}

One of the prime motivations for heavy-ion-collision experiments at a
broad range of collision energies is to gain insight into the properties
of the hot and dense medium consisting of strongly interacting
particles. At the highest beam energies, as investigated at the Large
Hadron Collider (CERN) and the Relativistic Heavy Ion Collider (RHIC) a
state of this matter can be investigated, which is close to the
situation in the very early universe (a few microseconds after the big
bang), consisting of a strongly coupled plasma of quarks and gluons
(QGP) at vanishing net-baryon density ($\mu_{\text{B}}=0$), which
rapidly expands and cools undergoing a transition from partonic to
hadronic ``relevant degrees of freedom''. In contradistinction to the
hadronic observables, which reflect the properties of the medium at
chemical freezeout, where the inelastic collision processes become
ineffective, (via the particle abundancies, which are well described by
the statistical hadronization model) and at thermal freezeout, from when
on the hadrons stream freely to the detectors, for electromagnetic
probes, i.e., photons and dileptons ($\e^+ \e^-$ or $\upmu^+ \upmu^-$
pairs), the QCD medium is transparent, and they are thus nearly
unaffected by final-state interactions during the entire dynamical
evolution of the hot and dense fireball. They are thus unique probes for
the in-medium properties of quarks, gluons, and the hot and dense
hadron-resonance gas. At lower beam energies, as investigated in the
beam-energy scan (BES) program at RHIC and at GSI and in the future at
FAIR, the fireball reaches lower temperatures and high net-baryon
densities (finite $\mu_{\text{B}}$), and with the BES one can hope to
find clear indications for the onset of the formation of a QGP and the
different phase transitions of the QCD medium expected from various
theoretical models. 

The phases of strongly interacting matter are thereby characterized by
the approximate chiral symmetry of QCD in the light-quark sector with
the quark condensate $\erw{\bar{q} q} \neq 0$ as an order parameter,
while the confinement-deconfinement transition, where the effective
degrees of freedom change between ``hadronic'' and ``partonic'' degrees
of freedom. For this transition one uses Polyakov loops as indicators
for the transition, although they are order parameters in the strict
sense only in ``quenched QCD'', i.e., when dynamical quarks are
neglected.
\begin{figure}[t]
\centering
\includegraphics[width=0.9\linewidth]{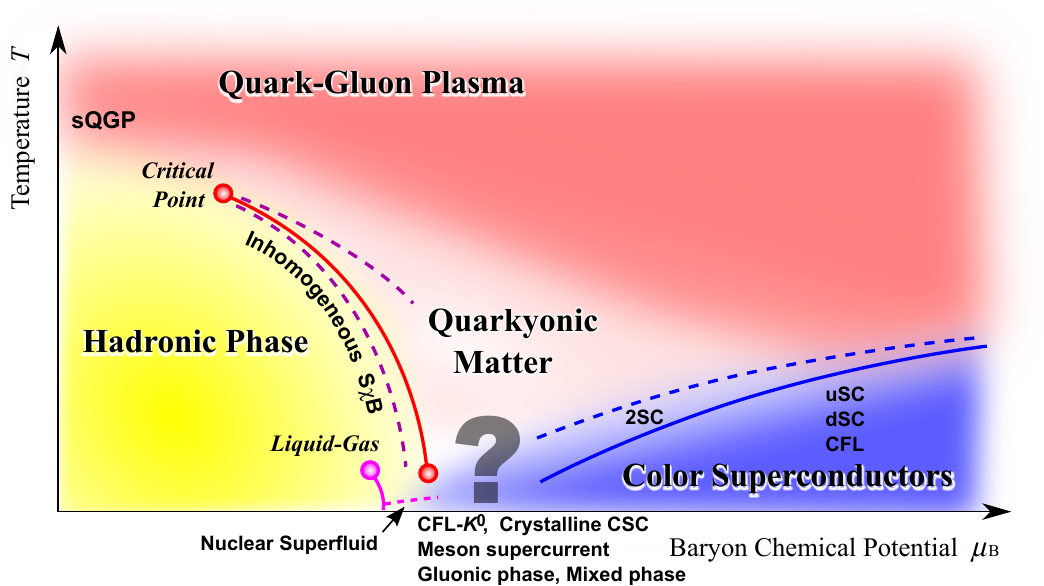}
\caption{Schematic QCD phase diagram with possible phase boundaries based
  on chiral symmetry. Figure taken from \cite{Fukushima:2010bq}.}
\label{fig.phase-diag}
\end{figure}

Theoretically the properties of the QCD medium at thermal equilibrium
can be addressed from first principles by lattice-QCD (lQCD)
calculations, which however is only applicable at $\mu_{\text{B}}=0$ or
at low $\mu_{\text{B}}$, due to the notorious ``sign problem'' at finite
$\mu_{\text{B}}$. In the region of the phase diagram close to the
temperature axis lQCD shows that the confinement-deconfinement as well
as the chiral transition are a crossover, both occuring at a temperature
of about 155 MeV \cite{kar01,Philipsen:2012nu,Borsanyi:2011zzc}.

Although considerable progress has been made to overcome the problem to
evaluate the QCD Equation of State (EoS) at $\mu_{\text{B}} \neq 0$ with
lQCD methods, here effective hadronic or quark-meson models based on
chiral symmetry have to be used to explore possible scenarios for phase
transitions from hadronic to partonic matter. Many of these models
indicate that the phase transition at $T=0$ is a first-order transition,
i.e., in the $T$-$\mu_{\text{B}}$ plane one expects that at lower
temperatures and large net-baryon densities a first-order transition
line should occur, which ends at a critical point, where the transition
becomes of $2^{\text{nd}}$ order. On the other hand some lQCD
calculations indicate that such a critical point also might not exist
\cite{Philipsen:2012nu}.

Recently, many effective theories, based on the approximate chiral
symmetry of QCD in the light-quark sector have been investigated at
finite temperature and baryochemical potential. For a realistic
description of the EoS also the gluonic degrees of freedom have to be
considered to account for the confinement properties of QCD. For this
purpose effective chiral models like the Nambu-Jona-Lasionio model
\cite{Meisinger:1995ih,Pisarski:2000eq,Fukushima:2003fw,Megias:2004hj,Ratti:2005jh,Ghosh:2006qh,Sasaki:2006ww,Kashiwa:2007hw}
or quark-meson models (i.e., a linear $\sigma$ model with quarks and
mesons as elementary degrees of freedom)
\cite{Kovacs:2006ym,Schaefer:2007pw,Mao:2009aq,Gupta:2009fg,Herbst:2010rf,Marko:2010cd}
have been generalized by the inclusion of Polyakov-loop degrees of
freedom (so-called Polyakov-loop extended Nambu-Jona-Lasinio (PNJL) or
quark-meson (PQM) models, respectively). These models can well reproduce
the findings in lQCD calculations at vanishing or small baryochemical
potential, and thus one may hope that they may also be successfully
extrapolated to higher net-baryon densities. Here one needs the
dependence of the deconfinement-transition temperature on quark-flavor
number and density in the Polyakov-Loop potential, which are estimated
with perturbative hard/dense-thermal-loop (HTL/DTL) techniques. E.g.,
within a PQM model \cite{Schaefer:2011pn} such techniques lead to the
location of the critical point at a quite large baryochemical potential,
$\mu_{\mathrm{B},\text{crit}} \simeq 300 \; \MeV$, and low temperature,
$T_{\text{crit}} \simeq 20 \; \MeV$. At the same time the region, where
chiral symmetry is restored but matter is still (effectively) confined
(``quarkyonic state'' predicted based on the large-$N_{\text{c}}$ limit
of QCD \cite{McLerran:2007qj,Hidaka:2008yy}) becomes quite small.

For a realistic description of the QCD-phase diagram, calculations going
beyond a pure mean-field model (Ginzburg-Landau approximation) are
necessary, as the functional renormalization group 
\cite{Jungnickel:1995fp,Schaefer:1999em,Tetradis:2003qa,Schaefer:2004en,Schaefer:2006ds,Schaefer:2006sr,Skokov:2010sf,Nakano:2009ps,Skokov:2010wb,Skokov:2010uh}.


The rest of this review is organized as follows: in Sect.\
\ref{sect:in-med-models} the general interpretation of
dilepton-production data in heavy-ion collisions as a probe for the
in-medium properties of the electromagnetic (e.m.)
curren-current-correlation function is given and its relation with the
(approximate) chiral symmetry of the light-quark sector of QCD is given.

In Sect.\ \ref{sect:dileps-hics} we review models to describe the
in-medium properties of quarks, gluons, and hadrons, the dynamical
evolution of this medium as created in heavy-ion collisions.

In Sect.\ \ref{sect:dileps-hics-results} we confront the before
described models for in-medium dilepton production with data from
heavy-ion collision experiments at a broad range of avialable beam
energies.

\section{General properties of dilepton emission from a QCD medium}
\label{sect:in-med-models}

For a realistic modelling of (thermal) dilepton production in heavy-ion
collisions one needs both a model for the in-medium properties of
strongly interacting constituents of the strongly interacting medium and
a realistic model for the dynamical evolution of the rapidly expanding
``fireballs'' of this medium created in the collisions.

\subsection{In-medium dilepton production rates}
\label{sect:McLerranToimela}

Following \cite{rw99} and \cite{Rapp:2009yu}, we briefly discuss
general ideas on the radiation of dileptons ($\e^+ \e^-$ and
$\upmu^+ \upmu^-$ pairs) from a thermal source of strongly interacting
particles. As indicated by the hadronic observables of heavy-ion
collisions at higher beam energies, particularly $p_{\text{T}}$-spectra and
anisotropic flow $v_2$, the hot and dense fireballs created in heavy-ion
collisions behave with good accuracy like a collectively expanding fluid
close to local thermal equilibrium, i.e., as a strongly coupled
many-body system as far as the strong interaction is concerned. On the
other hand the medium is transparent for leptons and photons that
interact with the medium only via the electromagnetic and weak
interactions. This implies that the radiation of electromagnetic probes
can be expressed in terms of the equilibrium \textbf{electromagnetic
  current-current correlation function}, with the average taken in the
fully interacting quantum field theory as far as the strong interaction
is concerned and in leading order of the electromagnetic interaction,
$\mathcal{O}(\alpha_{\text{em}})$ for photons and
$\mathcal{O}(\alpha_{\text{em}}^2)$ for dileptons
\cite{Feinberg:1976ua,McLerran:1984ay,Gale:1990pn}.

We derive the corresponding McLerran-Toimela formula for dileptons,
assuming a (locally) equilibrated medium with a fluid cell at rest. The
aim is to calculate the dilepton-production rate,
\begin{equation}
\label{mtform.1}
\frac{\dd R_{\ell^+ \ell^-}}{\dd^4k} = \frac{\dd N_{\ell^+
    \ell^-}}{\dd^4x \dd^4 k},
\end{equation}
i.e., the number of $\ell^+ \ell^-$-pairs per time and volume and pair
energy and momentum. To that end we work in an interaction picture with
the ``undisturbed Hamiltonian'' fully including the strong interactions
of quarks and gluons or hadrons and the ``interaction Hamiltonian''
given by the electromagnetic interaction
\begin{equation}
\label{mtform.2}
H_{\text{I}} = e \int_{\R^3} \dd^3 \vec{x} \; J_{\text{em}}^{\mu}(t,\vec{x})
A_{\mu}(t,\vec{x}).
\end{equation}
On the fundamental level the electromagnetic current
$J_{\text{em}}^{\mu}$ is given in terms of the charged leptons (charge
$-1$) and the quarks by
\begin{equation}
\begin{split}
\label{weak.44}
J_{\text{em}}^{\mu} = &
-(\overline{\psi}_{\ee},\overline{\psi}_{\mu},\overline{\psi}_{\tau})
\gamma^{\mu}
\vvv{\psi_{\e}}{\psi_{\mu}}{\psi_{\tau}} \\
&+ \frac{2}{3} (\overline{\psi}_u,\overline{\psi}_c,\overline{\psi}_t)
\gamma^{\mu} \vvv{\psi_{\text{u}}}{\psi_{\text{c}}}{\psi_t} 
-\frac{1}{3} (\overline{\psi}_d,\overline{\psi}_s,\overline{\psi}_b)
  \gamma^{\mu} \vvv{\psi_{\text{d}}}{\psi_{\text{s}}}{\psi_b}.
\end{split}
\end{equation}
Working in leading order of the electromagnetic interaction, a schematic
Feynman diagram for the transition-matrix element for the production of
one $\ell^+ \ell^-$ pair
$\ket{i} \rightarrow \ket{f'}=\ket{f,\ell^+ \ell^-(k)}$, where $i$ and
$f$ are arbitrary partonic or hadronic initial and final states, is
given by
\begin{equation}
\begin{split}
\label{mtform.3}
S_{f'i}
&= \parbox{7cm}{\includegraphics[width=7cm]{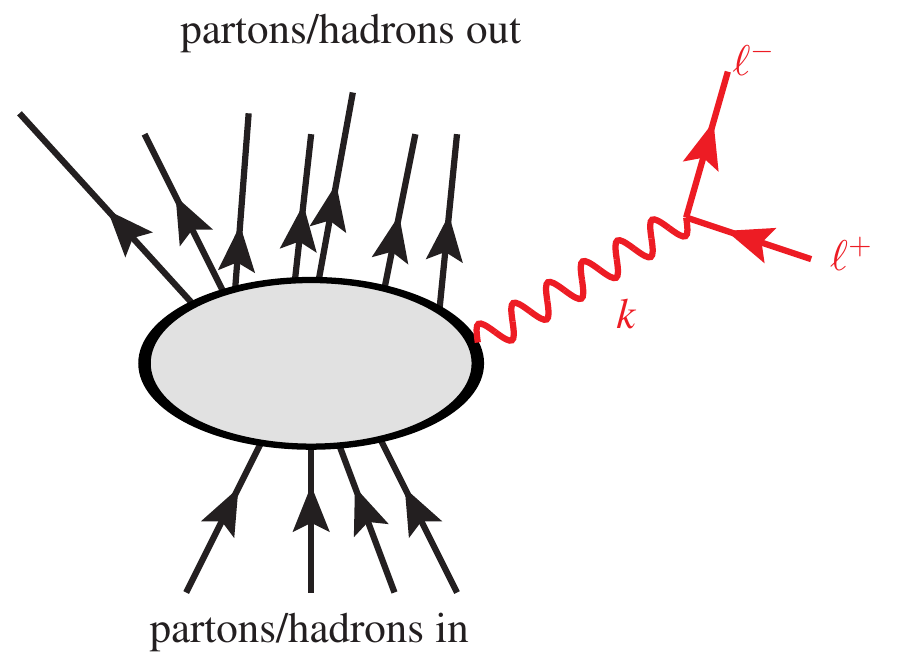}}
\\
&= e \matrixe{f}{\int \dd^4 x J_{\text{em}}^{\mu}(x)}{i} D_{\gamma}^{\mu
  \nu}(x,x') \overline{u_{\ell}}(x) \gamma_{\nu} v_{\ell}(x').
\end{split}
\end{equation}
Here $u_{\ell}$ and $v_{\ell}$ are the usual free Dirac spinors for the
lepton and the antilepton in the final state, and $D_{\gamma}^{\mu \nu}$
is the photon propagator in an arbitrary gauge, e.g., in the Feynman
gauge, where it reads in momentum representation
\begin{equation}
\label{mtform.4}
D_{\gamma}^{\mu \nu}(k)=-\frac{\eta^{\mu \nu}}{k^2+\ii 0^+}.
\end{equation}
In the momentum representation the matrix element reads
\begin{equation}
\label{mtform.5}
S_{f'i} = \ii T_{f'i} (2 \pi)^4 \delta^{(4)}(P_f+k-P_i)
\end{equation}
and the dilepton-production rate according to ``Fermi's trick'' \cite{ps95}
\begin{equation}
\label{mtform.6}
\frac{\dd R}{\dd^4 k} = (2 \pi)^4 \erw{\delta^{(4)}(P_f+k-P_i)|T_{f'i}|^2}
\end{equation}
with the expectation value taken over all partonic/hadronic initial
states with the (grand-canonical) equilibrium statistical operator,
written in the local rest frame of the fluid cell,
\begin{equation}
\label{mtform.7}
\op{\rho} = \frac{1}{Z} \exp \left[-\frac{1}{T}(\op{H}_{\text{QCD}}-\mu_{\B}
\op{Q}_{\B}) \right ], \quad Z= \Tr \exp \left[-\frac{1}{T}(\op{H}_{\text{QCD}}-\mu_{\B}
\op{Q}_{\B}) \right ].
\end{equation}
Here $T$ is the fluid cell's temperature and $\mu_{\B}$ the
baryochemical potential to take into account the conservation of the
net-baryon number under the strong and electromagnetic interactions. In
(\ref{mtform.7}) one also sums over all possible partonic/hadronic final
states and the spin-degrees of freedom of the lepton and
antilepton. Making use of the analytic properties of the retarded em.\
current-current-correlation function, the final result reads
\cite{Gale:1990pn}
\begin{equation}
\label{mtform.8a}
\frac{\dd R_{\ell^+ \ell^-}}{\dd^4 k} = \frac{\alpha_{\text{em}}^2}{6
  \pi^3} \frac{k^2+2m_{\ell}^2}{(k^2)^2} \sqrt{1-\frac{4
    m_{\ell}^2}{k^2}} \eta_{\mu \nu} \rho_{\text{em}}^{\mu \nu}(k) n_{\B}(k^0),
\end{equation} 
where the spectral function of the correlation function is given in
terms of the electromagnetic spectral function
\begin{equation}
\label{mtform.8b}
\rho_{\text{em}}^{\mu \nu}(k)=-2 \im \Pi_{\text{em,rest}}^{\mu \nu}(k)
\end{equation}
with the retarded em.\ current-current correlation function (i.e., the
in-medium photon polarization tensor), evaluated in the grand-canonical
equilibrium state (\ref{mtform.7}) and to any order concerning the
strong interaction,
\begin{equation}
\label{mtform.10}
\ii \Pi_{\text{em,ret}}^{\mu \nu}(k) = \int_{\R^4} \dd^4 x \exp(\ii k \cdot
x) \erw{\comm{\op{J}_{\text{em}}^{\mu}(x)}{\op{J}_{\text{em}}^{\nu}(0)}}_{T,\mu_{\B}} \Theta(x^0),
\end{equation}
and
\begin{equation}
\label{mtform.9}
n_{\B}(k^0)=\frac{1}{\exp(k^0/T)-1}
\end{equation}
denotes the Bose-Einstein distribution function. Thus, (\ref{mtform.8a})
implies that the measurement of the dilepton-production rate in
heavy-ion collisions provides information about the em.\ current-current
correlation function of strongly interacting matter in thermal
equilibrium (\ref{mtform.8b}). According to the above derivation of
(\ref{mtform.8a}) of course the measured dilepton spectra are only
accessible as a space-time weighted average over the entire fireball
evolution. This implies that modeling the dilepton production in
heavy-ion collisions to interpret the corresponding data must include a
comprehensive understanding of the relevant microscopic processes for
dilepton production in the medium as well as a detailed description of
the fireball evolution.

More recently also more differential aspects of the
dilepton-production spectra have become of interest and in reach of
experimental observation in terms of the polarization of thermal
dileptons from heavy-ion collisions.

Angular dependencies in the dilepton production rate can be unravelled
by resolving the angle, $\Omega_l=(\phi_l,\theta_l)$, of a single lepton
relative to the virtual photon's momentum in the latter's rest
frame~\cite{Bratkovskaya:1995kh,Baym:2017qxy,Speranza:2018osi}. It can
be shown \cite{Seck:2023oyt} that, in the local restframe of the fluid cell,
\begin{equation}
\frac{\dd R_{\ell^+ \ell^-}}{d^4k\, d\Omega_l} = \frac{\alpha^2}{32
  \pi^4 M^4} \sqrt{1-\frac{4 m_{\ell}^2}{M^2}}\rho_{\text{em}}^{\mu\nu}L_{\mu\nu} n_{\B}(k_0;T),
\end{equation}
with the lepton tensor 
\begin{equation}
L^{\mu\nu}=2(q^2 g^{\mu\nu} - q^\mu q^\nu + \Delta l^\mu \Delta l^\nu) \ ,
\end{equation}
where $\Delta l^\mu = l^{+\mu}-l^{-\mu}$, and $l^{\pm}$ are the lepton four-momenta. More explicitly, the angular distribution takes the form
\begin{align}
\label{angdist}
\frac{d N_{ll}}{d^4 x \, d^4q \,d\Omega_l}&\propto\Big(1+\lambda_\theta\cos^2\theta_l\\
&+\lambda_\phi \sin^2\theta_l\cos2\phi_l+\lambda_{\theta\phi}\sin2\theta_l\cos\phi_l \nonumber\\
&+\lambda^{\bot}_\phi \sin^2\theta_l\sin2\phi_l+\lambda^{\bot}_{\theta\phi}\sin2\theta_l\sin\phi_l \Big)\nonumber,
\end{align}
where the $\lambda$'s are the anisotropy coefficients. 

Even in an isotropic thermal medium, nontrivial anisotropies in the
angular distribution of the produced leptons occur, e.g., for basic
hadronic and partonic sources ($\pi\pi$ vs.\ $q\bar q$ annihilation,
respectively) at a few percent level~\cite{Speranza:2018osi}. In
addition, the anisotropy of lepton pairs in the $M=1$-$1.5 \, \GeV$ region
might be able to distinguish whether the so-called ``chiral mixing''
between $\rho$ and $a_1$ channels via $\pi a_1$ annihilation or
$q \bar q$ annihilation is the dominant source.

One should note that various ``frames of reference'' of the anisotropy
coefficients are used in high-energy particle physics
\cite{Faccioli:2022peq}.

For theoretical investigations for the case of a static thermal medium
the natural reference frame is the (local) rest frame of the fluid cell,
defined by its four-velocity, $u'=\gamma(1,\vec{\beta})=(1,0,0,0)$. In
this frame the rotational symmetry concerning the dilepton-emission rate
is only broken by the virtual photon's momentum direction. Taking the
polarization axis as the $z'$-axis in this frame of reference, defining
the helicity frame $\text{HX}'$, the only non-vanishing anisotropy is
given by
\begin{equation}
    \lambda_{\theta}^{\rm HX'} (M,|\vec{q}|) =
    \frac{ \varrho_{\rm T} - \varrho_{\rm L} } { \varrho_{\rm T} + \varrho_{\rm L} } \ , 
    \label{lambda_rest}
\end{equation}
where the photon-polarization tensor is given by the spatially
transverse and longitudinal components,
\begin{equation}
\rho_{\text{em}}^{\mu \nu} = \rho_{\T} \Theta_{\T}^{\mu \nu} + \rho_{\LH}
\Theta_{\LH}^{\mu \nu}
\end{equation}
with the decomposition of the four-transverse tensor wrt.\ the
four-momentum dilepton (virtual photons) of invariant mass $M$,
\begin{alignat}{2}
\label{1}
\Theta^{\mu\nu}(q) &=\eta^{\mu \nu}-\frac{q^{\mu} q^{\nu}}{M^2},\\
\label{2}
  \left (\Theta_{\text{T}}^{\mu \nu}(q) \right)&=\begin{pmatrix} 0 & \vec{0}^{\text{T}}
                                      \\
                                      \vec{0} & \left (-\delta^{jk} +
                                                q^j q^k/\vec{q}^2
                                                \right) 
                                    \end{pmatrix},\\
\label{3}
\Theta_{\text{L}}^{\mu \nu}(q) &=\Theta^{\mu
  \nu}(q)-\Theta_{\text{T}}^{\mu \nu}.
\end{alignat}
For measurements in heavy-ion experiments, of course one must define a
polarization observable with respect to an experimentally well-defined
reference frame. 

To define the socalled helicity frame (HX), we start from the virtual
photon with four-momentum $k^{\mu}=(k_0,0,0,k)$ in the center-momentum
system (CMS) of the collision, defining the $z$-axis along its
three-momentum, $\vec{k}$. In the helicity frame (HX) this defines the
polarization axis while the pertinent $y$-axis is defined by the normal
vector in the plane spanned by the beam momenta, and thus defines the
($xyz$) system. On the other hand, the $z'$ axis is defined in the
thermal rest frame, which is moving with the medium's flow velocity
$u^\mu$ in the CMS.

The photon four-momentum in this system, $q^\mu$, is obtained from the
Lorentz boost of $k^\mu$ using $u^\mu$, and determines the only
non-vanishing coefficient in this system, $\lambda_{\theta}^{\rm
  HX'}$. With the above definitions, one can then transform the angular
distribution into the HX system by a succession of three Euler
rotations:
\begin{enumerate}
\item[(i)] around the $z'$-axis by an angle $\psi$ to bring the $y$-axis
perpendicular to the $z$-axis; 
\item[(ii)] around the thus obtained $y''$ axis by an angle $\zeta$ to align
the $z'$-axis with the $z$-axis; and 
\item[(iii)] around the $z$-axis by an angle $\omega$ to align the $x'$-
  and $y'$-axes along the $x$- and $y$-axes, respectively.  In this way,
  all five coefficients in Eq.~(\ref{angdist}) can be determined from
  $\lambda_{\theta}^{\rm HX'}$ and the three rotation angles described
  above.
\end{enumerate}
In a similar way also the description of the measurements in the
Collins-Soper (CS) frame of reference have to be descriped. In the CS
frame the quantization axis is defined as the bisector between the
projectile and target momentum in the restframe of the dilepton (virtual
photon) \cite{Faccioli:2022peq}.

\begin{figure}[H]
\centerline{\includegraphics[width=0.9 \linewidth]{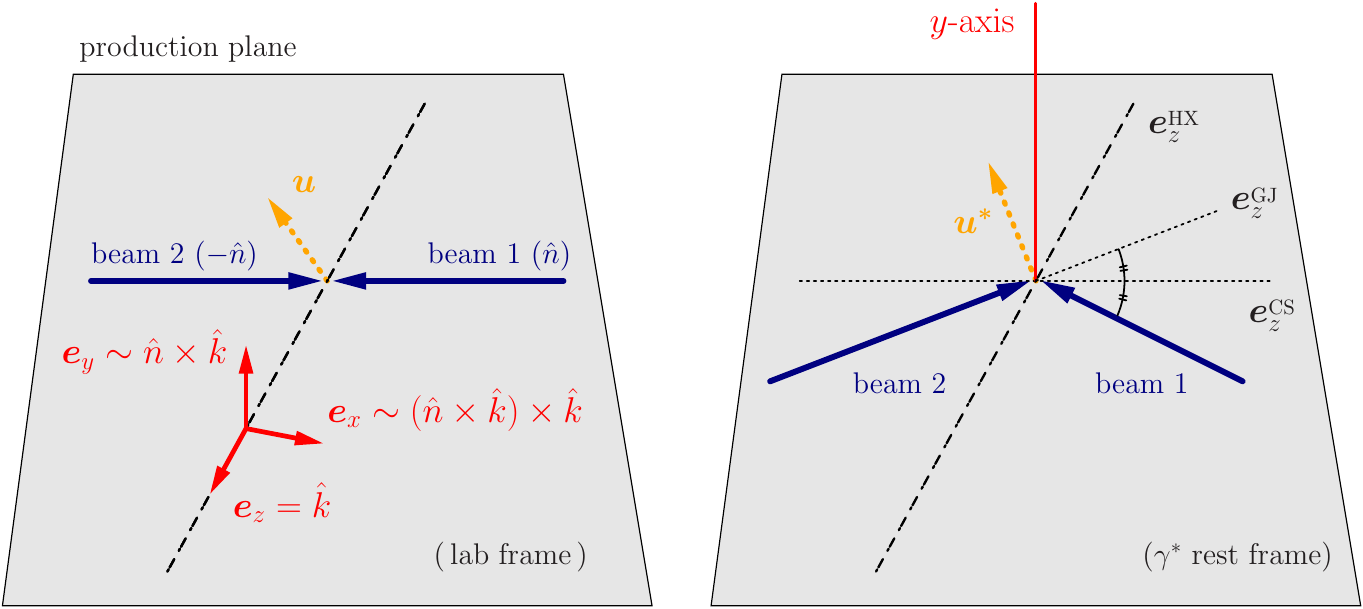}}
\caption{The definition of the various reference frames for defining
  polarization observables. The choice of the spin-quantization axis
  $\vec{e}_z$ is defines the various frames. Helicity frame (HX):
  $vec{e}_z$ is the direction of the virtual-photon ($\gamma^*$)
  momentum in the lab frame, Collins-Soper frame (CS): $\vec{e}_z$ is in
  the direction of the bisector of angle between the direction of one
  beam and the opposite of the direction of the other, as measured  in the $\gamma^*$
  rest frame , and Gottfried-Jackson frame (GJ): $\vec{e}_z$ is in the
  direction of one of the beams in the $\gamma^*$ fest frame (for
  details, see \cite{Faccioli:2022peq}). Figure
  taken from \cite{Gao:2026vxs}.}
\label{fig.pol-obs-frames}
\end{figure}

\subsection{Dileptons and chiral symmetry}

In this Section we discuss the relation of the electromagnetic
current-correlation function in a strongly interacting medium to the
approximate chiral
$\text{SU}(2)_{\LH} \times \text{SU}(2)_{\RH} \times \text{U}(1)_{\V}$
symmetry of QCD.

On the fundamental level of the Standard Model the key is the
decomposition of the light (u- and d-) quarks' electromagnetic current
\begin{equation}
\label{dil-chisy.1}
J_{\text{em,ud}}=\frac{2}{3} \overline{\psi_{\text{u}}} \gamma^{\mu}
\psi_{\text{u}}-\frac{1}{3} \overline{\psi_{\text{d}}} \gamma^{\mu} \psi_{\text{d}},
\end{equation}
cf.\ (\ref{weak.44}), in an isovector and an isoscalar component. To that
end we just have to remember that the electrically neutral component of
the isovector current is given by
\begin{equation}
\label{dil-chisy.2}
J_{\V}^{3 \mu}=\overline{\psi}_{\text{ud}} t^{3} \gamma^{\mu} \psi_{\text{ud}}
= \frac{1}{2} \left (\overline{\psi}_{\text{u}} \gamma^{\mu}
\psi_{\text{u}}-\overline{\psi}_d \gamma^{\mu} \psi_{\text{d}} \right),
\end{equation}
which carries the quantum numbers of the neutral $\rho(770)$ meson. The
isoscalar current reads
\begin{equation}
\label{dil-chisy.3}
j_{\V}^{\mu} = \overline{\psi}_{\text{u}} \gamma^{\mu}
\psi_{\text{u}}+\overline{\psi}_d \gamma^{\mu} \psi_{\text{d}},
\end{equation}
which carries the quantum numbers of the $\omega(770)$
meson. Now, as is immediately clear from
(\ref{dil-chisy.1}-\ref{dil-chisy.3}) the electromagnetic current
(\ref{dil-chisy.1}) can be written as
\begin{equation}
\label{dil-chisy.4}
J_{\text{em,ud}}^{\mu} = J_{\V}^{3\mu}+\frac{1}{6} j_{\text{V}}^{\mu}.
\end{equation}
To extend this pattern to also include the strange quark\footnote{Since
  also the strange-quark mass is small compared to the typical hadronic
  scale of 1\;\GeV one can extend the approximate chiral
  $\text{SU}(2)_{\LH} \times \text{SU}(2)_{\RH} \times \text{U}(1)_{\V}$
  symmetry of the ud-quark sector of QCD to the uds-quark sector with
  the symmetry group
  $\text{SU}(3)_{\LH} \times \text{SU}(3)_{\RH} \times
  \text{U}(1)_{\V}$. The symmetry is spontaneously broken to
  $\text{SU}(3)_{\V} \times \text{U}(1)_{\V}$ and of course also
  explicitly broken by the quark masses and the electroweak
  interactions. The pseudo-Goldstone bosons are grouped into the
  pseudoscalar SU(3) octet, consisting of the 3 pions, 4 kaons, and one
  $\eta^0$.} we only have to add
$J_{\text{em,s}}^{\mu}=-1/3 \overline{\psi_{\text{s}}} \gamma^{\mu}
\psi_{\text{s}}$ to (\ref{dil-chisy.1}). Then we can write the
electromagnetic current of the three light quarks in the form
\begin{equation}
\label{dil-chisy.5a}
\begin{split}
J_{\text{em,uds}} = \frac{1}{\sqrt{2}} \Bigg [ & \frac{1}{\sqrt{2}} \left (\overline{\psi}_{\text{u}} \gamma^{\mu}
\psi_{\text{u}}-\overline{\psi}_d \gamma^{\mu} \psi_{\text{d}} \right)
  \\
&+ \frac{1}{3 \sqrt{2}}  \left (\overline{\psi}_{\text{u}} \gamma^{\mu}
\psi_{\text{u}}+\overline{\psi}_d \gamma^{\mu} \psi_{\text{d}} \right)
-\frac{\sqrt{2}}{3} \overline{\psi}_{\text{s}} \gamma^{\mu}
  \psi_{\text{s}} \Bigg ].
\end{split}
\end{equation}
It is suggestive to associate the three terms in the bracket with the
corresponding light vector mesons $\rho$, $\omega$, and $\phi$. The
relative weights in the electromagnetic current correlation function
is thus 9:1:2. Empirically the partial decay widths of the light vector
mesons to dielectrons are
$\Gamma_{\rho \rightarrow \e^+ \e^-}/\Gamma_{\omega \rightarrow \e^+
  \e^-} \simeq 10.5$ and
$\Gamma_{\phi \rightarrow \e^+ \e^-} /\Gamma_{\omega \rightarrow \e^+
  \e^-} \simeq 1.9$ \cite{pdb14}, which is not too far from the naive
parton argument based on (\ref{dil-chisy.5a}). 

In hadronic models the assumption that the hadronic electromagnetic
current is proportional to the neutral vector-meson fields, the
so-called \textbf{vector-meson dominance model (VMD)}
\cite{sak60,gounaris:1968,klz67} leads to a quite successful description
of hadronic electromagnetic transition form factors, particularly the
electromagnetic form factor of the pion.

\begin{figure}
\centerline{\includegraphics[width=0.75 \linewidth]{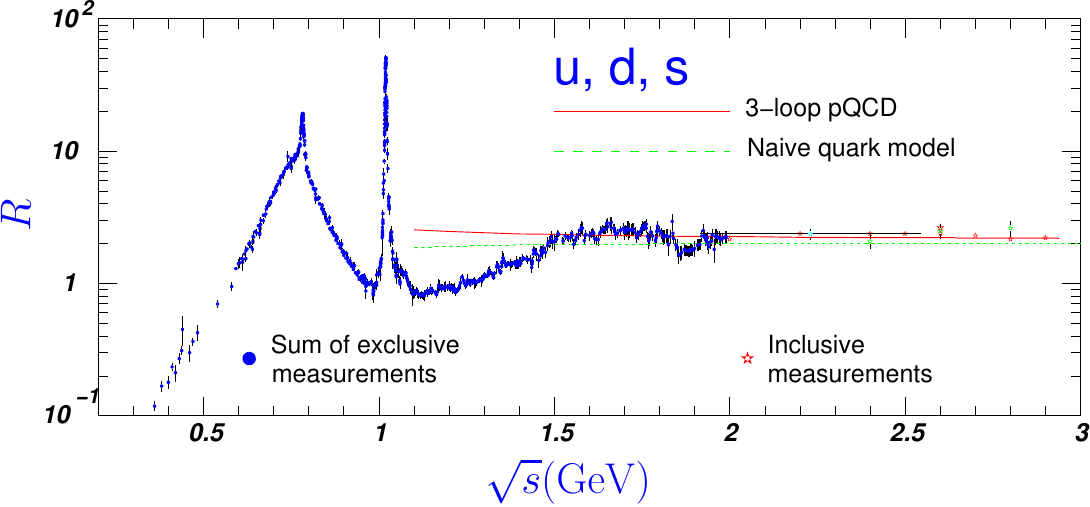}}
\caption{The ratio
  $\sigma_{\ee^+ + \ee^- \rightarrow \text{hadrons}}/\sigma_{\ee^+ + \ee^-
    \rightarrow \mu^+ + \mu^-}$
  in the low- and intermediate-mass region. Figure taken from
  \cite{pdb06}.}
\label{fig.R-epl-emi-to-hadrons}
\end{figure}

Very accurate measurements of the hadronic em.\
current-current-correlation function are provided by the inclusive
hadron production in the reaction
$\e^+ + \e^- \rightarrow \text{hadrons}$, which is usually depicted in
terms of the ratio
\begin{equation}
\label{dil-chisy.5b}
R=\frac{\sigma_{\e^+ + \e^- \rightarrow \text{hadrons}}}{\sigma_{\e^+ + \e^-
    \rightarrow \mu^+ + \mu^-}},
\end{equation}
as shown in Fig.~\ref{fig.R-epl-emi-to-hadrons} as a function of the
invariant mass of the electron-positron pair $M=\sqrt{s}$. The low-mass
region $2m_{\e} \leq M \lesssim m_{\phi}$ is dominated by the three
light vector mesons, $\rho$, $\omega$, and $\phi$, followed in the
intermediate-mass region $m_{\phi} \lesssim M \lesssim M_{J/\psi}$ by a
broader vector-meson resonance $\rho'$ and a continuum that is
well-described in the naive parton model, where the ratio is given by
\begin{equation}
\label{dil-chisy.6}
R_{\text{parton model}} = N_c \sum_{f \in \{\text{u},\text{d},\text{s}
  \}} q_f^2 = 3 \left (\frac{4}{9}+2 \cdot \frac{1}{9} \right ) = 2.
\end{equation}
From the electroweak sector of the Standard Model it is known that the
vector and axial-vector strong-isovector current-correlation functions
are directly related to the charged electroweak current, which is of the
clean ``$V-A$ structure'' (see, e.g., \cite{Nachtmann:1990}). Taking
into account parity conservation of the strong interaction it is clear
that the semileptonic decay $\tau \rightarrow \nu_{\tau}+\text{pions}$
of $\tau$-leptons together with the known weak coupling constant (or
equivalently the Fermi constant in the effective four-fermion model) and
the relevant CKM-matrix element $|V_{ud}|$ allows for an accurate
quantitative separation of the current-correlation functions into the
isovector- and axial-vector channel by exclusive measurements of the
partial $\tau$-lepton decay widths into an even or odd number of pions
respectively.
\begin{figure}
\centering
\begin{minipage}[b]{0.55\linewidth}
\includegraphics[width=\textwidth]{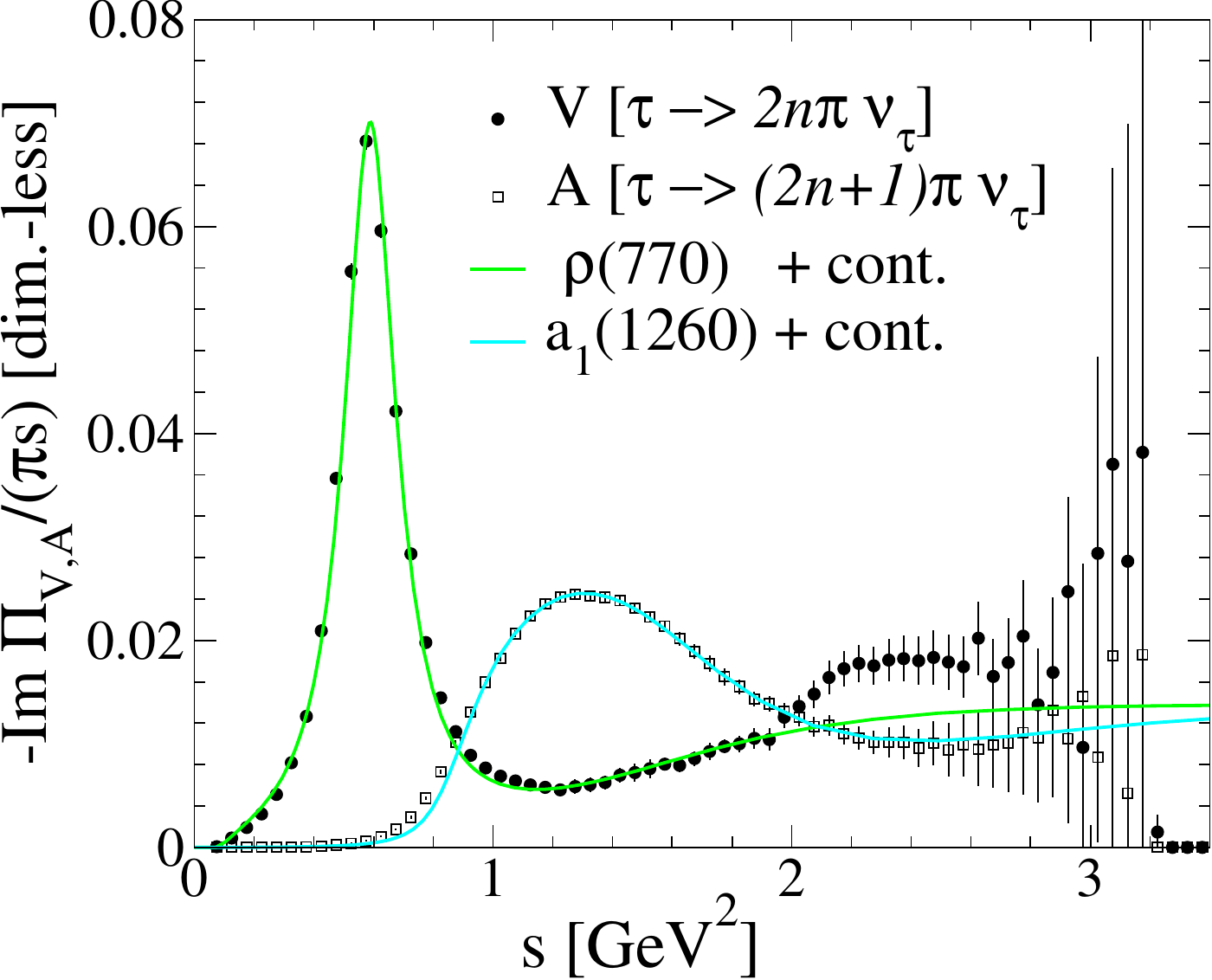}
\end{minipage}\hfill
\begin{minipage}[b]{0.43\linewidth}
\includegraphics[width=0.9\textwidth]{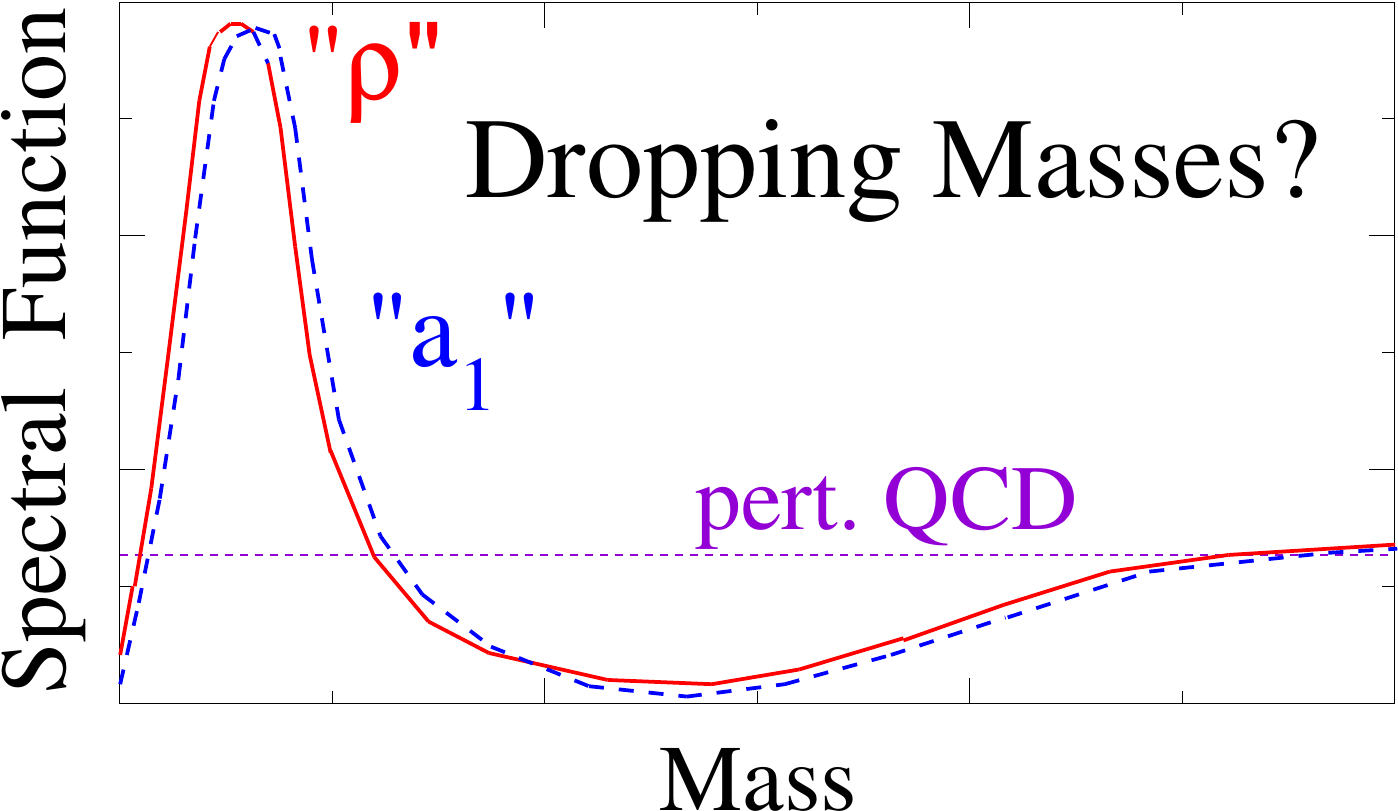}\\
\includegraphics[width=0.9\textwidth]{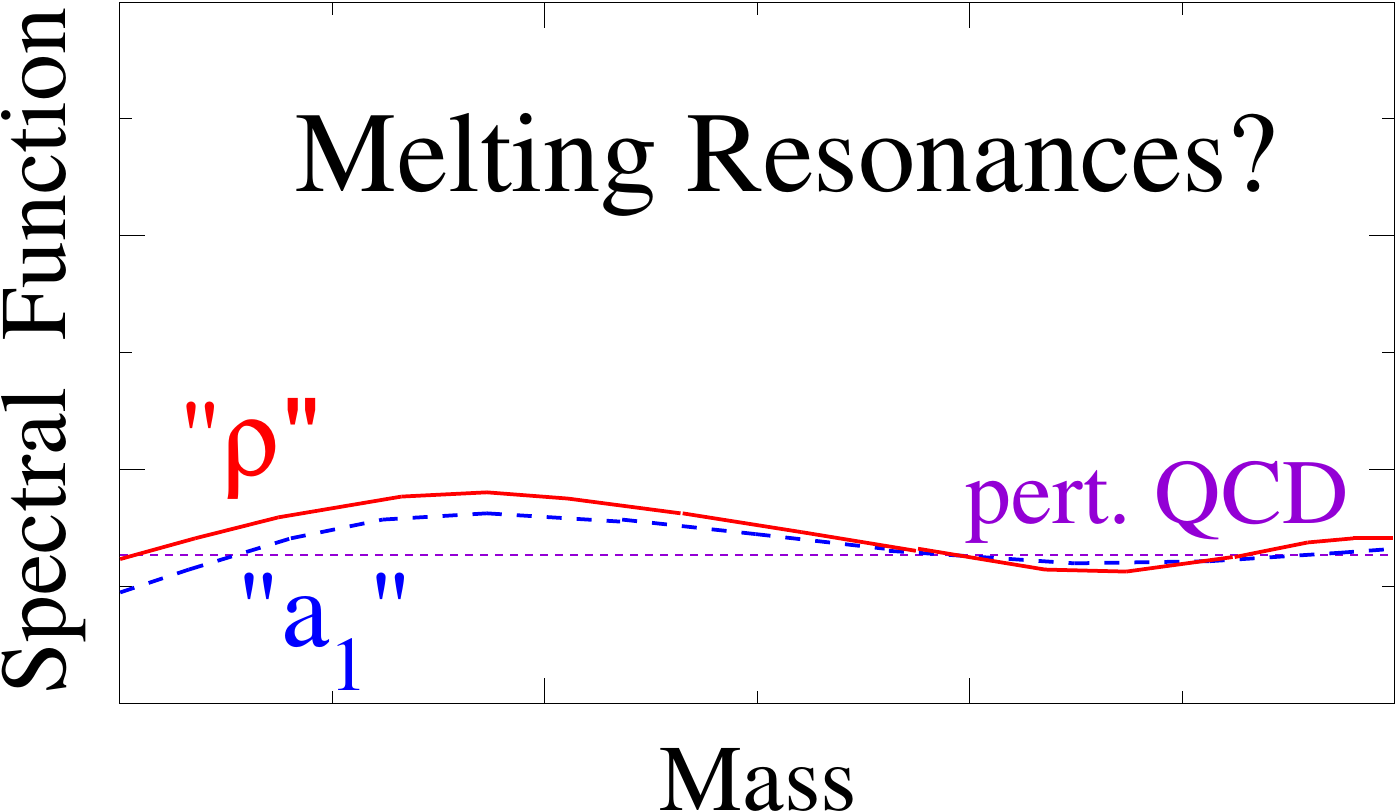}
\end{minipage}
\caption{Left panel: The isovector-vector and -axial-vector current
  correlation functions as a function of the total center-momentum
  energy of the pions ($s=E_{\mathrm{cm}}^2$) as extracted from
  $\tau$-lepton decay to an even and odd number of pions \cite{aleph98}
  together with model fits with vacuum $\rho$- and $\text{a}_1$-meson
  spectral functions and a perturbative continuum \cite{rap02b}. Right
  panel: Possible scenarios for the effect of chiral-symmetry
  restoration to the vector-axial-vector spectral functions in the hot
  and dense medium. Figure taken from {\protect \cite{Rapp:2009yu}}.}
\label{fig.tau-decay-data}
\end{figure}
As shown in Fig.~\ref{fig.tau-decay-data} these data can be interpreted
as an accurate demonstration of the spontaneous breaking of the
(approximate) chiral symmetry of QCD in the light-quark sector.

One of the fundamental questions addressed with the accurate measurement
of dileptons in heavy-ion collisions is to learn about how this chiral
symmetry of QCD is realized at low energies, i.e., in which way the
dynamical generation of hadron masses by the strong interaction comes
about, at least for the $\rho$ meson, which dominates the
electromagnetic current-current correlation function in the low-mass
region, as explained above. As the detailed analysis of effective
hadronic models shows, chiral symmetry of the light vector bosons can be
realized in (at least) two possible ways. E.g.\ in terms of
``hidden-local-symmetry models'', which describe the massive vector
($\rho$) and in its generalized version also the axial-vector
($\text{a}_1$) meson with a ``Higgsed'' additional local chiral gauge
symmetry (usually realized in a non-linear way)
\cite{Bando:1984ej,Bando:1987br,Harada:2003jx}. It can be realized
either by introducing only the $\rho$-meson as a gauge field,
corresponding to the unbroken part $\mathrm{SU}(2)_{\V}$ of the
$\mathrm{SU}(2)_{\LH} \times \mathrm{SU}(2)_{\RH}$ or both a $\rho$ and
an $\mathrm{a}_1$ meson. In this kind of models chiral symmetry can be
realized in the so-called vector manifestation, where in the model
introducing only the $\rho$-meson the longitudinal component of the
$\rho$ meson becomes the chiral partner of the pions, which leads to a
dropping mass towards the chiral phase transition. On the other hand,
even in the same class of models, also the realization of chiral
symmetry is possible, where the chiral partners are $\rho$ and $a_1$,
and the mass spectra of these two mesons become degenerate by a large
broadening of their spectral functions in the medium
\cite{Harada:2005br,Harada:2008hj}.

\section{Models for electromagnetic probes in heavy-ion collisions}
\label{sect:dileps-hics}

To address the production of dileptons in heavy-ion collisions, the
medium modification of the em.\ current-current correlator entering the
production rate as discussed in Sect.\ \ref{sect:McLerranToimela}, cf.\
(\ref{mtform.8a}) has to be evaluated. At higher
collision energies the produced matter becomes hot and dense enough to
enter the deconfined phase of a quark-gluon plasma and then evolves as
an expanding and cooling fireball undergoing the transition to a hot and
dense hadron-resonance gas that finally decouples to freely streaming
hadrons observed in the detectors.

As a comparison of the found particle abundances and spectra with
relativistic hydrodynamic simulations shows, that the hot and dense
fireball in this evolution is well described by a medium close to local
thermal equilibrium. This allows the use of the equilibrium in-medium
electromagnetic current-current correlation function (\ref{mtform.8a})
to describe the dilepton-production rate in heavy-ion collisions. Since
the electromagnetic probes are emitted during the entire evolution of
the medium, both a detailed description of this collective dynamics of
the medium as well as the production rates are necessary over a wide
range of temperatures and baryochemical potentials. In the following we
first briefly describe the quantum-field theoretical models for the
in-medium production rates and then the bulk-evolution models.

\subsection{Electromagnetic radiation from the QGP}
\label{sect:em-rad-qgp}

\begin{figure}[t]
\centering
\includegraphics[width=0.6\linewidth]{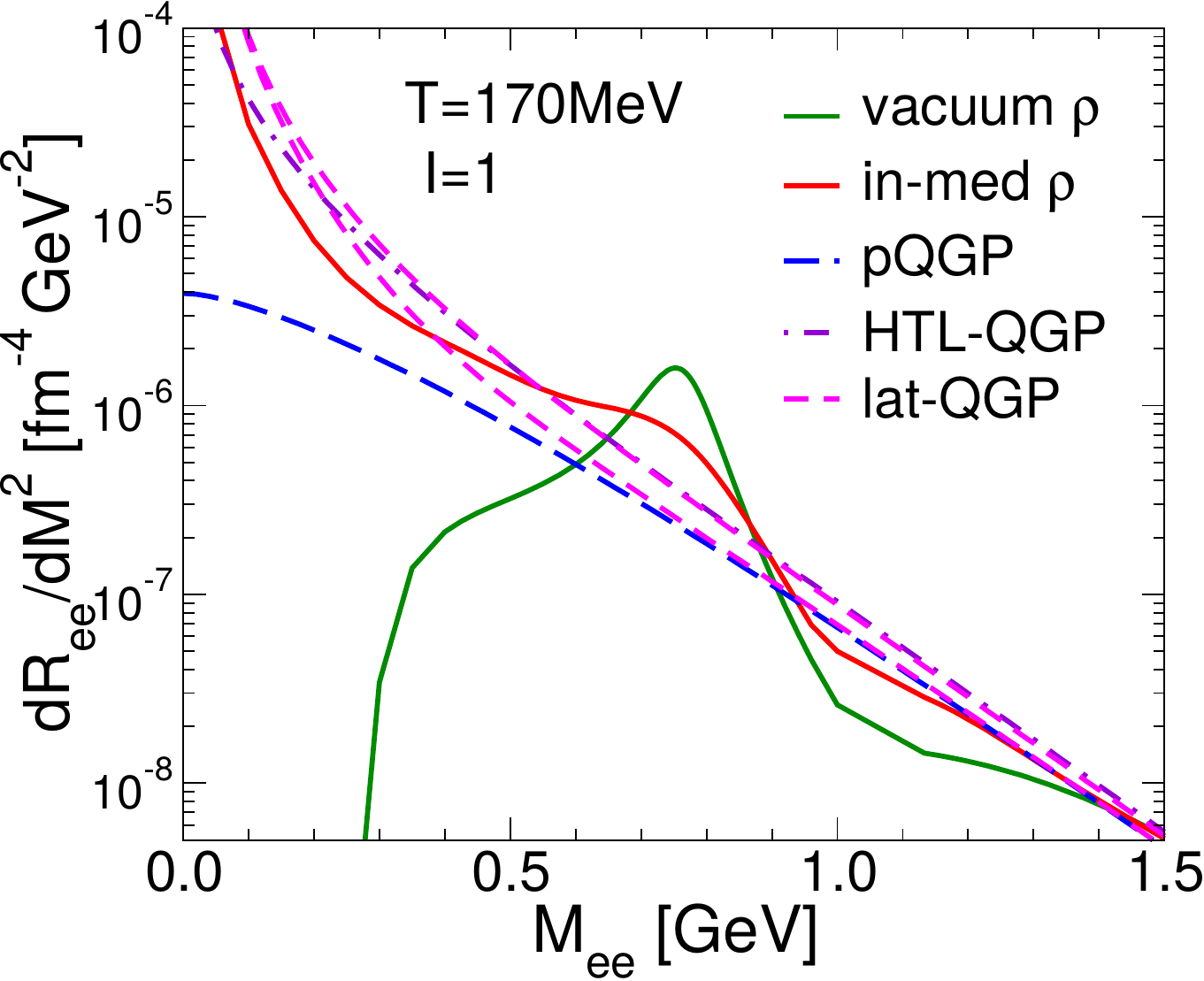}
\caption{Dilepton-emission rate as a function of invariant mass at $T=170
  \; \MeV$ from $q\overline{q} \rightarrow \ell^+ \ell^-$ annihilation in
  leading order perturbation theory (blue line), a leading-order
  hard-thermal loop calculation (violet line), and a lattice-QCD (red
  line) in comparison to the calculation from hadronic effective theory
  with and without medium effects. Fig.\ taken from
  \cite{Rapp:2013nxa}.}
\label{fig:htl-lqgp-rate-comparison}
\end{figure}
At leading order (LO) in $\alpha_{\text{em}}$, the basic process for
dilepton production in the partonic phase is $q$-$\overline{q}$
annihilation, $q+\overline{q} \rightarrow \ell^-+\ell^+$. In terms of
the current-current correlation function, which in quantum-field
theoretical view is just given by the photon polarization function, this
refers to the one-loop diagram,

\begin{equation*}
\parbox{0.3 \linewidth}{\includegraphics[width=\linewidth]{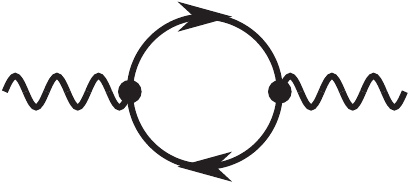}}\;.
\end{equation*}

Its evaluation leads to the dilepton production rate from the QGP at
temperature $T$ and quark-chemical potential
$\mu_{\text{q}}=\mu_{\text{B}}/3$ \cite{Cleymans:1986na}.
\begin{equation}
\begin{split}
\label{qgp-dil.1}
\im \Pi_{\text{em}}^{(\text{QGP})}(k) &= -\sum_f q_f^2 \frac{N_c}{12 \pi}
\frac{T k^2}{\left |\vec{k} \right|} \ln \left
  (\frac{\{x_-+\exp[-(k^0+\mu)/T]\}[x_+ +\exp(-\mu/T)]}{\{x_++\exp[-(k^0+\mu)/T]\}
    [x_-+\exp(-\mu/T)]} \right) \\
&= \frac{C_{\text{em}} N_c}{12 \pi} M^2 \hat{f}_2(k_0,k;T)
\end{split}
\end{equation}
with
\begin{equation}
\label{qgp-dil.2}
x_\pm=\exp \left (-\frac{E_{\pm}}{T} \right), \quad E_{\pm} =
\frac{1}{2} (k^0 \pm \left |\vec{k} \right|).
\end{equation}
However, at lower invariant dilepton masses the rate is tremendously
enhanced by taking into account the leading-order $\alpha_{\text{s}}$
corrections employing the hard-thermal-loop (HTL) resummation techniques
to dress the quark propagators \cite{bpy90}: Although both, quarks and
gluons, acquire a thermal mass $\sim gT$ bremsstrahlung contributions
within the HTL resummation leads to an enhancement of the
dilepton-production rates down to the two-lepton threshold (cf.\ Fig.\
\ref{fig:htl-lqgp-rate-comparison}). 

The dilepton-production rates have also been calculated in thermal
lattice QCD \cite{Ding:2010ga,Brandt:2012jc,Ding:2013qw}. Here the
Euclidean-time vector-current correlation function,
\begin{equation}
\label{qgp-dil.3}
\Pi_B(\tau,q;T) = \int_0^{\infty} \frac{\dd q_0}{2 \pi} \rho_V(q_0,q,T)
\frac{\cosh[q_0(\tau-1/2T)]}{\sinh(q_0/2T)}, 
\end{equation}
is evaluated in quenched QCD for $T=1.45 T_{\text{c}}$ at $q=0$. The
spectral function is obtained from these numerical results by fitting
the parameters $\Gamma$ and $\kappa$ in the ansatz
\begin{equation}
\label{qgp-dil.4}
\rho_V(q_0)=S_{\text{BW}} \frac{q_0 \Gamma/2}{q_0^2+\Gamma^2/4} +
\frac{C_{\text{em}} N_{\text{c}}}{2\pi} (1+\kappa) q_0^2 \tanh(q_0/2T) 
\end{equation}
for the spectral function to the numerical results for
(\ref{qgp-dil.3}).

To use the corresponding dilepton rates in calculations to describe
heavy-ion collisions an extrapolation to finite three-momentum is
needed. In \cite{Rapp:2013nxa} such an extrapolation is provided by
using the transverse part of the electromagnetic spectral function from
the leading-order pQCD photon rate for the three-momentum dependence of
the spectral function (\ref{qgp-dil.4}), replacing the Breit-Wigner
piece. This finally leads to the parameterization
\begin{equation}
\begin{split}
\label{qgp-dil.5}
\im \Pi_{\text{T}} &= -\frac{C_{\text{em}}}{12 \pi} M^2 \left
  [\hat{f}_2 \left (k_0,\left |\vec{k} \right|;T \right) + 2 \pi \alpha_{\text{s}} \frac{T^2}{M^2} K F(M^2)
  \ln \left (1+\frac{2.912 k_0}{4 \pi \alpha_{\text{s}} T} \right) \right] \\
&= \frac{C_{\text{em}} N_c}{12 \pi} M^2 \left
  [\hat{f}_2 \left(k_0,\left |\vec{k} \right|;T \right) +
  Q_{\text{lat}}^{(\text{T})} \left(k_0,\left |\vec{k} \right| \right) \right].
\end{split}
\end{equation}
Here $K=2$ is introduced to account for the enhancement of the photon
rate over the LO rate and to better reproduce the low-energy limit of
the lattice-QCD spectral function. To accommodate the behavior at higher
energies, an additional form factor $F(M^2)=\Lambda^2/(\Lambda^2+M^2)$
with $\Lambda =2 T$ has been used. In Fig.\
\ref{fig:htl-lqgp-rate-comparison} the upper (lower) dashed line
corresponds to this parameterization with (without) this form
factor. Finally one has to also reconstruct the longitudinal part of the
current-correlation function. This is achieved by using the standard
construction of gauge-invariant s-wave $\rho$-baryon interactions,
$\Pi_L=(M^2/q_0^2) \Pi_T$ \cite{Rapp:1999ej}.

\subsection{Electromagnetic radiation from a hot/dense hadron gas}
\label{sect:em-rad-had}

At lower temperatures and densities the low-mass dilepton spectrum is
governed by the in-medium spectral functions of the light vector mesons,
$\rho$, $\omega$, and $\phi$ in the sense of the vector-meson dominance
model. For the description of this contribution to the dilepton rate in
heavy-ion collisions we use the ``Rapp-Wambach Model'' based on
\cite{Rapp:1995zy,Rapp:1997fs,ubrw98,rubw98,gale-rapp99} as summarized
in \cite{Rapp:1999us}. The evaluation of the spectral properties of the
vector mesons in a dense and hot hadronic medium consists in the
determination of their self-energies in thermal and chemical
equilibrium. Here we concentrate on the $\rho$ meson. Microscopically
the corresponding in-medium contributions consist (i) the modification
of the pion loop in the $\rho \pi \pi$ loop and
(ii) via direct couplings of the $\rho$ meson to both mesons and
baryons. As will become clear, particularly the baryon contributions
play a crucial role in describing the dilepton spectra in heavy-ion
collisions at all energies. Although at the highest available collision
energies at RHIC and LHC the net-baryon density is low
($\mu_{\text{B}} \simeq 0$), the total baryon density
$n_{\text{B}}+n_{\overline{\text{B}}}$ is high, and thus leads to
significant modifications of the $\rho$-meson spectral function,
contributing particularly in the low-mass tail of the dilepton
invariant-mass spectrum.

The $\rho \pi \pi$ interaction is based on the interaction Lagrangian
\cite{Rapp:1995zy,Rapp:1997fs}
\begin{equation}
\label{had-dil.1}
\Lag_{\rho \pi \pi}=g_{\rho \pi \pi} (\vec{\pi} \times \partial^{\mu}
\vec{\pi}) \cdot \vec{\rho}_{\mu}
\end{equation}
with $\vec{\pi}$ and $\vec{\rho}_{\mu}$ the isovector pion and isovector
$\rho$-meson fields, respectively. To account for medium modifications
of the pion, the pion self-energy $\Sigma_{\pi}$ is evaluated employing
particle-hole excitations \cite{ew88,Rapp:1993bi}, where the pions
interact with nucleons and $\Delta$ resonances through particle-hole
excitations of the type $NN^{-1}$, $\Delta N^{-1}$, $N \Delta^{-1}$, and
$\Delta \Delta^{-1}$. In order to guarantee the transversality of the
$\rho$ self-energy the corresponding vertex corrections to restore the
pertinent Ward-Takahashi identities have to be taken into account
\cite{ubrw98}. The needed $\rho N$ and $\rho\pi N$ couplings are
obtained from the pionic Lagrangian by minimal coupling to the $\pi N$
couplings,
\begin{equation}
\begin{split}
\label{had-dil.2}
\Lag_{\rho N} &= -\frac{g}{2} \overline{\psi} \fslash{\rho} \tau_3 \psi,
\\
\Lag_{\rho \pi N} &= \ii g \frac{f_N}{m_{\pi}} \overline{\psi} \gamma^5
\fslash{\rho} \vec{\tau} \psi \cdot T_3 \vec{\pi},
\end{split}
\end{equation}
and the $\pi \Delta$ couplings,
\begin{equation}
\begin{split}
\label{had-dil.3}
\Lag_{\rho \Delta} &= g \overline{\psi}_{\mu} \fslash{\rho} T_3^{(3/2)}
\psi^{\mu} - \frac{g}{3} \overline{\psi}_{\mu} (\gamma^{\mu} \rho_{\nu}
+ \gamma_{\nu} \rho^{\mu}) T_3^{(3/2)} \psi^{\nu} \\
& \quad  + \frac{g}{3}
\overline{\psi}_{\mu} \gamma^{\mu} \fslash \rho T_3^{(3/2)} \gamma_{\nu}
\psi^{\nu},\\
\Lag_{\rho \pi N \Delta} &= -\ii g \frac{f_{\Delta}}{m_{\pi}}
\overline{\psi} T^{\dagger} \psi_{\mu} \rho^{\mu} \cdot T_3 \vec{\phi} +
\text{h.c.}
\end{split}
\end{equation}
One way to fix the various parameters in the model is to aim at a
description of data on photon absorption on nucleons and nuclei. For
that purpose the direct coupling of the $\rho$ meson to various baryon
resonances has to be addressed \cite{rubw98}. The corresponding
interaction Lagrangians are given by p-wave couplings of positive parity
states and s-wave couplings of negative-parity states to the $\rho N$
system, which read in the here employed non-relativistic limit
\begin{equation}
\begin{split}
\label{had-dil.4}
\Lag_{\rho BN}^{(\text{p-wave})} &= \frac{f_{\rho BN}}{m_{\rho}}
\Psi_B^{\dagger} (\vec{s} \times \vec{q}) \cdot \vec{\rho}_a t_a \Psi_N +
\text{h.c.}, \\
\Lag_{\rho BN}^{(\text{s-wave})} &= \frac{f_{\rho BN}}{m_{\rho}}
\Psi_B^{\dagger} (q_0 \vec{s} \cdot \vec{\rho}_a - \rho_a^0 \vec{s} \cdot
\vec{q} ) t_a \Psi_N+\text{h.c.}
\end{split}
\end{equation}
Here, $\vec{s}$ denote spin (transition) operators for $J=1/2$ and $J=3/2$ and $t$
the isospin (transition) operators for $I=1/2$ and $I=3/2$, depending on the quantum
numbers of the baryons, $B=\text{N}(939)$, $\Delta(1232)$, $\text{N}(1720)$
(positive parity) as well as $B=\text{N}(1520)$, $\Delta(1620)$,
$\Delta(1700)$ (negative parity). For the spin-5/2 $\Delta(1905)$ a
tensor coupling of the type $(R_{ij} q_i \rho_{j,a} T_a)$ is employed.

\begin{figure}[t]
\includegraphics[width=0.46 \linewidth]{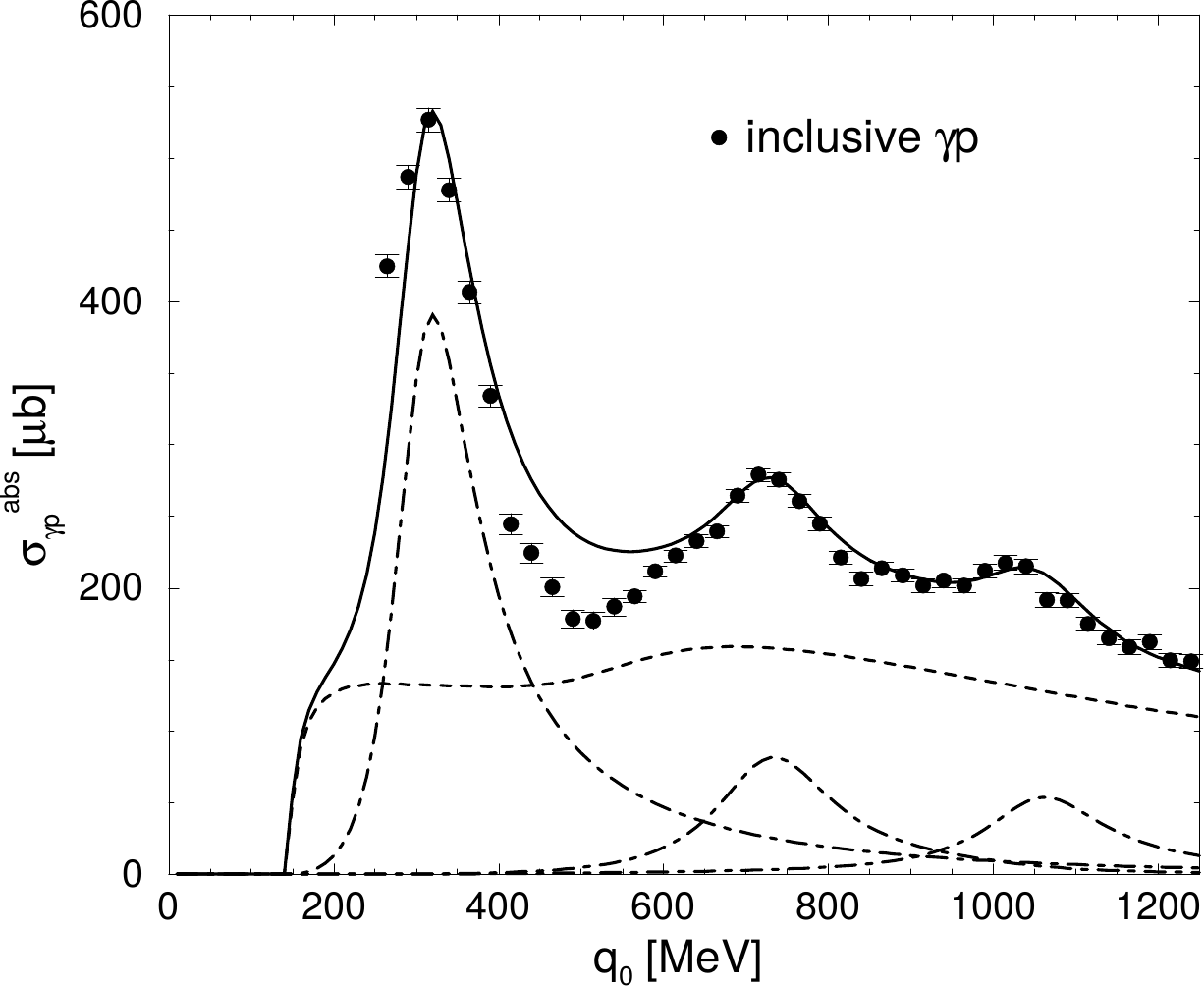}
\hfill
\includegraphics[width=0.46 \linewidth]{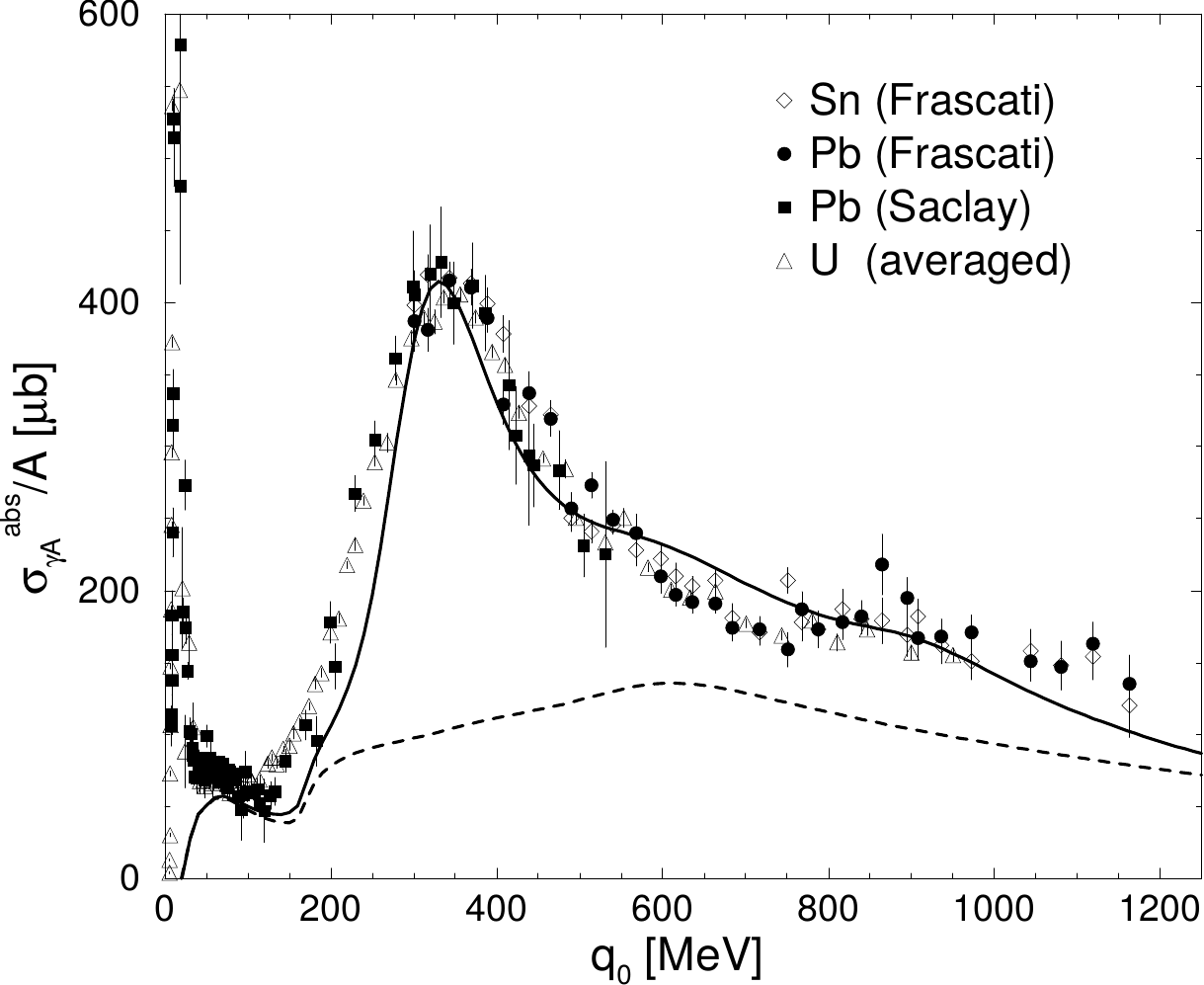}
\caption{Left panel: Total photoabsorption cross section on the proton. The solid
line represents the full result of the fit of the model; the dashed
lines show the $\pi \pi$ non-resonant background and the three dominant
$\rho N$ resonances $\Delta(1232)$, N(1520), and N(1720). The data are
from \cite{Armstrong:1971ns}. Right panel: Total photoabsorption cross section
on various nuclei. The solid line represents the full result of the
in-medium model calculation and the short-dashed line the $\pi \pi$
background, both calculated at a nuclear density $\overline{\varrho}_{\text{N}}=0.8
\varrho_0$. The long-dashed line corresponds to a linear-density
approximation, reflecting the line shape of the left figure. The data
are taken from
\cite{Lepretre:1978me,Ahrens:1984ve,Ahrens:1986hn,Frommhold:1992um,Bianchi:1992ze,
  Bianchi:1995vb}. The figures are taken from \cite{rubw98}.}
\label{fig.photoabs-urban}
\end{figure}

The in-medium self-energies from the resulting baryon nucleon-hole loop
diagrams are of the form
\begin{equation}
\label{had-dil.5}
\Sigma_{\rho \alpha}^{(0),T}(q_0,q) = - \left (\frac{f_{\rho \alpha}
    F_{\rho \alpha}(q)}{m_{\rho}} \right) \text{SI}(\rho\alpha) Q^2 \phi_{\rho \alpha}(q_0,q),
\end{equation}
where $Q^2=q^2,q_0^2$ for p- and s-wave couplings, respectively. Also a
monopole form factor
$F_{\rho \alpha}(q)=\Lambda_{\rho}^2/(\Lambda_{\rho}^2+q^2)$ has been
introduced, and $\text{SI}(\rho \alpha)$ denotes the spin-isospin factor;
$\phi_{\rho \alpha}(q_0,q)$ is the Lindhard function corresponding to
the one-loop self-energy diagram.
\begin{table}
\centering
\begin{tabular}{ccccccc}
 B & $l_{\rho N}$ & $\text{SI}(\rho BN^{-1})$ & $\Gamma^0_{\rho N}$ (MeV) &  
$\left(\frac{f_{\rho BN}^2}{4\pi}\right)_{\text{est}}$ & 
$\left(\frac{f_{\rho BN}^2}{4\pi}\right)_{\text{fit}}$ & 
$\Gamma^{\text{med}}$ [MeV] \\
\hline
N(939)         & p & 4    & --   & 4.68  & 5.8  & 0  \\
$\Delta$(1232) & p & 16/9 & --   & 18.72 & 23.2 & 15   \\
$N$(1520)      & s &  8/3 & 24   & 6.95  & 5.5  & 250 \\
$\Delta$(1620) & s &  8/3 & 22.5 & 1.01  & 0.7  & 50  \\
$\Delta$(1700) & s & 16/9 & 45   & 1.2   & 1.2  & 50  \\
$N$(1720) & p &  8/3 & 105  & 8.99  & 9.2  & 50  \\
$\Delta$(1905) & p &  4/5 & 210  & 17.6  & 18.5 & 50  \\
\end{tabular}
\caption{Properties of the $\rho BN$ vertices as derived from the 
  interaction Lagrangians (\ref{had-dil.4}); columns from left to
  right: baryon resonance, relative angular momentum in the $\rho N$
  decay, spin-isospin factor, partial decay width into $\rho N$
  as extracted from \cite{Effenberger:1996im,Peters:1997va}, 
  coupling constant as estimated from $\Gamma^0_{\rho N}$ (for N(939) and 
  $\Delta$(1232) we have indicated the values from the BONN 
  potential \cite{Machleidt:1987hj} which uses  
  somewhat harder form factors), coupling constant as in the fit to
  photoabsorption data, in-medium correction to the total decay width
  (cf. Fig.~\ref{fig.photoabs-urban}). Table taken from \cite{rubw98}.}
\label{tab-urban.1}
\end{table}
The photoabsorption cross section within the strict vector-meson
dominance model reads
\begin{equation}
\label{had-dil.6}
\frac{\sigma_{\gamma A}^{\text{abs}}}{A}=-\frac{4\pi\alpha_{\text{em}}}{q_0} \ 
\frac{(m_\rho^{(0)})^4}{g^2} \frac {1}{\rho_N} \im D_\rho^T(q_0,\vec q ;\rho_N).
\end{equation}
On the other hand it is known that this strict realization of the
vector-meson dominance ansatz tends to overestimate the
$B \rightarrow N\gamma$ branching fractions with the hadronic couplings
determined from the corresponding $B \rightarrow \rho N$ decay
widths. This can, however, be corrected by using the extended
realization of VMD in \cite{klz67}, which allows to adjust the
BN$\gamma$ coupling at the photon point independently \cite{fp97}. As
shown in Fig.\ \ref{fig.photoabs-urban}, making use of this freedom a
satisfactory fit to the data on photoabsorption on protons as well as
nuclei can be achieved, resulting in the values of the various coupling
constants given in Table~\ref{tab-urban.1}.

Finally the in-medium modifications of the $\rho$ meson due to direct
interactions with various mesons has to be taken into account
\cite{gale-rapp99}. At temperatures relevant for the hadron-gas phase of
the medium created in heavy-ion collisions the most abundant hadrons are
the pseudoscalar pseudo-Goldstone mesons $P=\pi$, K. So the heavier
mesons can be treated as ``$\rho P$ states'', i.e., vector
($V$) and axial-vector mesons ($A$). The corresponding interaction
Lagrangians, obeying chiral symmetry and (via the VMD conjecture)
electromagnetic gauge invariance read
\begin{equation}
\begin{split}
\label{rho-meson-int.1}
\Lag_{\rho PA} &= G_{\rho PA} A_{\mu} (\eta^{\mu \nu} q_{\alpha}
p^{\alpha}-q^{mu} p^{\nu}) \rho_{\nu} P,\\
\Lag_{\rho P V} &= G_{\rho P V} \epsilon_{\mu \nu \rho \sigma} k^{\mu}
V^{\nu} q^{\rho} \rho^{\sigma} P, 
\end{split}
\end{equation}
with the four-momenta $p^{\mu}$, $q^{\mu}$, and $k^{\mu}$ of the
pseudoscalar, $\rho$, and vector or axial-vector meson, respectively. The
$\rho P$ scattering through a vector-meson resonance is due to the
Wess-Zumino anomaly.
\begin{table}[t]
\centering
\begin{tabular}{c|ccccc}
 $R$ & $I^GJ^P$ & $\Gamma_{\text{tot}}$ (MeV) & $\rho h$ Decay &
$\Gamma^0_{\rho h}$ (MeV) & $\Gamma^0_{\gamma h}$ (MeV) \\
\hline
$\qquad\omega(782)\qquad$ & $0^-1^-$ & 8.43 & $\rho\pi$ & $\sim 5$ & 0.72 \\
$h_1(1170)$   & $0^-1^+$ & $\sim 360$  & $\rho\pi$ & seen  &   ?  \\
$a_1(1260)$   & $1^-1^+$ & $\sim 400$  & $\rho\pi$ & dominant & 0.64 \\
$K_1(1270)$   & $\frac{1}{2}1^+$ & $\sim 90$ & $\rho K$  & $\sim 60$ &   ?  \\
$f_1(1285)$   & $0^+1^+$        & 25 & $\rho\rho$ & $\le$8   & 1.65  \\
$\pi'(1300)$  & $1^-0^-$        & $\sim 400$ & $\rho\pi$ & seen   & ?  \\
\end{tabular}
\caption{Mesonic Resonances $R$ with masses $m_R\le 1300 \; \MeV$
  and substantial branching ratios
  into final states involving direct $\rho$'s (hadronic)
  or $\rho$-like photons (radiative). Table taken from \cite{gale-rapp99}.}
\label{tab.rapp-gale-mesons.1}
\end{table}

Further the $\rho P$ scattering can also be mediated by a pseudoscalar
resonance, $\rho \pi \rightarrow \pi'(1300)$ with the interaction
Lagrangian
\begin{equation}
\label{rho-meson-int.2}
\Lag_{\rho P P'} = G_{\rho P P'} P' (k \cdot q p_{\mu} - p \cdot q
k_{\mu}) \rho^{\mu} P.
\end{equation}
Finally, there are $\rho VA$ vertices related to anomaly terms,
\begin{equation}
\label{rho-meson-int.3}
\Lag_{\rho VA} = G_{\rho VA} \epsilon_{\mu \nu \rho \sigma} p^{\mu}
V^{\nu} \rho^{\rho \alpha} k_{\alpha} A^{\sigma} -\frac{\lambda}{2}
(k_{\beta} A^{\beta})^2,
\end{equation}
with the field-strength tensor
$\rho_{\mu \nu} = q_{\mu} \rho_{\nu} - q_{\nu} \rho_{\mu}$. The second
term on the right-hand side is a gauge-fixing term for the axial-vector
field, and the Feynman gauge is chosen by setting $\lambda=1$.

The considered vector and axial-vector mesons as well as the heavy
pseudoscalar $\pi'(1300)$ with their corresponding decay properties to
$\rho h$ and $\gamma h$ partial decay widths are summarized in
Table~\ref{tab.rapp-gale-mesons.1}. These widths are calculated with the
above defined interaction vertices, taking into account the finite width
of the $\rho$ meson. Additionally hadronic dipole-form factors,
\begin{equation}
\label{rho-meson-int.4}
F_{\rho PR}(q_{\text{cm}})=\left(\frac{2\Lambda^2_{\rho P}+m_R^2}
{2\Lambda^2_{\rho P}
+\left[\omega_\rho(q_{\text{cm}})+\omega_P(q_{\text{cm}})\right]^2}\right)^2,
\end{equation}
are introduced. 
\begin{table}
\centering
\begin{tabular}{c|ccccc}
 $R$ & $IF(\rho hR)$ & $G_{\rho hR}$ (GeV$^{-1}$) & $\Lambda_{\rho hR}$ (MeV) &
$\Gamma^0_{\rho h}$ (MeV) & $\Gamma^0_{\gamma h}$ (MeV) \\
\hline
$\qquad\omega(782)\qquad$ & 1 & 25.8  & 1000 & 3.5 & 0.72 \\
$h_1(1170)$               & 1 & 11.37 & 1000 & 300 & 0.60 \\
$a_1(1260)$               & 2 & 13.27 & 1000 & 400 & 0.66 \\
$K_1(1270)$               & 2 & 9.42  & 1000 &  60 & 0.32 \\
$f_1(1285)$               & 1 & 35.7  &  800 &   3 & 1.67  \\
$\pi'(1300)$               & 2 & 9.67  &  1000 & 300 & 0  \\
\end{tabular}
\caption{Results of the fit to the decay properties of $\rho$-$h$
induced mesonic resonances $R$ with masses $m_R\le 1300 \; \MeV$ (the
$f_1(1285)$ and $\pi'(1300)$ coupling constants are in units of
$\GeV^{-2}$). $IF(\rho hR)$ denotes the isospin factor in the
decay-matrix element. Table taken from \cite{gale-rapp99}.}
\label{tab.rapp-gale-mesons.2}
\end{table}

\begin{figure}[t]
  \raisebox{-0.5\height}{\includegraphics[width=0.45
    \linewidth]{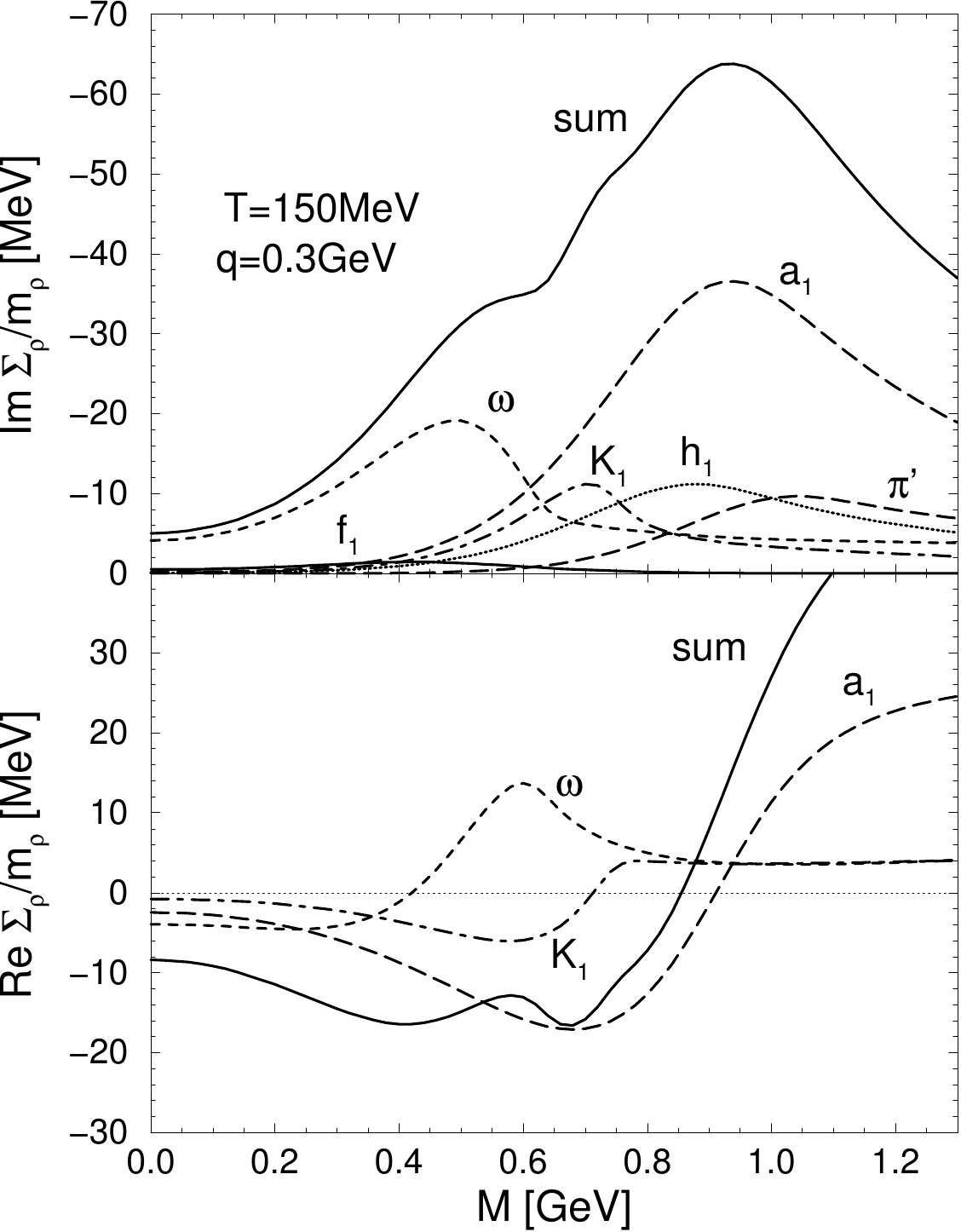}} \hfill
  \raisebox{-0.5\height}{\includegraphics[width=0.45
    \linewidth]{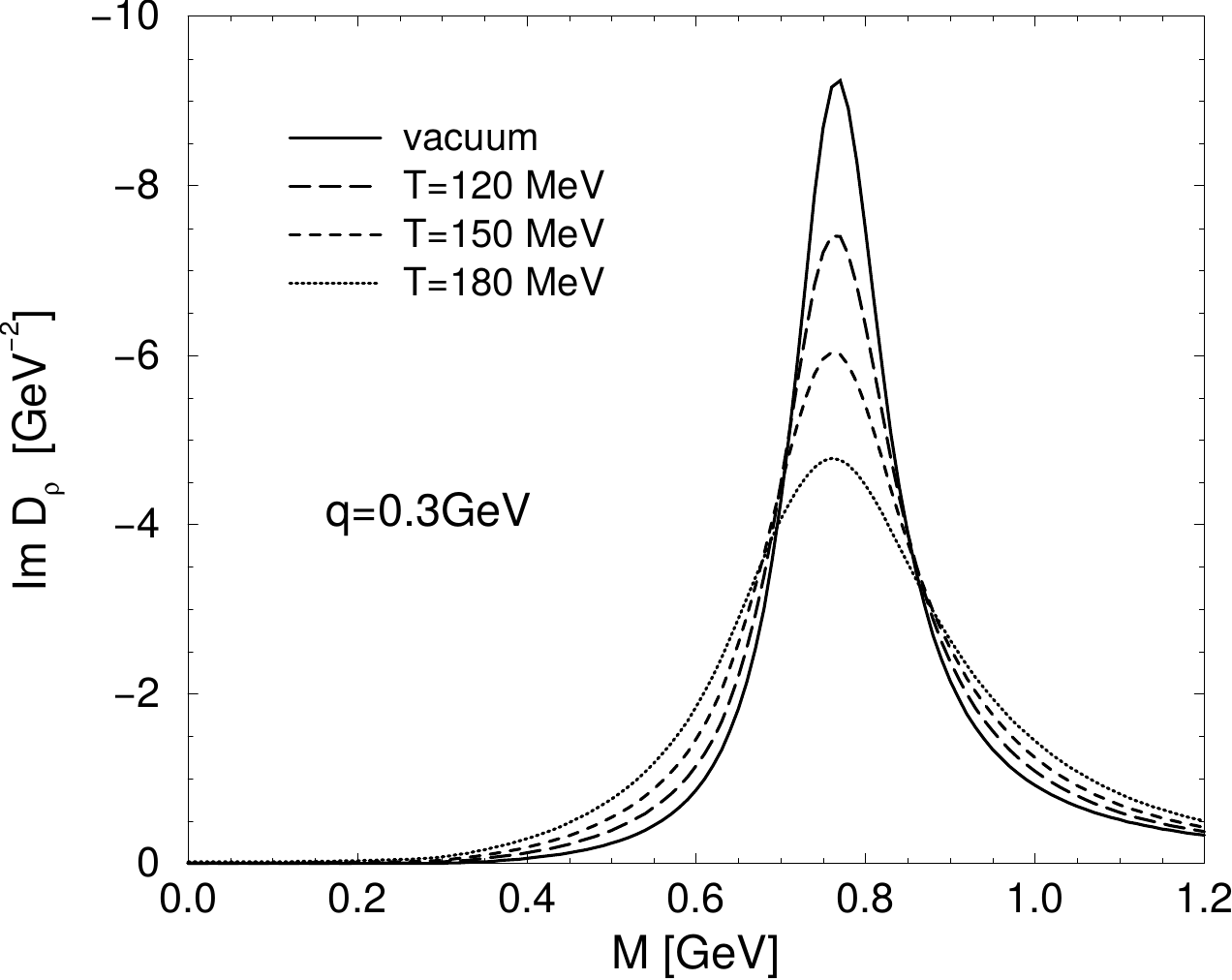}} 
  \caption{The polarization-averaged in-medium $\rho$-meson self-energy
    (left) and spectral function (right). Figures taken from
    \cite{gale-rapp99}.}
\label{fig.rapp-gale-mesons.2}
\end{figure}

The medium modifications of the $\rho$-meson self-energy and the
corresponding spectral function is then evaluated with these model
parameters. As is nicely demonstrated in
Fig.~\ref{fig.rapp-gale-mesons.2}, any process adds to the imaginary
part of the self-energy, i.e., the $\rho$-meson width, while the
contribution to the real part around the vacuum mass can be positive or
negative, depending on whether the effective interaction due to the
involved meson resonances is repulsive or attractive, respectively. The
net result is a considerable broadening of the $\rho$-meson's spectral
shape with only moderate mass shifts.
The same qualitative feature is also seen when taking all the in-medium
effects on the $\rho$-meson self-energy into account, i.e., the
modification of the pion cloud in the $\rho \pi \pi$ loop as well as the
just discussed direct interactions of the $\rho$ with mesonic and
baryonic resonances, one finds a tremendous broadening of the spectral
function with only small mass shifts \cite{Rapp:1999us}. It is important
to note that the baryons provide a lot of spectral strength in the
low-mass tail, which within the VMD model leads to a considerable
enhancement of the dilepton-production rate down to the $\ell^+ \ell^-$
threshold. As already shown in Fig.~\ref{fig:htl-lqgp-rate-comparison}
the full in-medium dilepton-production rates close to the
pseudo-critical temperature for the chiral phase transition, obtained by
extrapolating the pertinent partonic rates down and the hadronic rates
up to temperatures around $T_{\text{c}} \simeq 160 \; \MeV$ become very
similar, hinting at a restoration of chiral symmetry through ``melting
of the vector-meson resonances'' merging smoothly into the corresponding
QCD continuum.

In the intermediate-mass region,
$m_{\phi} \lesssim M_{\ell^+ \ell^-} \lesssim m_{J/\psi}$, medium
modifications can be addressed approximately using a
low-temperature/density expansion, leading to a mixing of the vector and
axial-vector current-correlation functions via the presence of thermal
pions \cite{Dey:1990ba},
\begin{equation}
\begin{split}
\label{chiral-mixing.1}
\Pi_V(q) &= (1-\epsilon) \Pi_V^{\text{vac}}(q) + \epsilon
\Pi_A^{\text{vac}}(q), \\
\Pi_A(q) &= \epsilon \Pi_V^{\text{vac}}(q) + (1-\epsilon) \Pi_A^{\text{vac}}(q).
\end{split}
\end{equation}
The mixing coefficient $\epsilon$ is given by pion tadpole diagrams via
a thermal loop integral,
\begin{equation}
\label{chiral-mixing.2}
\epsilon=\frac{2}{f_{\pi}^2} \int \frac{\dd^3 \vec{k}}{(2 \pi)^3
  \omega_k^{\pi}} f_{\pi}(\omega_k^{\pi};T),
\end{equation}
where $\omega_k^{\pi}=\sqrt{m_{\pi}^2+k^2}$ is the on-shell energy of
the pion and $f_{\pi}=93 \;\MeV$ the pion-decay constant. For
$\epsilon \rightarrow 1/2$ this leads to a full restoration of chiral
symmetry around $T=T_{\mathrm{c}} \simeq 160 \; \MeV$, i.e., degeneracy
in the vector and axial-vector correlation function. In the intermediate
mass region this leads to a smooth distribution resembling the QCD
continuum.

\begin{figure}[t]
\centering
\includegraphics[width=0.3 \linewidth]{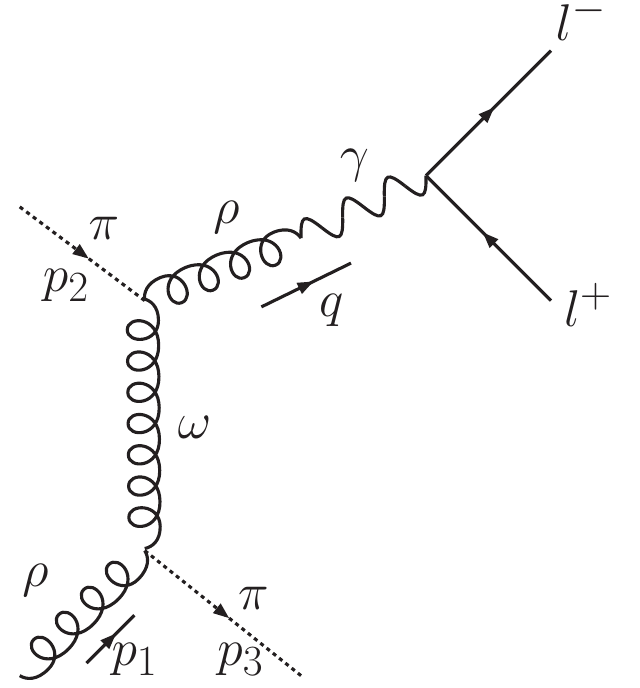}
\caption{Feynman diagram describing the annihilation of a $\rho$ meson,
  $\rho \pi \rightarrow \pi \ell^+ \ell^-$ via the $t$-channel exchange
  of an $\omega$ meson. Figure taken from {\protect
    \cite{vanHees:2007th}}.}
\label{fig.om-t-exch}
\end{figure}

In \cite{vanHees:2006ng,vanHees:2007th} this chiral-mixing effect is
implemented taking into account the presence of an effective pion
chemical potential, which ensures the conservation of the pion number
after chemical freezeout (in Boltzmann approximation), using
(\ref{chiral-mixing.1}) with a mixing parameter
$\hat{\epsilon}=\frac{1}{2}
\epsilon(T,\mu_{\pi})/\epsilon(T_{\mathrm{c}},0)$ and with a fugacity
factor $z_{\pi}=\exp(\mu_{\pi}/T)$ in (\ref{chiral-mixing.2}). Here it
is important to avoid double counting with the above described in-medium
evaluations of the $\rho$ self-energy due to interactions with vector
and axial-vector mesons, i.e., the two-pion piece and the part
corresponding to the $a_1 \rightarrow \rho \pi$ decay have to be
omitted. A detailed analysis, based on the chiral-reduction formalism
\cite{syz96,syz97}, finally leads to
\begin{equation}
\label{chiral-mixing.3}
\Pi_V = (1-\hat{\epsilon}) z_{\pi}^4 \Pi_{V,4 \pi}^{\text{vac}} +
\frac{\hat{\epsilon}}{2} z_{\pi}^3 \Pi_{A,3 \pi}^{\text{vac}} +
\frac{\hat{\epsilon}}{2} (z_{\pi}^4+z_{\pi}^5) \Pi_{A,5 \pi}^{\text{vac}}.
\end{equation}
As a further source of dileptons from a hot and dense hadronic medium,
relevant at higher three-momenta, in \cite{vanHees:2007th} also
contributions from the annihilation of a $\rho$ meson through
$\omega$-meson $t$-channel exchange has been taken into account (cf.\
Fig.\ \ref{fig.om-t-exch}).
The $\rho$-$\gamma$ mixing vertex has been implemented by the Lagrangian
$\Lag_{\rho \gamma}=-C m_{\rho}^2 A_{\mu} \rho^{\mu}$, using the
strict VMD value $C=e/g_{\rho \pi \pi}=0.052$ \cite{gale-rapp99}. For
the intermediate $\rho$-meson line in Fig.\ \ref{fig.om-t-exch} the full
in-medium propagator of the $\rho$ meson has been used, and for the
incoming $\rho$-meson line a weight
$(-2 m_{\rho}/\pi) \im D_{\rho}(m_{\rho})$. The coupling
$g_{\rho \omega \pi}=25.8 \;\GeV^{-1}$ is fixed in
\cite{gale-rapp99} by a simultaneous fit to the hadronic and radiative
$\omega$ decays, taking into account a hadronic dipole-form factor,
\begin{equation}
\label{om-t-ex.1}
F(t) = \left (\frac{2 \Lambda^2}{2 \Lambda^2-t} \right)^2
\end{equation}
with $\Lambda=1 \; \GeV$. To ensure gauge invariance, the form factor is
taken out of the integral, introducing an average momentum transfer
$\overline{t}$ via \cite{Turbide:2003si}
\begin{equation}
\begin{split}
\label{om-t-ex.2}
\frac{F^4(\overline{t})}{(m_{\omega}^2-\overline{t})^2} = \frac{1}{2}
\int_{-1}^{1} \dd x \frac{F^4[t(x)]}{[m_{\omega}^2-t(x)]^2},\\
t(x)=M^2+m_{\pi}^2 - 2(M^2+q^2-2 p \cdot q x).
\end{split}
\end{equation}
Finally, to avoid double counting with the $s$-channel $\omega$-exchange
contribution, which is already included in the evaluation of the
in-medium $\rho$ self-energy, in evaluating the corresponding
dilepton-emission rate from standard kinetic theory the integral has
been restricted to the kinematic region $t<0$ (for details, see
\cite{vanHees:2007th}).

For a realistic description of the dilepton production in heavy-ion
collisions one has to take into account also non-thermal sources: $\rho$
decay after thermal freeze-out as well as decay of high-momentum $\rho$ mesons
and Drell-Yan pairs produced in primordial hard collisions.

For the decay of $\rho$ mesons after thermal freezeout the usual
Cooper-Frye description \cite{vanHees:2007th},
\begin{equation}
\label{fo-rhos.1}
\dd N=q_{\mu} \dd \sigma^{\mu} \frac{\dd^3 \vec{q}}{(2 \pi)^3 q^0} \fB \left
  (\frac{q_{\mu} u^{\mu}}{T} \right),
\end{equation} 
is used, where $q_{\mu}$ is the four-momentum of the particle,
$\dd \sigma^{\mu}$ a freezeout-surface element, and $u^{\mu}$ the local
four-velocity of the fluid cell under consideration. To take into
account the finite width of the $\rho$ meson one uses the substitution
\begin{equation}
\label{fo-rhos.2}
\frac{\dd^3 \vec{q}}{q^0} = \dd^4 q 2 \Theta(q^0) \delta (q_{\mu}
q^{\mu}-m^2) \rightarrow \dd^4 q \frac{A_{\rho}}{\pi}, \quad
A_{\rho}=-\frac{2}{3} \im (D_{\rho})^{\mu}_{\mu}.
\end{equation}
Within the VMD model the corresponding dilepton rate is given by the
matrix element for the process $\rho \rightarrow \gamma^* \rightarrow
\ell^+ \ell^-$,
\begin{equation}
\label{fo-rho2.3}
\frac{\dd N_{\ell \ell}}{\dd^3 \vec{x} \dd^4 q} = \frac{\alpha_{\text{em}}^2
  m_{\rho}^4}{g_{\rho}^2 M^2} \frac{A_{\rho}}{2 \pi^3} L(M) \fB \left
  (\frac{q \cdot u}{T} \right) \frac{q^0}{M \Gamma_{\rho}^{(\text{fo})}},
\end{equation}
where
\begin{equation}
\label{fo-rho2.4}
L(M)=\left (1+\frac{2 m_{\ell}^2}{M^2} \right) \sqrt{1-\frac{4m_{\ell}^2}{M^2}}
\end{equation}
is the dilepton phase-space factor. One should note that compared to the
emission from a thermal source the spectral shape is modified by the
life-time dilation factor $\gamma=q^0/M=\sqrt{\vec{q}^2+M^2}/M$.

To estimate the contribution to the dilepton rate from $\rho$ mesons
produced in hard initial collisions, which do not fully equilibrated
with the bulk medium, one starts from a phenomenological $q_{\text{T}}$
spectrum in pp collisions,
\begin{equation}
\label{prim-rho.1}
\frac{1}{q_{\text{T}}} \frac{\dd N_{\text{prim}}}{\dd q_{\text{T}}} =
\frac{A}{(1+B q_{\text{T}}^2)^a}
\end{equation}
with $B=0.525 \; \GeV^{-2}$ and $a=5.5$ from fitting pp-scattering data
\cite{AguilarBenitez:1991yy} (for $400 \;\GeV$ pp collisions). The total
number of primordial $\rho$ mesons in AA collisions is estimated from
the empirical freeze-out systematics of light-hadron production
\cite{vanHees:2007th}. To take cold-nuclear matter effects into
account the Cronin effect is implemented by a ``Gaussian smearing'' of
the spectrum (\ref{prim-rho.1}),
\begin{equation}
\label{prim-rho.2}
\frac{\dd N_{\text{prim}}^{\text{cron}}}{\dd^2 q_{\text{T}}} = \int
\frac{\dd^2 q_{\text{T}}}{\pi \Delta k_{\text{T}}^2} \frac{\dd
  N_{\text{prim}}}{\dd^2 k_{\text{T}}} \exp \left
  [-\frac{(q_{\text{T}}-k_{\text{T}})^2}{\Delta k_{\text{T}}^2} \right],
\end{equation} 
with $\Delta k_{\text{T}}^2=0.2 \;\GeV^2$ based on direct-photon spectra
in pA collisions \cite{Turbide:2003si}. Finally the absorption of
primordial $\rho$ mesons through traversing the medium (``jet
quenching'') is evaluated by estimating the escape probability
\begin{equation}
\label{prim-rho.3}
P=\exp \left [-\int \dd t \sigma_{\rho}^{\text{abs}}(t) \varrho(t)
\right ],
\end{equation}
where 
\begin{equation}
\label{prim-rho.4}
\sigma_{\rho}^{\text{abs}} = \begin{cases} 
\sigma_{\text{ph}}=0.4 \;\text{mb} & \text{for} \quad t<q_0/m_{\rho}
\tau_{\text{f}},\\
\sigma_{\text{had}}=5 \;\text{mb} & \text{for} \quad t>q_0/m_{\rho}
\tau_{\text{f}},
\end{cases}
\end{equation}
with the $\rho$-meson formation time $\tau_{\text{f}}=1 \; \fm$;
$\sigma_{\text{ph}}$ and $\sigma_{\text{had}}$ are the absorption cross
sections for pre-hadrons and hadrons, respectively, and $\varrho(t)$
denotes the partonic or hadronic particle density of the medium.

Finally, the contribution to the dilepton yield from the Drell-Yan (DY)
process, i.e., the quark-antiquark annihilation in hard initial
collisions, is estimated using
\begin{equation}
\label{dy.1}
\left . 
\frac{\dd N_{\text{DY}}^{AA}}{\dd M \dd y} \right |_{b=0}=\frac{3}{4\pi
 R_0^2} \, A^{4/3} \, \frac{\dd \sigma_{\text{DY}}^{NN}}{\dd M \dd y} 
\end{equation}
for central $AA$ collisions with the root-mean squared radius
$R_0 \simeq 1.05 \;\fm$ originating from folding over a Gaussian
thickness function. The DY cross section in nucleon-nucleon collisions
is given in leading order
$\mathcal{O}(\alpha_{\text{s}}^0 \alpha_{\text{em}}^2)$ by
\begin{equation}
\label{dy.2}
\frac{\dd \sigma_{\text{DY}}^{NN}}{\dd M \dd y}=K \frac{8\pi\alpha_{\text{em}}}{9sM}
\sum\limits_{q=u,d,s} e_q^2 \left[ q(x_1) \overline{q}(x_2) +
\overline{q}(x_1) q(x_2) \right].
\end{equation}
where $q(x_{1,2})$ and $\overline{q}(x_{1,2})$ denote the GRV94LO
parton-distribution functions for quarks and antiquarks
\cite{Gluck:1994uf}. The $K$ factor takes into account higher-order
corrections in $\alpha_{\text{s}}$ with $K \simeq 1.5$ as inferred from
data on DY production in p$A$ collisions
\cite{Spieles:1997ih}. Higher-order effects also lead to a non-zero
dilepton-$q_{\text{T}}$, which is adopted from the procedure by the NA50
Collaboration \cite{Abreu:1999jr,Abreu:2000nj}, according to which in
both p$A$ and $AA$ collisions the $q_{\text{T}}$ dependence of the DY
dileptons can be described by a Gaussian distribution,
\begin{equation}
\label{dy.3}
\frac{\dd N_{\text{DY}}}{\dd M \dd y \dd q_{\text{T}}^2} =  \frac{\dd
  N_{\text{DY}}}{\dd M \dd y} \, 
\frac{\exp(-q_{\text{T}}^2/2\sigma_{q_{\text{T}}}^2)}{2\sigma_{q_{\text{T}}}^2}  
\end{equation}
with $\sigma_{q_{\text{T}}} \simeq 0.8$-$1\;\GeV$.

The DY contribution in the region of low invariant mass and momentum,
$M,q \lesssim 1.5 \;\GeV$ is problematic. In $AA$ collisions it is small
compared to the emission from thermal sources, but becomes significant
at higher $q_{\text{T}} \gtrsim 1 \; \GeV$. There the additional
constraint by the photon point $M \rightarrow 0$ allows an extrapolation
of the DY spectrum to lower mass \cite{vanHees:2007th}: Comparing the emission rate for
dileptons from a thermal source (\ref{mtform.8a}) with that of photons we see that for $q \gg M$
\begin{equation}
\label{dy.4}
q_0 \frac{\dd N_{\ell \ell}}{\dd M \dd^3 \vec{q}} = q_0 \frac{\dd
  N_{\gamma}}{\dd^3 \vec{q}} \frac{2 \alpha_{\text{em}}}{3 \pi M}.
\end{equation}
Thus we evaluate the DY-$q_T$ spectrum at a mass
$M_{\text{cut}} =0.8$-$1\;\GeV$ and extrapolate it down in mass by
$M_{\text{cut}}/M$.

Finally, for photon production in addition to the already discussed
hadronic model for the in-medium electromagnetic current correlator,
some additional meson-exchange reactions ($\pi$, $K$, $\rho$, $K^*$, and
$a_1$) in a meson gas become relevant at photon momenta
$q \gtrsim 1 \; \GeV$ \cite{Turbide:2003si}. Here a massive-Yang-Mills
model based on a nonlinear
$\mathrm{U}_{\mathrm{L}}(3) \times \mathrm{U}_{\mathrm{R}}(3)$-$\sigma$
model has been employed:
\begin{equation}
\begin{split}
\label{photons.1}
  \Lag = & \frac{1}{8} F_\pi^2 {\rm Tr} D_\mu U D^\mu U^\dag + \frac{1}{8}
  F_\pi^2 {\rm Tr} M (U + U^\dag -2) \\
  &  - \textstyle{\frac{1}{2}}
  {\rm Tr} \left(F_{\mu \nu}^L
    {F^L}^{\mu \nu} + F_{\mu \nu}^R {F^R}^{\mu \nu} \right) + m_0^2 {\rm Tr}
  \left(A_\mu^L {A^L}^{\mu \nu} + A_\mu^R {A^R}^\mu\right)+
  \gamma \Tr F_{\mu \nu}^L U F^{R \mu \nu}U^\dag 
  \\
  & -i\xi \Tr \left(D_\mu UD_\nu U^\dag F^{L \mu \nu}
    +D_\mu U^\dag D_\nu U F^{R \mu \nu}\right)\ .
\end{split}
\end{equation}
Here
\begin{equation}
\begin{split}
\label{photons.2}
&U = \exp \left( \frac{2 i}{F_\pi} \sum_i \frac{\phi_i
\lambda_i}{\sqrt{2}}\right) = \exp\left( \frac{2 i}{F_\pi} 
\phi \right)\ ,\  \\
 &A_\mu^L = \textstyle{\frac{1}{2}}(V_\mu + A_\mu)\ ,  \\
 &A_\mu^R = \textstyle{\frac{1}{2}}(V_\mu - A_\mu)\ ,  \\
 &F_{\mu \nu}^{L, R}  = \partial_\mu A_\nu^{L, R} - \partial_\nu A_{\mu}^{L, R} -
i g_0 \left[A_{\mu}^{L, R}, A_\nu^{L, R} \right]\ , \\
 &D_\mu U = \partial_\mu U - i g_0 A_\mu^L U + i g_0 U A_\mu^R\ , \\
 &M = \frac{2}{3} \left[ m_K^2 + \frac{1}{2} m_\pi^2\right] - \frac{2}{\sqrt{3}}
(m_K^2 - m_\pi^2) \lambda_8
\end{split}
\end{equation}
with $F_{\pi} = 135 \; \MeV$ and the Gell-Mann matrices $\lambda_i$;
$\phi$, $V_{\mu}$, and $A_{\mu}$ are the pseudoscalar, vector and
axial-vector meson matrices, respectively.

In the non-strange sector, including pions, $\rho$, and $a_1$ mesons, in
\cite{Song:1993ae} two parameter sets for the four free parameters have
been fitted to the masses and widths of the $\rho$ and $a_1$ mesons,
\begin{equation} 
\begin{split}
\label{photons.3}
I:& \qquad \tilde{g} = 10.3063, \quad \gamma = 0.3405,\ 
\xi = 0.4473,\ m_0 = 0.6253 \GeV, \\
II:& \qquad \tilde{g} = 6.4483, \quad \gamma = -0.2913,\ 
\xi = 0.0585,\ m_0 = 0.875 \GeV.
\end{split}
\end{equation}
In \cite{Gao:1997vm} the D- and S-wave content in the
$a_1 \rightarrow \rho \pi$ decay has been found to be $D/S=0.36$ and
$D/S=-0.099$ for parameter sets I and II, respectively. Given the
experimental finding of $D/S=-0.107 \pm 0.016$, in the following
parameter set II has been used, employing the kinetic-theory expression
for a photon-production process, $1+2 \rightarrow 3+\gamma$,
\begin{equation}
\begin{split}
\label{photons.4}
q_0 \frac{\dd R_\gamma}{\dd^3 \vec{q}} =\int \frac{\dd^3 \vec{p}_1}{2(2\pi)^3E_1}
\frac{\dd^3 \vec{p}_2}{2(2\pi)^3E_2}\frac{\dd^3 \vec{p}_3}{2(2\pi)^3E_3}(2\pi)^4 &
\delta^{(4)}(p_1+p_2\rightarrow p_3+q) \\
\times \left|\mathcal{M} \right|^2\frac{f(E_1)f(E_2)[1\pm f(E_3)]}{2(2\pi)^3}.
\end{split}
\end{equation}
For the matrix elements, Born graphs in all possible $s$-, $t$-, and
$u$-channels for reactions of the type $X+Y \rightarrow Z+\gamma$,
$\rho \rightarrow Y+Z+\gamma$, and $K^* \rightarrow Y+Z+\gamma$, where
for $X$, $Y$, and $Z$ all combinations of $\rho$, $\pi$, $\mathrm{K}$,
and $\mathrm{K}^*$ mesons allowed by the conservation of charge,
isospin, strangeness, and $G$ parity have been considered. In addition
also the same model for $\omega$-$t$-channel exchange has been used for
photons as for dileptons (analogous to Fig. \ref{fig.om-t-exch} with a
real photon instead of a dilepton, $\ell^+\ell^-$ in the final
state). At all vertices hadronic dipole form factors (\ref{om-t-ex.1})
have been employed, ensuring gauge invariance by the averaging procedure
defined in (\ref{om-t-ex.2}).

\begin{figure}[t]
\begin{center}
\includegraphics[width=0.95\linewidth]{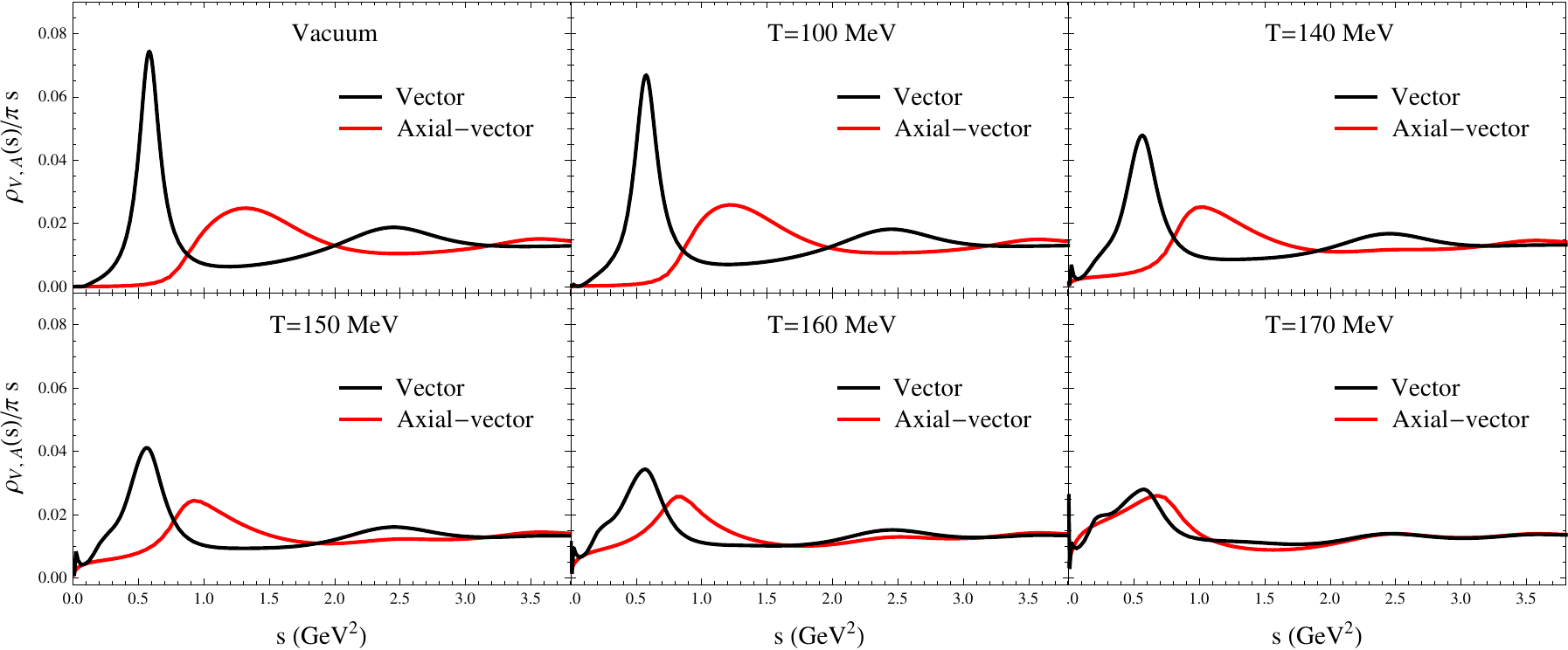}
\end{center}
\caption{The contruction of the axial-vector spectral function from the
  vector channel provided by the model for the in-medium spectral
  function of the $\rho$ meson, demonstrating the compatibility of the
  resulting ``broadening-mass scenario'' for chiral-symmetry
  interpretaion. Figure taken from \cite{Hohler:2013eba}.}
\label{fig.chir-sum-rule}
\end{figure}

One should note that the here described phenomenological model, which
cocentrates on the vector-isovector channel (light vector mesons) to
describe dilepton production in heavy-ion collisions, is compatible with
chiral symmetry as is demonstrated in \cite{Hohler:2013eba} by
constructing the axial-vector channel via QCD and chiral (Weinberg) sum
rules. This study clearly demonstrates that the ``broadening-resonance
scenario'' following from the phenomenological model is compatible with
chiral-symmetry restoration, i.e., the mass spectra of the vector and
axial-vector curren-current correlation function become degenerate at a
temperature of about $170 \, \MeV$ (cf.\ Fig.\ \ref{fig.chir-sum-rule}).

\subsection{Bulk-evolution models}
\label{sect:bulk-evolution-mods}

To evaluate the dilepton spectra in relativistic heavy-ion collisions in
addition to the above discussed models for the various dilepton sources
a reliable description of the evolution of the hot and dense medium
created in such collisions is necessary. In the here discussed work
simple \textbf{thermal-fireball parameterizations} (``blast-wave''
models) as well as \textbf{coarse-grained transport simulations} have
been employed.

The thermal-fireball parameterizations are motivated by the observation
that at the higher collision energies the bulk evolution of the medium
is well-described by hydrodynamical expansion of a hot medium. In its
simplest form the fireball volume is taken as an expanding cylinder with
volume \cite{Rapp:1999us}
\begin{equation}
\label{blast.1}
  V_{\mathrm{FB}}(t) = \pi \left (r_{\perp,0}+\frac{1}{2} a_\perp t^2
  \right)^2 \left (z_0+v_{z,0} t + \frac{1}{2} a_z t^2 \right ).
\end{equation}
The initial transverse radius $r_{\perp,0}$ is determined by the
centrality of the collision. The initial longitudinal size $z_0$
reflects the formation time $\tau_0 =1 \; \fm/c$ of the thermal medium,
which translates into $z_0 \simeq \tau_0 \Delta y =1.8 \; \fm$, where
$\Delta y=1.8$ is the rapidity width of a thermal fireball. The values
for the transverse acceleration are determined together with the
fireball lifetime by the transverse-momentum spectra of hadrons, which
are affected by the ``Doppler blueshift'' due to the radial flow. In
accordance with hydrodynamical calculations the radial flow-velocity is
assumed to grow linearly with $r_{\perp}$,
\begin{equation}
\label{blast.1b}
v_{\perp}(t,r_{\perp}) = \frac{r_{\perp}}{r_{\perp,0} + a_{\perp}t^2/2}
a_{\perp} t.
\end{equation}
With given expansion parameters, the fireball life-time is determined by
the condition for thermal freeze-out which can also be determined from
measurements of the hadronic spectra (see, e.g., for SPS energies
\cite{Bearden:1996dd,Appelshauser:1997rr,Antinori:2001yi,Antinori:2007jr,Adamova:2002wi}). The
time evolution of the temperature is determined by the simplifying
assumption that the temperature is constant throughout the fireball
volume (i.e., taken as an average temperature) and that the expansion is
isentropic in accordance with ideal hydrodynamics, using an EoS. For
$T>T_{\text{c}}$ an ideal partonic EoS and a hadron-resonance-gas (HRG)
EoS for $T<T_{\text{c}}$ has been used. In
\cite{vanHees:2006ng,vanHees:2007th} a first-order transition has
been assumed, connecting the QGP and HRG phases with a mixed phase at
$T=T_c=\text{const}$ with the hadron-gas fraction,
\begin{equation}
\label{blast.2}
f_{\rm HG}(t)=\frac{s_c^{\rm{QGP}}-s(t)}{s_c^{\text{QGP}}-s_c^{\text{HG}}}.
\end{equation}
Of course, also other EoS with a cross-over transition, which is more
adequate at higher beam energies, like latPHG \cite{He:2011zx} can be
easily implemented. After chemical freeze-out the system falls off
chemical equilibrium, and besides the usual baryochemical potential,
$\mu_{\text{B}}$, chemical potentials like $\mu_{\pi}$ and
$\mu_{\text{K}}$ are introduced to keep the particle abundances
fixed. The chemical potential of resonances is then determined through
their (finally stable) decay products; e.g., for the $\Delta$ resonance,
decaying mostly into $\pi N$, $\mu_{\Delta}=\mu_{\text{B}}+\mu_{\pi}$ or
for the $\rho$ meson $\mu_{\rho}=2 \mu_{\pi}$, etc.

\begin{figure}[!t]
\begin{minipage}{0.48\linewidth}
\includegraphics[width=\textwidth]{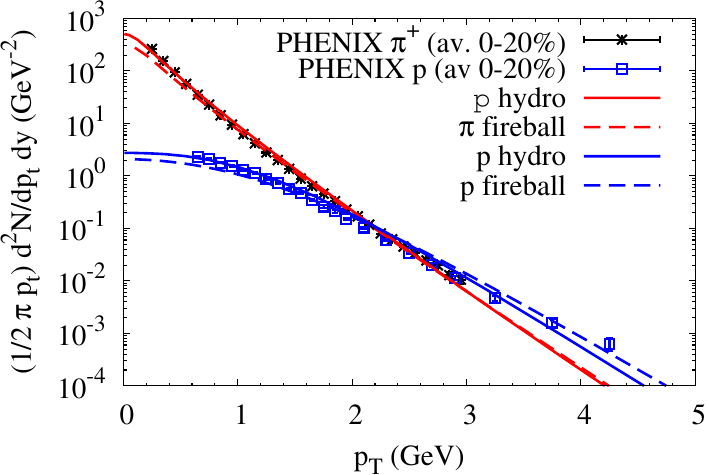}
\end{minipage}\hfill
\begin{minipage}{0.48\linewidth}
\includegraphics[width=\textwidth]{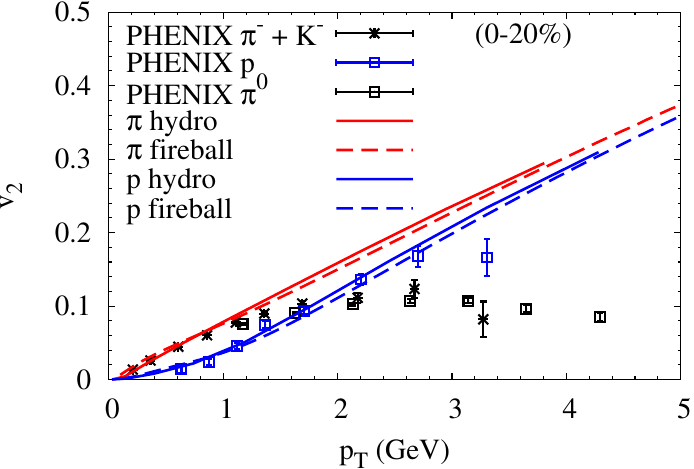}
\end{minipage}
\begin{minipage}{0.48\linewidth}
\includegraphics[width=\textwidth]{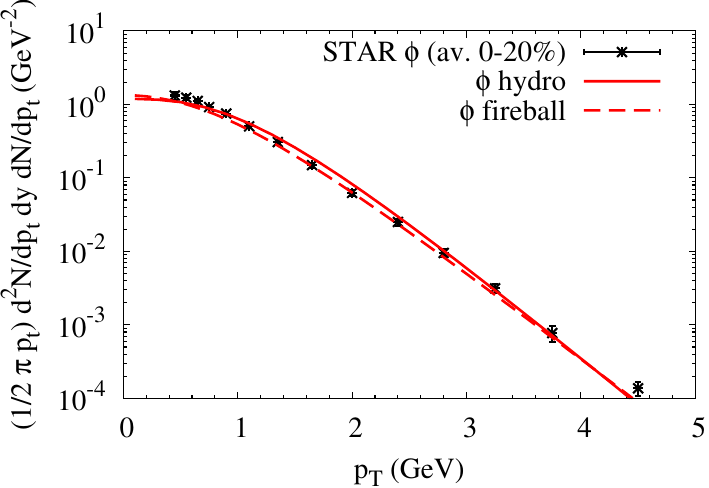}
\end{minipage}\hfill
\begin{minipage}{0.48\linewidth}
\includegraphics[width=\textwidth]{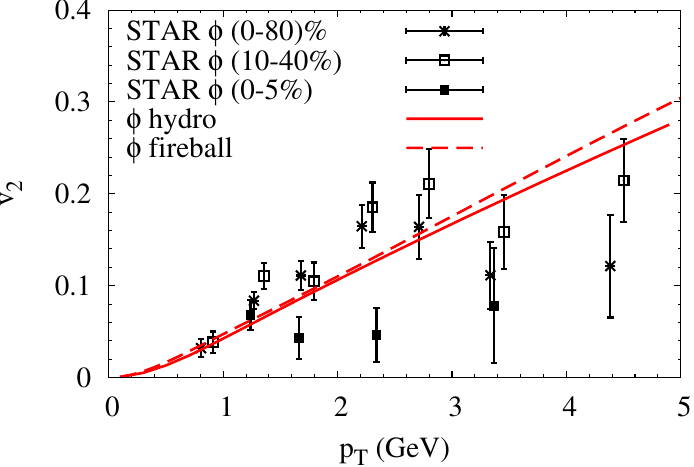}
\end{minipage}
\caption{Fits to the spectra and elliptic flow of light
  hadrons ($T_\mathrm{fo}\simeq110\;\MeV$, upper two panels) and $\phi$
  mesons ($T_\mathrm{fo}=160\;\MeV$, lower two panels) in
  Au-Au($\sqrt{s}=200\;A \GeV$) collisions, using EoS \textit{latPHG} \cite{He:2011zx}
  within either the fireball (dashed lines) or ideal hydrodynamic model 
  (solid lines). The data are taken from
  Refs.~\cite{Abelev:2007rw,Adler:2003kt,Adler:2003cb,Adams:2003xp}. Figure
taken from {\protect \cite{vanHees:2014ida}}.}
\label{fig:ell-fb-hadron-fit}
\end{figure}

To also take into account the elliptic flow of hadrons in order to
address the resulting elliptic flow of photons, in
\cite{vanHees:2011vb,vanHees:2014ida} an elliptic blast-wave
description has been developed. Also the parameterization of the radial
flow of the fireball boundary in (\ref{blast.1}) has been substituted by
the corresponding relativistic motion of constant proper acceleration,
i.e., the velocities and lengths of the major axes of the ellipse follow
the time evolution
\begin{equation}
\begin{split}
\label{blast.3}
v_a(t) &=\frac{a_a t}{\sqrt{1+(a_a t)^2}}, \quad v_b(t)=\frac{a_b
  t}{\sqrt{1+(a_b t)^2}}, \\
a(t) &= a_0 + \frac{\sqrt{1+(a_a t)^2}-1}{a_a}, \quad b(t) = b_0 +
\frac{\sqrt{1+(a_b t)^2}-1}{a_b}.
\end{split}
\end{equation}
To define the flow field, confocal elliptic coordinates,
\begin{equation}
\label{blast.4}
\vec{x}_{\perp} = r_0 (\sinh u \cos v, \cosh u \sin v)
\end{equation}
are introduced to parameterize the ellipse at each time, $t$, and the
radial-flow velocity field is parameterized as
\begin{equation}
\label{blast.5}
\vec{v}_{\perp} = \frac{r}{r_{\text{max}}} (v_b \cos v,v_a \sin v).
\end{equation}
The accelerations $a_a$ and $a_b$ are chosen differently in the QGP and
hadronic phases and determined such that the transverse-momentum spectra
and elliptic flow of hadrons (in the low-$p_T$ range) are well
described. It is assumed that multi-strange hadrons (like, e.g., the
$\phi$ meson) freeze out kinetically already at the phase transition,
while the light hadrons freeze out at the nominal kinetic freezeout (see
Fig. \ref{fig:ell-fb-hadron-fit}).

For smaller beam energies the description of the bulk evolution of the
medium created in heavy-ion collisions becomes questionable, and
transport simulations become more reliable. On the other hand, a fully
self-consistent off-equilibrium treatment of the in-medium properties of
hadrons, which is necessary for a successful description of dilepton and
photon production in heavy-ion collisions, is very challenging. In
\cite{Endres:2014zua,Endres:2015egk,Endres:2015fna,Endres:2016tkg} a
\textbf{coarse-grained transport approach} has been developed. Here, the
bulk evolution is simulated using the established transport simulation
Ultrarelativistic Quantum Molecular Dynamics (UrQMD)
\cite{Bass:1998ca,Bleicher:1999xi,Petersen:2008kb}. Averaging over
several runs, the phase-space distribution is mapped to a local
equilibrium description within a space-time grid, using an appropriate
EoS. In this way the thermal quantum-field theoretical results for the
in-medium dilepton and photon production rates described above become
applicable.

\begin{figure}[t]
\includegraphics[width=0.48 \linewidth]{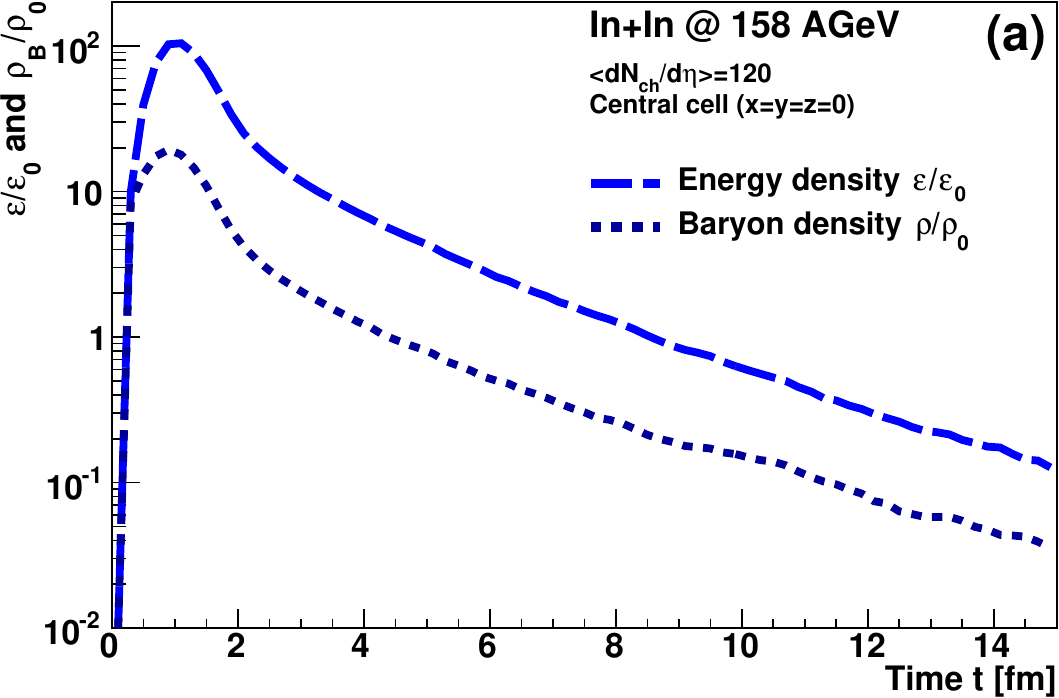} \hfill
\includegraphics[width=0.48 \linewidth]{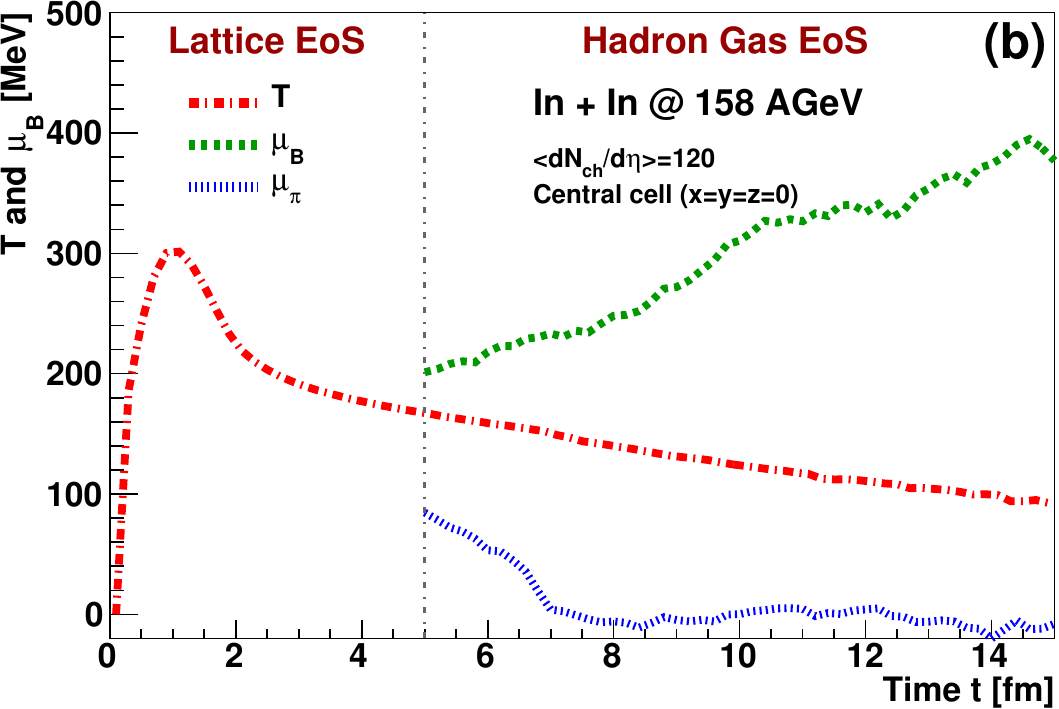}
\caption{\label{fig.density-temperature-cg} Left panel: Time evolution of
  the baryon density $\rho_{\mathrm{B}}$ (short dashed) and energy
  density $\varepsilon$ (long dashed) for the cell at the center of the
  coarse-graining grid ($x=y=z=0$). The results are given in units of
  the ground-state densities $\varepsilon_{0}$ and $\rho_{0}$. Right
  panel: Time evolution of the temperature $T$ (red dash-dotted), baryon
  chemical potential $\mu_{\mathrm{B}}$ (green short dashed) and the
  pion chemical potential $\mu_{\pi}$ (blue dotted) in the central
  cell. The thin grey line indicates the transition from the Lattice EoS
  to the Hadron Gas EoS at the transition temperature of
  $T=170\;\MeV$. Figure taken from {\protect \cite{Endres:2014zua}}.}
\end{figure}
The grid of space-time cells is defined by
$\Delta x=\Delta y = \Delta z=0.7$-$0.8\;\fm$ and $\Delta
t=0.2$-$0.6 \;\fm/c$, and the energy-momentum tensor $T_{\mu \nu}$ and
net-baryon four-flow $j_{\mu}^{\text{B}}$ in each cell are evaluated via
\begin{equation}
\begin{split}
\label{coarse-graining.1}
 T^{\mu\nu}&=\int\dd^{3}p\frac{p^{\mu}p^{\nu}}{p^{0}}f(\vec{x},\vec{p},t)
  =\frac{1}{\Delta V}\left\langle \sum\limits_{i=1}^{N_{h} \in \Delta %
      V} \frac{p^{\mu}_{i}\cdot p^{\nu}_{i}}{p^{0}_{i}}\right\rangle, \\
   j^{\mu}_{\mathrm{B}}&=\int\mathrm{d}^{3}p\frac{p^{\mu}}{p^{0}} f_{\mathrm{B}}(\vec{x},\vec{p},t)
  =\frac{1}{\mathrm{\Delta} V}\left\langle
    \sum\limits_{i=1}^{N_{\mathrm{B}/\overline{\mathrm{B}}} \in \Delta
      V}\pm\frac{p^{\mu}_{i}}{p^{0}_{i}}\right\rangle,
\end{split}
\end{equation}
where the averaging is understood as averaging over several UrQMD
events. The four-velocity of the fluid cell is defined using the Eckart
definition, i.e., according to the flow of the net-baryon number,
\begin{equation}
\label{coarse-graining.2}
u^{\mu} = \frac{j_{\text{B}}^{\mu}}{\sqrt{j_{\text{B}} \cdot
    j_{\text{B}}}}=(\gamma,\gamma \vec{v}).
\end{equation}
The transport simulations show that the assumption of isotropic thermal
equilibrium in the fluid cells is not justified in the early stages of
the collision. This kinetic off-equilibrium situation leads to the
ansatz for the energy-momentum stress tensor within the anisotropic
hydrodynamics approach \cite{Florkowski:2010cf,Florkowski:2012pf,Molnar:2016gwq},
\begin{equation}
\label{coarse-graining.3}
T^{\mu \nu} = \left( \varepsilon  + P_{\perp}\right) u^{\mu}u^{\nu} -
P_{\perp} \, \eta^{\mu\nu} - (P_{\perp} - P_{\parallel}) v^{\mu}v^{\nu},
\end{equation}
where $\varepsilon$ is the energy density, $P_{\perp}$ and
$P_{\parallel}$ the pressures perpendicular and parallel to the beam
direction; $u^{\mu}$ is the four-velocity of the fluid cell, and
$v^{\mu}$ the four-vector of the beam direction. Then an effective
energy density is obtained by the generalized EoS of a Boltzmann-like
system via
\begin{equation}
\label{coarse-graining.4}
\varepsilon_{\text{eff}} = \frac{\varepsilon}{r(x)}
\end{equation}
with the relaxation function
\begin{equation}
\label{coarse-graining.5}
r(x) =\begin{cases} 
        \frac{x^{-1/3}}{2}\left(1+\frac{x \artanh
        \sqrt{1-x}}{\sqrt{1-x}}\right) & \text{for } x \leq 1  \\ 
        \frac{x^{-1/3}}{2}\left(1+\frac{x \arctan
        \sqrt{x-1}}{\sqrt{x-1}}\right) & \text{for } x \geq 1
    \end{cases}, \quad x=\left (\frac{P_{\parallel}}{P_{\perp}} \right)^{3/4}.
\end{equation}
It turns out that $\varepsilon_{\text{eff}}$ deviates from the nominal energy
density,
\begin{equation}
\label{coarse-graining.6}
\varepsilon=u_{\mu} u_{\nu} T^{\mu \nu},
\end{equation}
only in the first 1-$2\;\fm/c$ of the time evolution, where the pressure
anisotropy is large. The effective energy density and net-baryon density
are used to determine the temperature and baryochemical potential as
well as the pion and kaon chemical potentials to take into account
chemical off-equilibrium, matching the EoS in the QGP phase based on
lattice calculations \cite{He:2011zx} with a hadron-resonance-gas EoS
including the hadronic degrees of freedom implemented in UrQMD
\cite{Zschiesche:2002zr,Petersen:2008dd} based on a hadronic chiral
model \cite{Papazoglou:1998vr,Zschiesche:2006rf}. As an example Fig.\
\ref{fig.density-temperature-cg} shows the time evolution of the
net-baryon density, temperature, $\mu_{\text{B}}$, and $\mu_{\pi}$ for
the central cell of the medium created in $158 \; A\GeV$ In-In
collisions at the CERN SPS, as investigated in the NA60 experiment.

Another transport model used to describe dilepton production in
heavy-ion collisions is Parton-Hadron-String Dynamics (PHSD)
\cite{Linnyk:2011vx,Linnyk:2011hz,Linnyk:2012pu}. In the partonic phase
it uses the dynamical quasiparticle model (DQPM) \cite{Cassing:2008sv}
to describe the transport of broad and massive quarks and gluons in the
medium. Dilepton production is implemented using the processes
$q \overline{q} \rightarrow \gamma^*$ (quark annihilation)
$q+g \rightarrow q + \gamma^*$,
$\overline{q}+g \rightarrow \overline{q} + \gamma^*$ (gluo-Compton
scattering), and $q \overline{q} \rightarrow \gamma^*+g$ (gluon
bremsstrahlung). In the hadronic phase the (off-shell) transport model
is identical with the hadron-string-dynamics model (HSD)
\cite{Ehehalt:1996uq,
  Bratkovskaya:1996qe,Cassing:1999es,Bratkovskaya:2007jk}. The dilepton
sources include $\pi$-, $\eta$-, $\eta'$, $\omega$, $\Delta$,
$a_1$-Dalitz as well as $\rho,\omega, \phi \rightarrow \e^+ \e^-$ decays.

\section{Dilepton production at various beam energies}
\label{sect:dileps-hics-results}

In this Section the results of simulations for heavy-ion collisions at
various beam energies is summarized. All calculations are based on the
microscopic models for dilepton production described in Sects.\
\ref{sect:em-rad-qgp} and \ref{sect:em-rad-had} for the emission from a
QGP and a hot and dense hadron gas in the deconfined and confined phases
of the evolution of the medium using the models discussed in
Sect. \ref{sect:bulk-evolution-mods}.

\subsection{Dielectron production at GSI-SIS energies}
\label{sect:dileps-SIS}

\begin{figure}[t]
\begin{minipage}{0.45\linewidth}
\includegraphics[width=\textwidth]{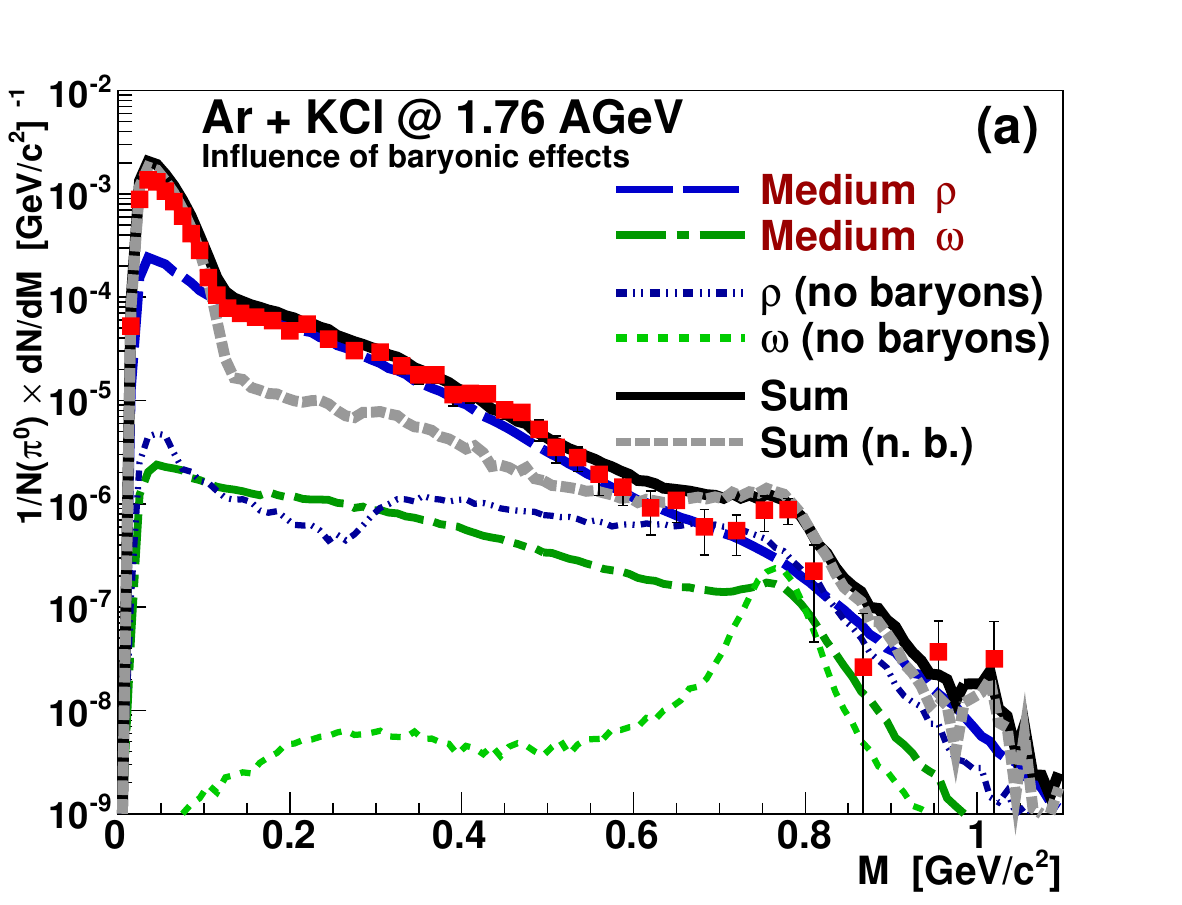} 
\end{minipage}\hspace*{3mm}
\begin{minipage}{0.45\linewidth}
\includegraphics[width=\textwidth]{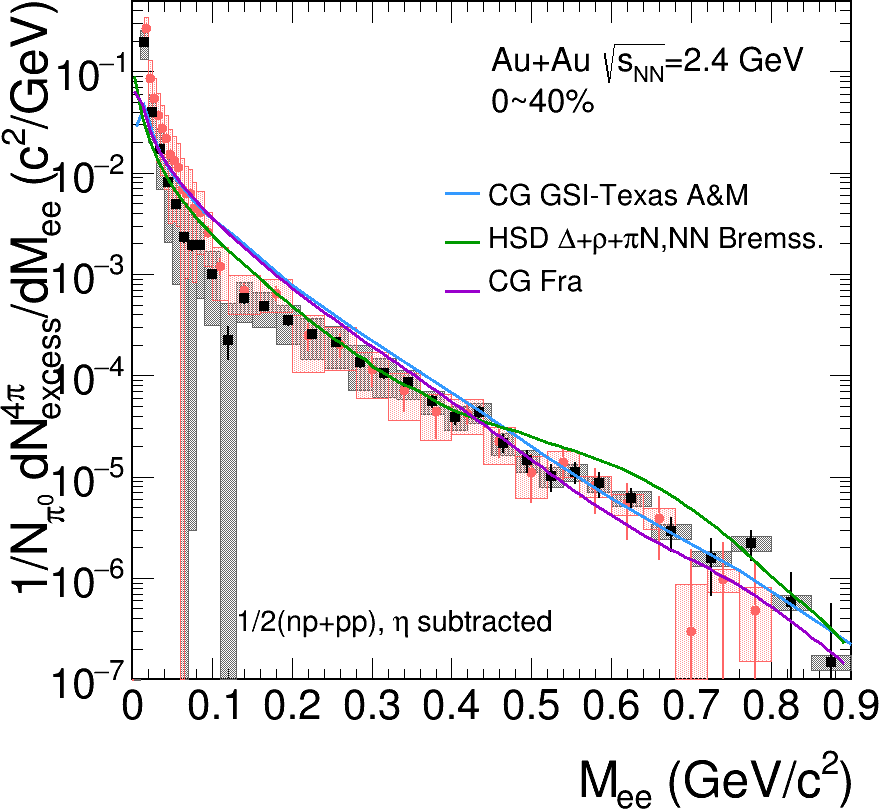}
\end{minipage}
\centerline{\includegraphics[width=0.48\linewidth]{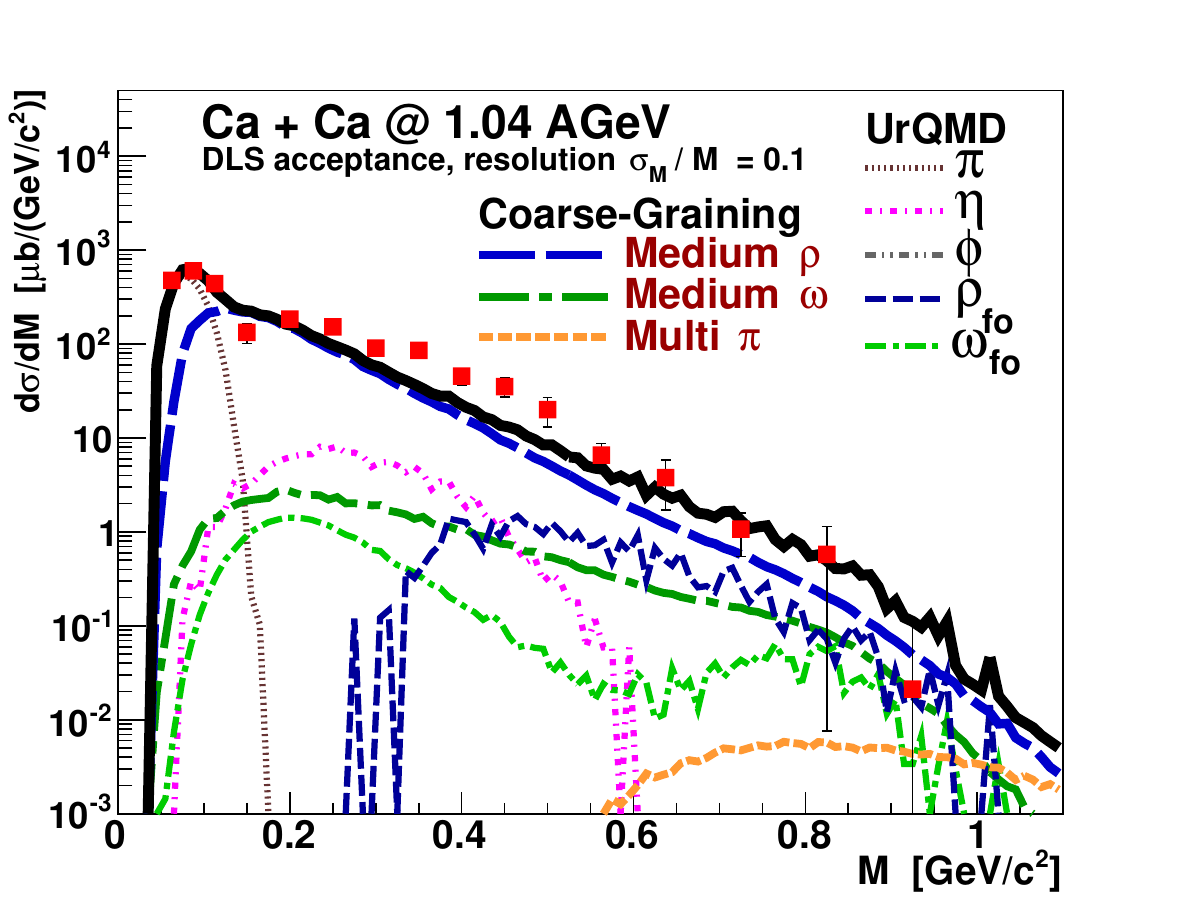}}
\caption{Upper panels: Comparison of invariant mass
  spectra with the full spectral function and for the case of no
  baryonic effects (i.e., for $\rho_{\mathrm{eff}}=0$). The dielectron
  yields for Ar+KCl collisions at $E_{\text{lab}}=1.76\; A \GeV$ (a) and
  for Au+Au at $E_{\mathrm{lab}}=1.23\;A \GeV$ (b) are shown within
  HADES acceptance and normalized to the average number of produced
  $\pi^{0}$. Note that the UrQMD contributions are included in the sum,
  but the different single yields are not shown explicitly for reasons
  of lucidity. Lower panel: Invariant-mass spectrum of the
  dielectron yield for Ca+Ca collisions at
  $E_{\mathrm{lab}}=1.04 \;A \GeV$ within the experimental acceptance. The
  result is compared to the data from the DLS Collaboration
  \cite{Porter:1997rc}. Figures taken from {\protect
    \cite{Endres:2015fna}}.}
\label{fig.hades-diel-ArKCl-AuAu}
\end{figure}
In \cite{Endres:2015fna} the production of dielectrons in heavy-ion
collisions at GSI-SIS energies as measured by the HADES collaboration
has been simulated using the coarse-grained transport approach as
described in Sect. \ref{sect:bulk-evolution-mods}. An ensemble of 1000
UrQMD events has been used to obtain sufficient statistics, particularly
for the contributions from non-thermal $\rho$ and $\omega$ mesons. The
impact-parameter distribution has been adapted to the HADES trigger
conditions for Ar+KCl reactions at a beam energy of $1.76\;A \GeV$
\cite{Agakishiev:2010rs} and Au+Au collisions at $1.23\;A \GeV$
\cite{GalatyukPC,Galatyuk:2014vha}, using a Woods-Saxon-type fit, which
in both cases approximately corresponds to a selection of the $0$-$40\%$
most central collisions. For the Ar+KCl case, the number of neutral
pions per event, used for the normalization of the spectra
$N_{\pi^0}^{\text{sim}} \simeq 3.9$ agrees well with the experimental
finding, $N_{\pi_0}^{\text{exp}} \simeq 3.5$. For the Au+Au collisions
the simulation predicts $N_{\pi_0}^{\text{sim}} \simeq 8.0$. In the
simulation the overall normalization of the dilepton yield uses the
simulated $\pi_0$ yields. To compare the simulated spectra with the
experimental result the HADES acceptance filter \cite{HADESwiki} as well
as the appropriate momentum cuts have been employed. To also confront
the model with the data from the DLS collaboration on Ca+Ca collisions
at a beam energy of $1.04 \; A \GeV$ the DLS acceptance filter (version
4.1) \cite{DLSfil} is used as well as an RMS smearing of 10\% to account
for the detector resolution. In this case a minimum-bias simulation has
been employed since for DLS no impact-parameter distributions are
available. The final invariant-mass spectrum is normalized to the total
cross section of a Ca+Ca reaction.

In Fig.\ \ref{fig.hades-diel-ArKCl-AuAu} the results for dielectron
production in Ar+KCl collisions are compared to the data from the HADES
collaboration and predictions for Au+Au collisions within the HADES
acceptance are made. The observed enhancement of dileptons over the
``hadronic cocktail'' is well explained by the medium modifications of
the $\rho$ and $\omega$ meson. To underline the importance of the
baryonic medium the calculation has also been performed neglecting
$\rho$- and $\omega$-baryon interactions. In addition also contributions
from the decays of $\rho$- and $\omega$-mesons in ``non-thermal cells''
(i.e., for cells of the space-time grid, for which a temperature
$T<50 \; \MeV$ results from the coarse-graining procedure) are
included. Here the microscopic transport-theoretical cross sections as
implemented in UrQMD are used (for details, see
\cite{Endres:2015fna}).

\begin{figure}[H]
\begin{center}
    \vspace{0.3cm}
    \includegraphics[width=0.6\textwidth]{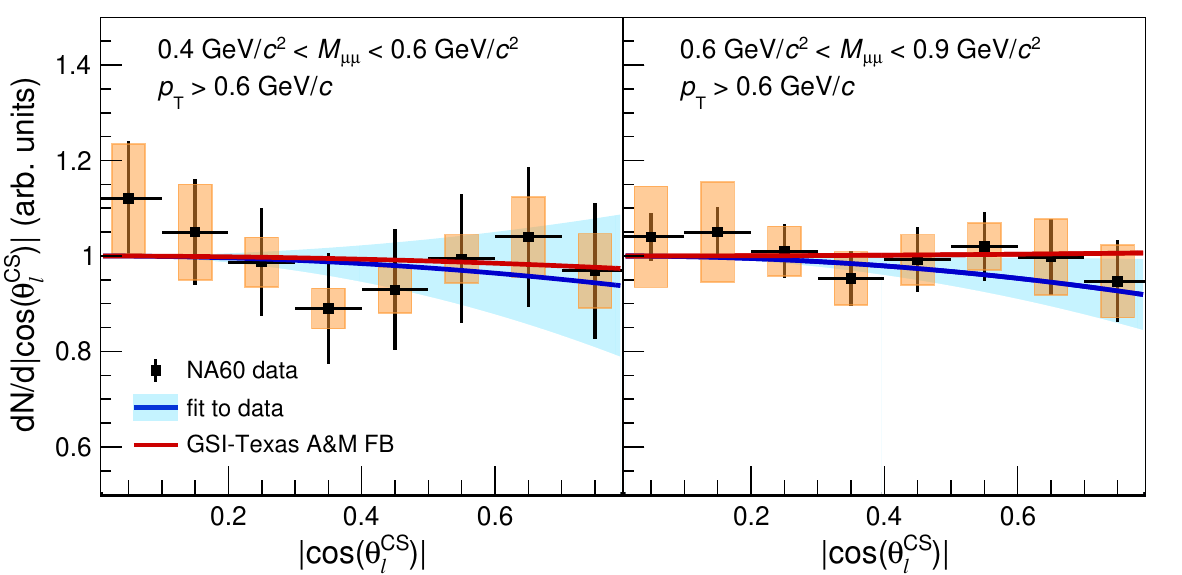}\\
    \vspace{0.3cm}
    \includegraphics[width=0.6\textwidth]{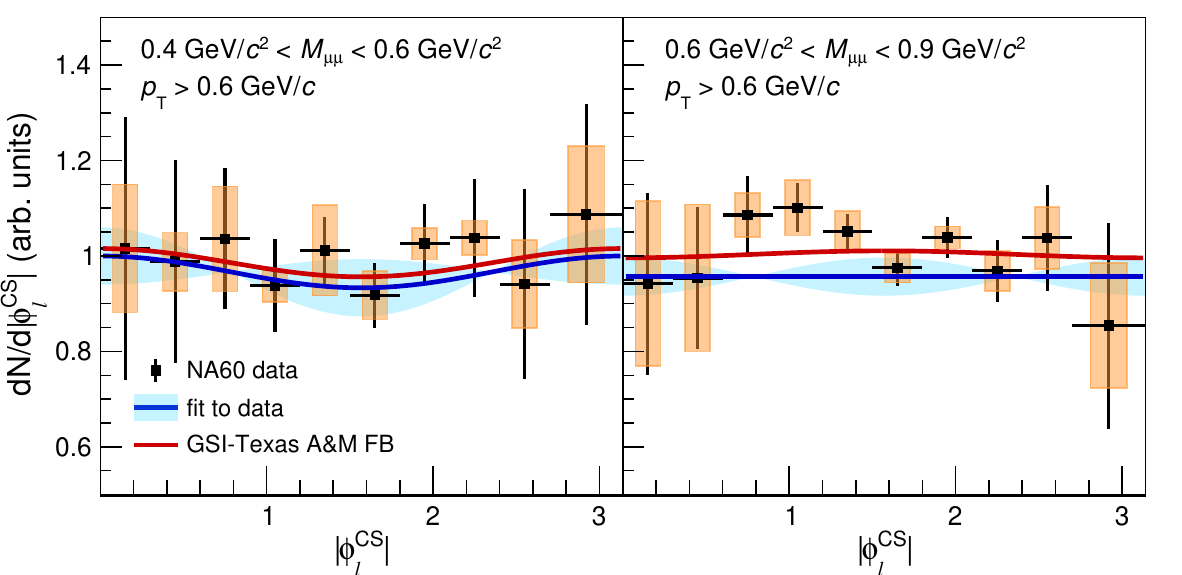}
\includegraphics[width=0.6\linewidth]{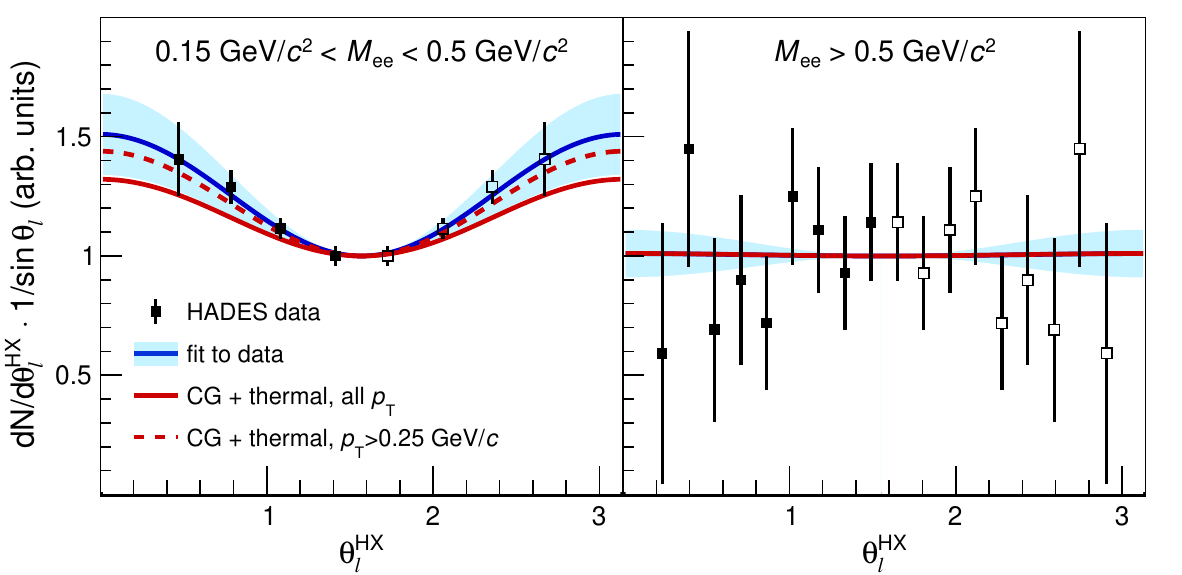}
\caption {Upper two rows: Dimuon angular distribution in the
  Colins-Soper frame calculated employing a coarse-graining approach
  using UrQMD simulations of the fireball evolution for implementing the
  in-medium thermal dilepton rates as described in \cite{Seck:2023oyt}
  in comparison to NA60 data in $158 A\,\GeV$ In+In collisions
  \cite{NA60:2008iqj}. The blue lines represent fits by NA60 with
  extracted anisotropy parameters of $\lambda_{\theta}$= -$0.10\pm0.24$
  (upper left) and -$0.13\pm0.12$ (upper right), compared to the
  calculated values of -0.04 and 0.01, respectively, and likewise
  $\lambda_{\phi}$=$0.05\pm0.09$ (lower left) and $0.00 \pm 0.06$ (lower
  right), compared to 0.04 and -0.01 from the calculation
  \cite{Seck:2023oyt}. Last row: Comparison of calculated angular
  distributions (red lines) as defined in the helicity frame, integrated
  over two dielectron mass bins \cite{Seck:2023oyt}, to HADES data in
  Ar(1.76\,GeV)+KCl collisions~\cite{HADES:2011nqx}. Fits by the HADES
  collaboration (blue lines with bands) yield anisotropy parameters of
  $\lambda_{\theta}$=$0.51 \pm 0.17$ (left panel) and $0.01\pm 0.1$
  (right panel) for the lower and higher mass window, respectively,
  compared to 0.34 and 0.01 from the calcuations. To illustrate effects
  of a finite HADES acceptance, theoretical results with a cut on
  transverse pair momentum in the lab of
  $p_T$$>$0.25\,GeV are also shown (dashed red line).}
\label{fig.2}
\end{center}
\end{figure}
More recently in addition also first assessments of the polarization of
the dilepton pairs in both experiment
\cite{NA60:2008iqj,Agakishiev:2011vf} and theory \cite{Seck:2023oyt} has
been achieved (cf.\ the last line of Fig.\ \ref{fig.2}). This probes the
difference between the transverse and longitudinal components of the
in-medium current-correlation function, which has some sensitivity to
the underlying in-medium scattering processes.

\begin{figure}[H]
\centerline{\includegraphics[width=0.97\linewidth]{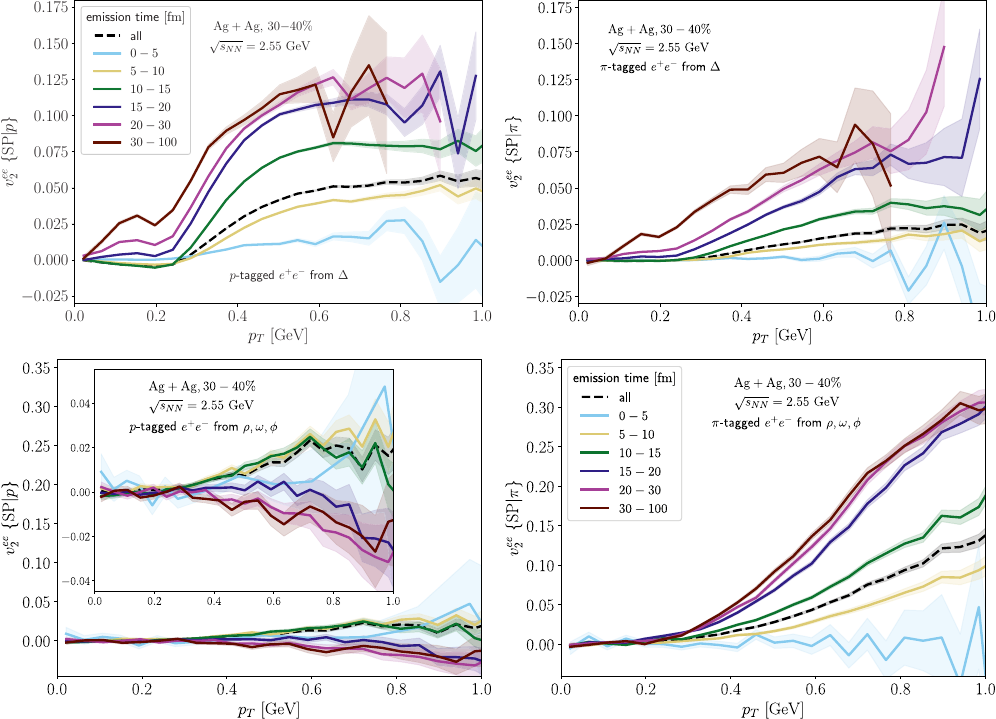}}
\caption{Scalar product flow of dielectrons radiated by (upper)
  $\Delta$ baryons (upper panel) and vector mesons (lower panel). Solid lines show the flow
  signal at different time intervals, and the dashed-black lines show
  the full time integrated contribution of the source. The reference
  plane is constructed with (left) protons or (right) pions. Figure
  taken from \cite{Goes-Hirayama:2024aqz}.}
\label{fig.dilep-h-tagged-v2}
\end{figure}

In \cite{Goes-Hirayama:2024aqz} the elliptic flow,
$v_2$ of di-electrons at HADES energies is investigated with the
hadronic transport model SMASH \cite{SMASH:2016zqf}. First measurements
of the HADES collaboration show $v_{2}^{(\e^+ \e^-)} \simeq
0$ (except in the very low-mass region, where the dileptons dominately
origin from pion-Dalitz decays and thus follow the negative
$v_2$ of the pions, which is due to ``squeeze-out'')
\cite{Galatyuk:2020lvg,Schild:2024ywo}. In \cite{Goes-Hirayama:2024aqz}
it is shown that the ``null result'' of the HADES experiment is due to
the cancellation of the
$v_2$ of dileptons from different, competing hadronic sources, and this
can be empirically validated by using the ``tagging method'' to
determine the dilepton-elliptic flow, using the cross correlation of the
event-flow vector of a specific particle species $X$,
\begin{equation}
q_n^X(\Omega)=\frac{\int_{\Omega} \dd \Omega \int_0^{2 \pi} \dd \phi \frac{\dd
    N^X}{\dd \Omega \dd \phi} \exp(\ii n \phi)}{\int_{\Omega} \dd \Omega \int_0^{2 \pi} \dd \phi \frac{\dd
    N^X}{\dd \Omega \dd \phi}},
\end{equation}
and the corresponding quantity for the dileptons,
\begin{equation}
v_n^{(\e^+ \e^-)}\{\text{EP}|h\}(\Omega) = \text{Cov}[q_n^{\e^+
  \e^-}(\Omega),q_n^h(\Omega)/|q_n^h(\Omega)|],
\end{equation} 
the socalled scalar-product flow. Using different hadron species,
$h$, for ``tagging'' it might be possible to distinguish the varying
$v_n$ of dileptons depending on their origin from different hadronic
sources (see Fig.\ \ref{fig.dilep-h-tagged-v2}).

\subsection{Dimuon production at top CERN-SPS energy}

\begin{figure}[t]
\centering
\begin{minipage}{0.4 \linewidth}
\includegraphics[width=0.99 \linewidth]{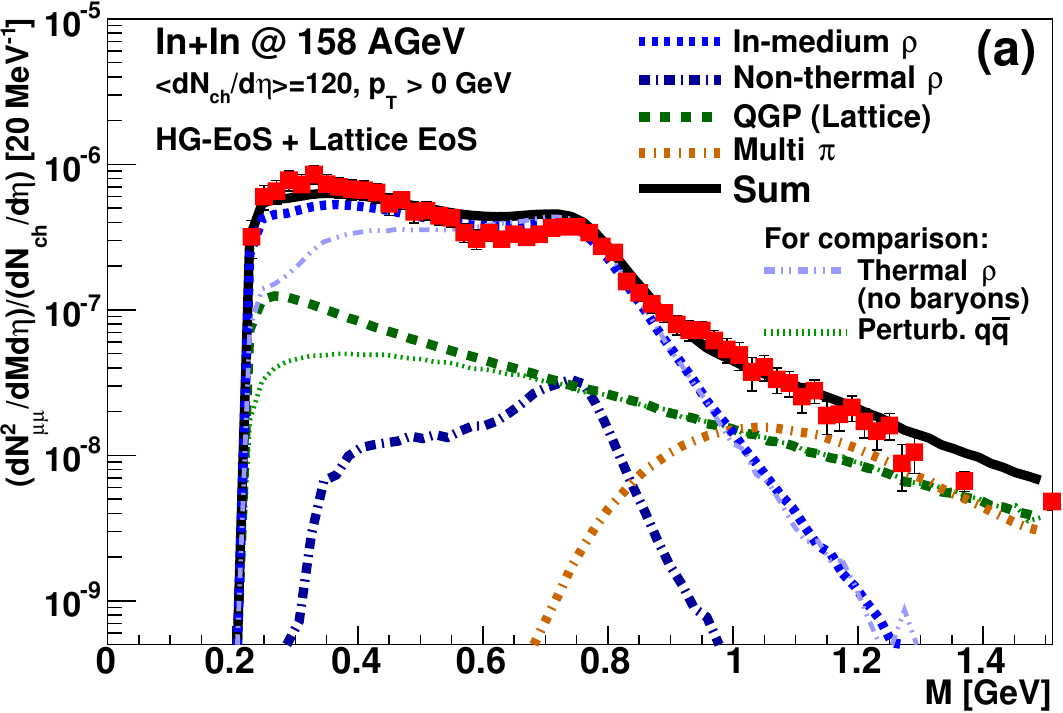}
\end{minipage} \hspace*{3mm}
\begin{minipage}{0.4 \linewidth}
\includegraphics[width=0.99 \linewidth]{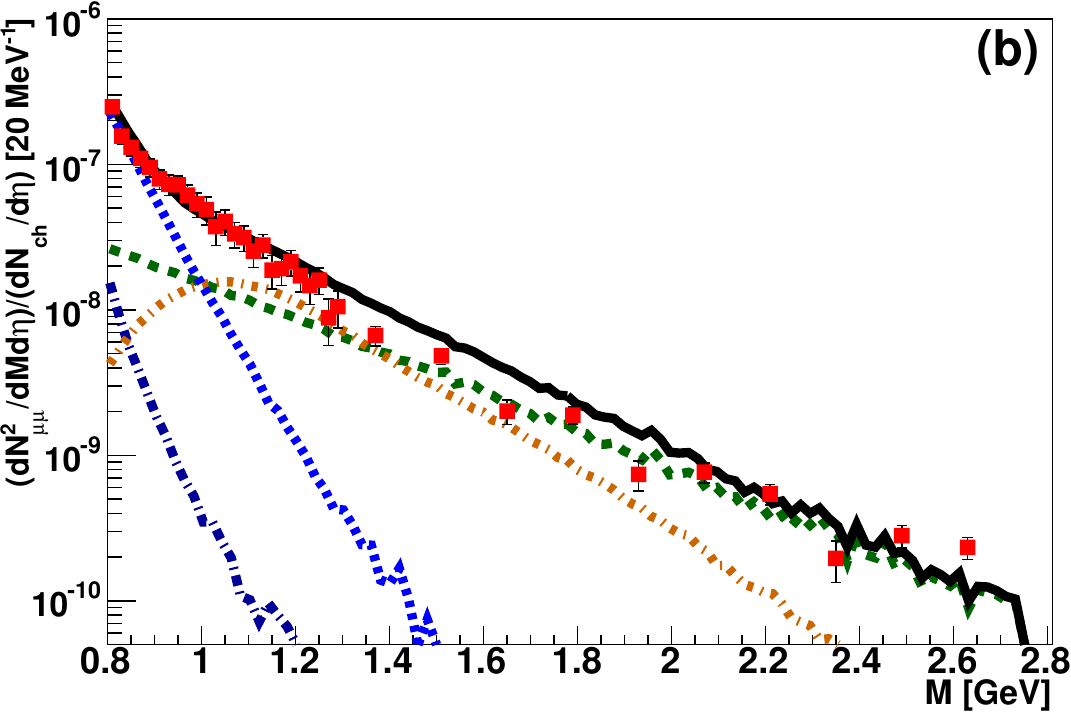}
\end{minipage}
\vspace*{3mm}
\centerline{\includegraphics[width=0.46 \linewidth]{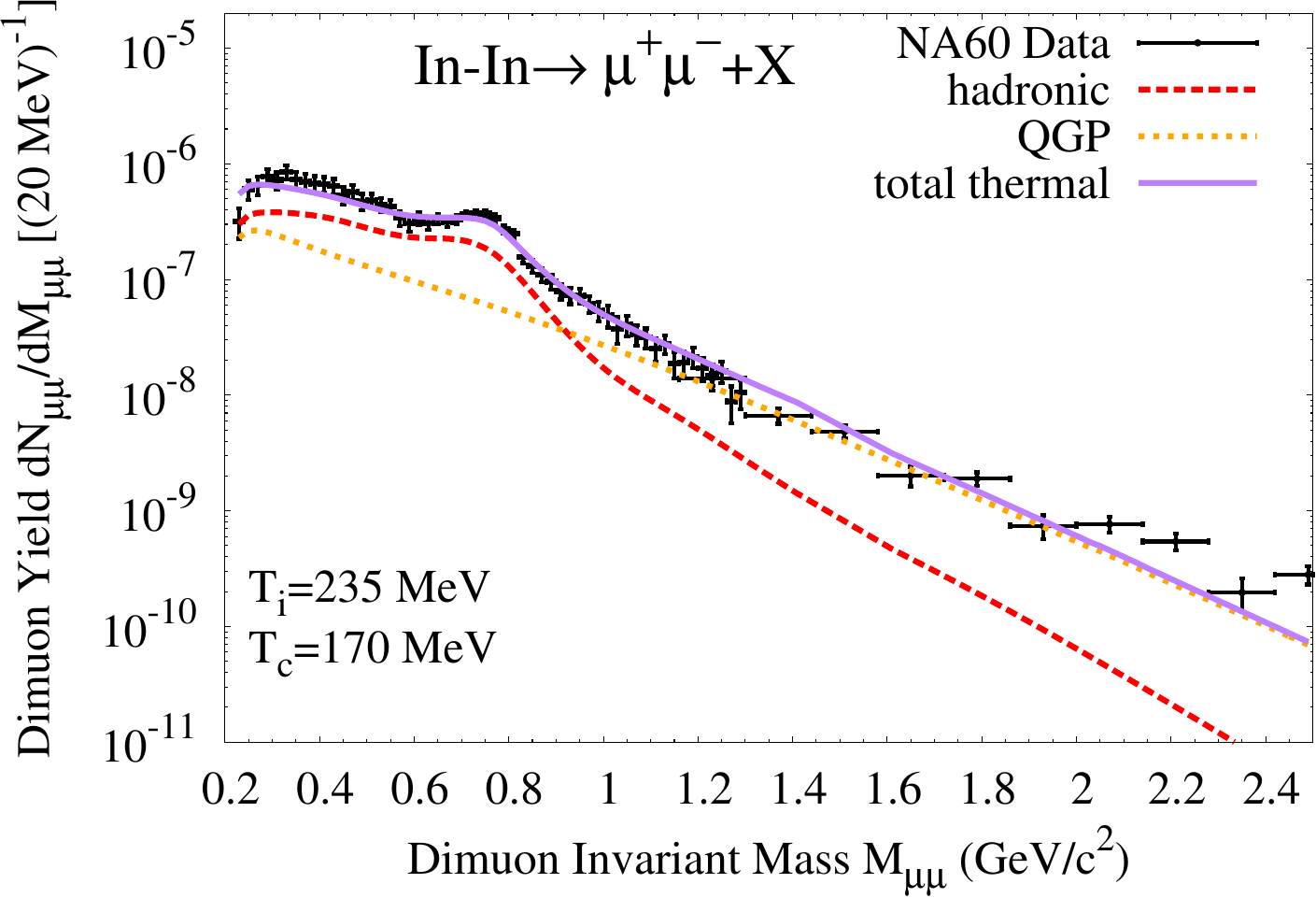}}
\caption{Upper panels: Invariant mass spectra of the
  dimuon excess yield in In+In collisions at a beam energy of
  $158 \;A \GeV$, for the low-mass region up $1.5\;\GeV$ (a) and the
  intermediate-mass regime up to $2.8\;\GeV$ (b) using the
  coarse-grained transport description of the medium. We show the
  contributions of the in-medium $\rho$ emission (blue short dashed),
  the contribution from the Quark-Gluon Plasma, i.e.,
  $q\overline{q}$-annihilation, according to lattice rates
  \cite{Ding:2010ga,Rapp:2013nxa} (green dashed) and the emission from
  multi-pion reactions (orange dash-dotted). Additionally a non-thermal
  transport contribution for the $\rho$ is included in the yield (dark
  blue dash-dotted). Only left plot: For comparison the thermal $\rho$
  without any baryonic effects, i.e. for $\rho_{\text{eff}}=0$, is shown
  (violet dash-double-dotted) together with the yield from pure
  perturbative $q\overline{q}$-annihilation rates (green dotted). The
  results are compared to the experimental data from the NA60
  Collaboration \cite{Arnaldi:2008er, Specht:2010xu,
    Arnaldi:2008fw}. Figures taken from {\protect
    \cite{Endres:2014zua}}. Lower panel: The same model for
  the dimuon-production rates, using the blast-wave-fireball
  parameterization for the medium evolution. Figures taken from
  {\protect \cite{Rapp:2014hha}}).}
\label{fig:na60-cg-urqmd-fireball}
\end{figure}
In \cite{vanHees:2006ng,vanHees:2007th} and \cite{Endres:2014zua}
the dimuon production in $158\;A\GeV$ In-In collisions as measured by
the NA60 collaboration
\cite{Arnaldi:2006jq,Arnaldi:2007ru,Damjanovic:2007qm,Arnaldi:2008er,Arnaldi:2008fw,Specht:2010xu}
at the CERN SPS have been calculated. To describe the medium evolution
both blast-wave parameterizations and the coarse-grained transport
approach have been applied.

As shown in Fig.\ \ref{fig:na60-cg-urqmd-fireball} both descriptions of
the fireball evolution lead to an excellent description of the data. It
is important to note that here the hadronic cocktail has been subtracted
from the data by the NA60 collaboration, i.e., a fully acceptance
corrected \emph{excess} spectrum is shown. Also the contribution from
decays of correlated $\D$ and $\overline{\D}$ mesons and, for invariant
masses $M_{\mu^+ \mu^-} > 1.2 \; \GeV$, Drell-Yan processes is
subtracted. In the low-mass region
$2m_{\mu} <M_{\mu^+ \mu^-} \lesssim 1 \; \GeV$ the dominant contribution
to the excess yield is from the decay of the tremendously broadened
$\rho$ mesons from the ``thermal'' medium, while at higher masses the
contribution from the QGP is the leading contribution. Again the
importance of the medium modifications becomes evident by comparing the
full result to the case, where the baryon contributions to the
$\rho$-meson self-energy are neglected. Particularly the enhancement in
the very-low-mass region towards the two-muon threshold is mostly due to
the interactions of the $\rho$ meson with baryons (and anti-baryons),
i.e., due to Dalitz decays of baryon resonances, which are included in
the thermal quantum-field theoretical evaluation of the in-medium
$\rho$-self-energy. As shown in \cite{Endres:2014zua} the model also
successfully describes the $q_{\mathrm{t}}$ dependence as well as the
mass spectra in various $q_{\mathrm{t}}$ bins. It should also be noted
that in the invariant-mass region
$m_{\phi} \lesssim M_{\mu^+ \mu^-} \lesssim M_{J/\psi}$ after the above
mentioned subtraction of contributions from decays of correlated
$\D$-$\overline{\D}$ pairs and the Drell-Yan process, a basically purely
thermal contribution from the resonance-free region of the dilepton
emission allows a direct determination of the space-time weighted
average of the temperature. In this intermediate-mass region
($T \ll M_{\mu^+ \mu^-}$) the emission from the earlier hot stages of
the fireball evolution dominates, as is also reflected in the model
calculation which identifies the thermal emission from the QGP as the
main source. Indeed, a fit of the experimental data leads to a
temperature $T=205 \pm 12 \;\MeV$ \cite{Specht:2010xu}, in accordance
with the model.

\subsection{Dileptons at FAIR and RHIC-BES energies}

\begin{figure}[t]
\centering 
\includegraphics[width=0.45\linewidth]{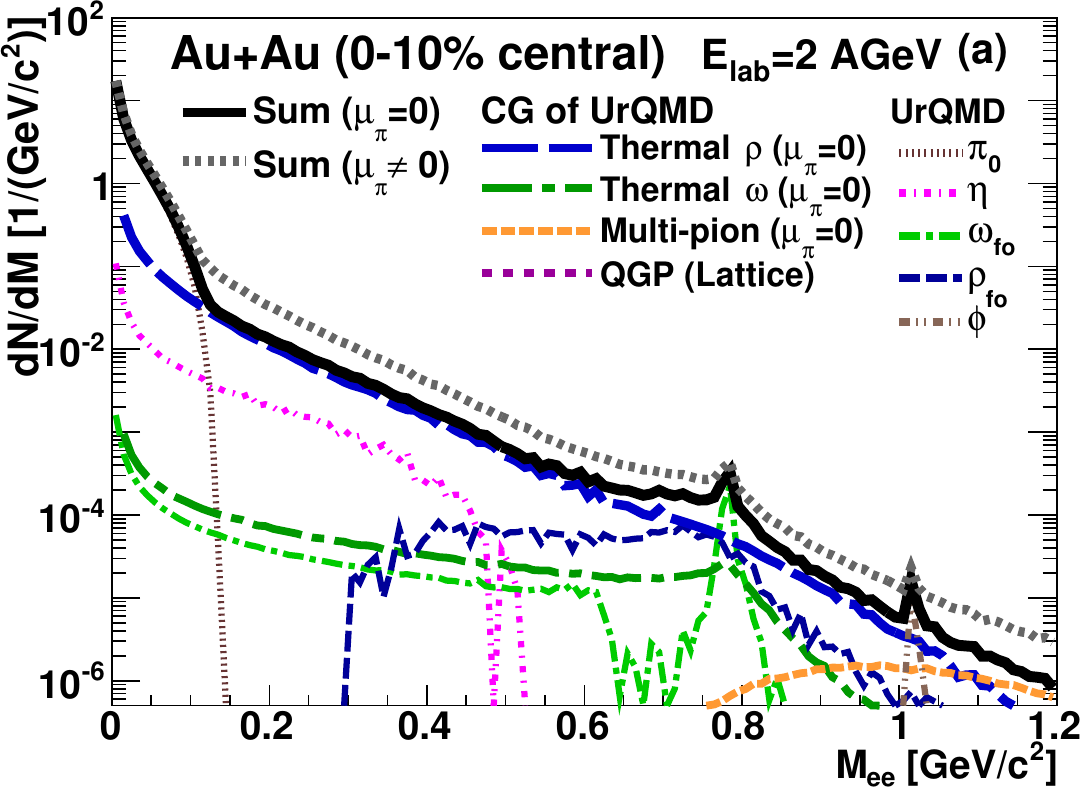} \hspace{7mm}
\includegraphics[width=0.45\linewidth]{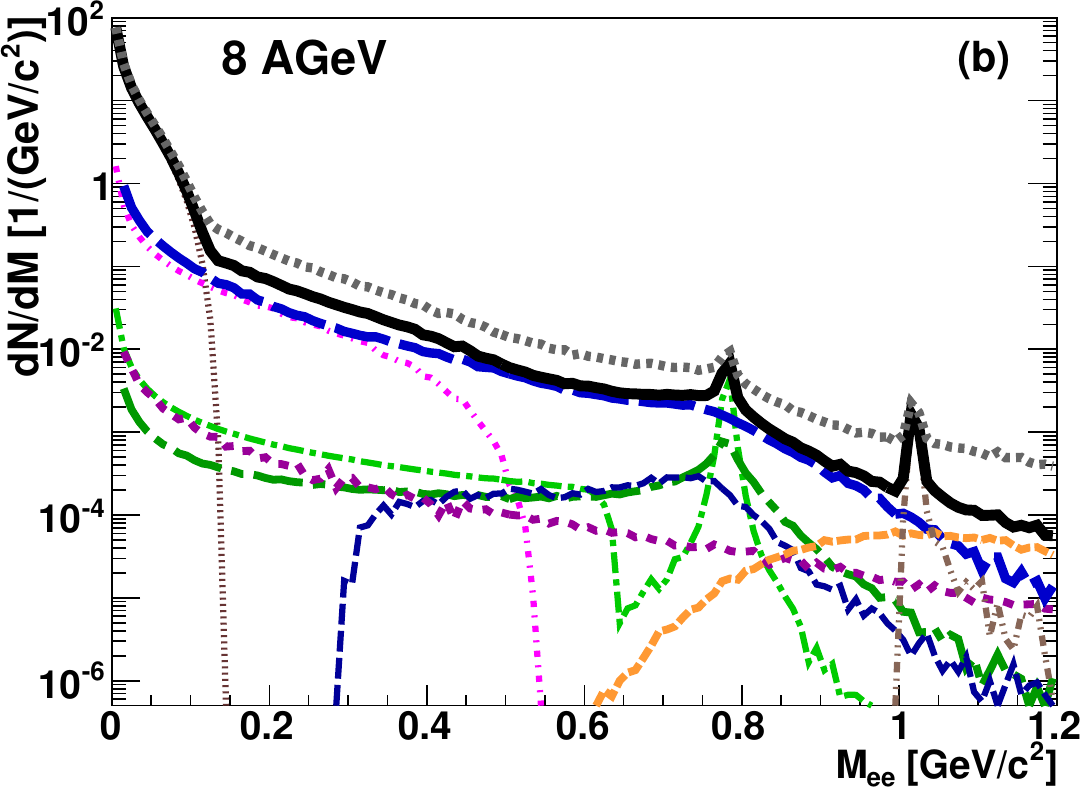}
\\[1mm]
\centering
\includegraphics[width=0.45 \linewidth]{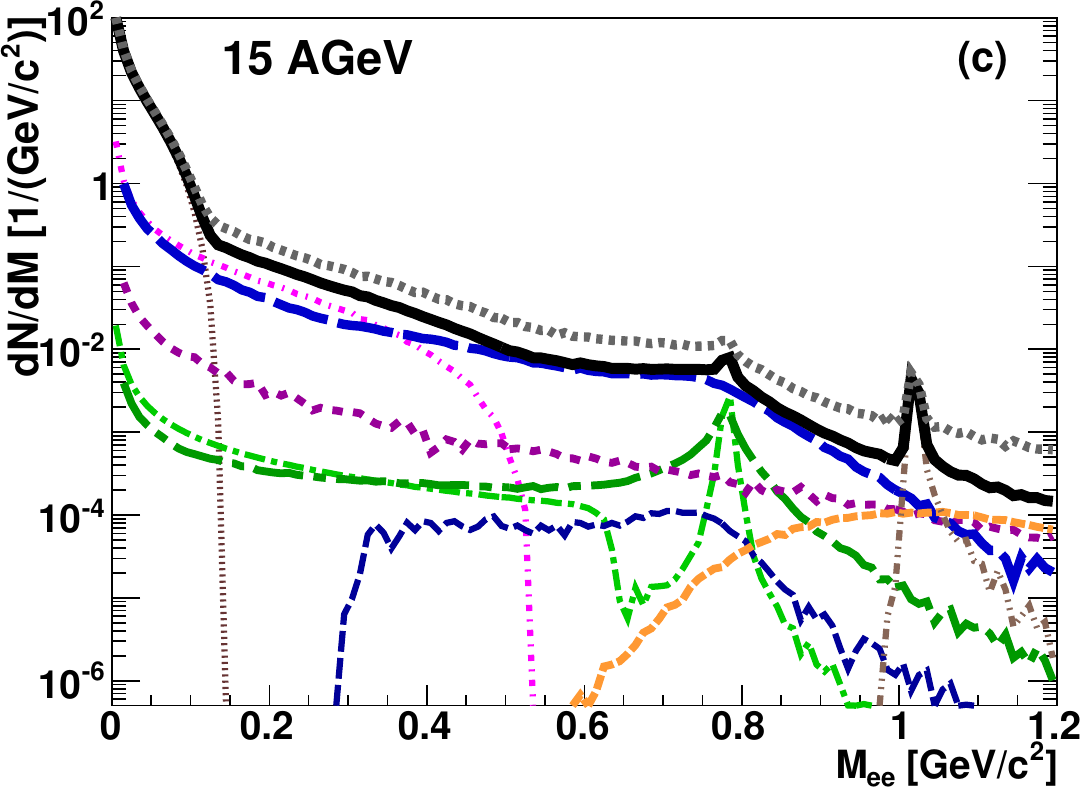} \hspace{7mm}
\includegraphics[width=0.45 \linewidth]{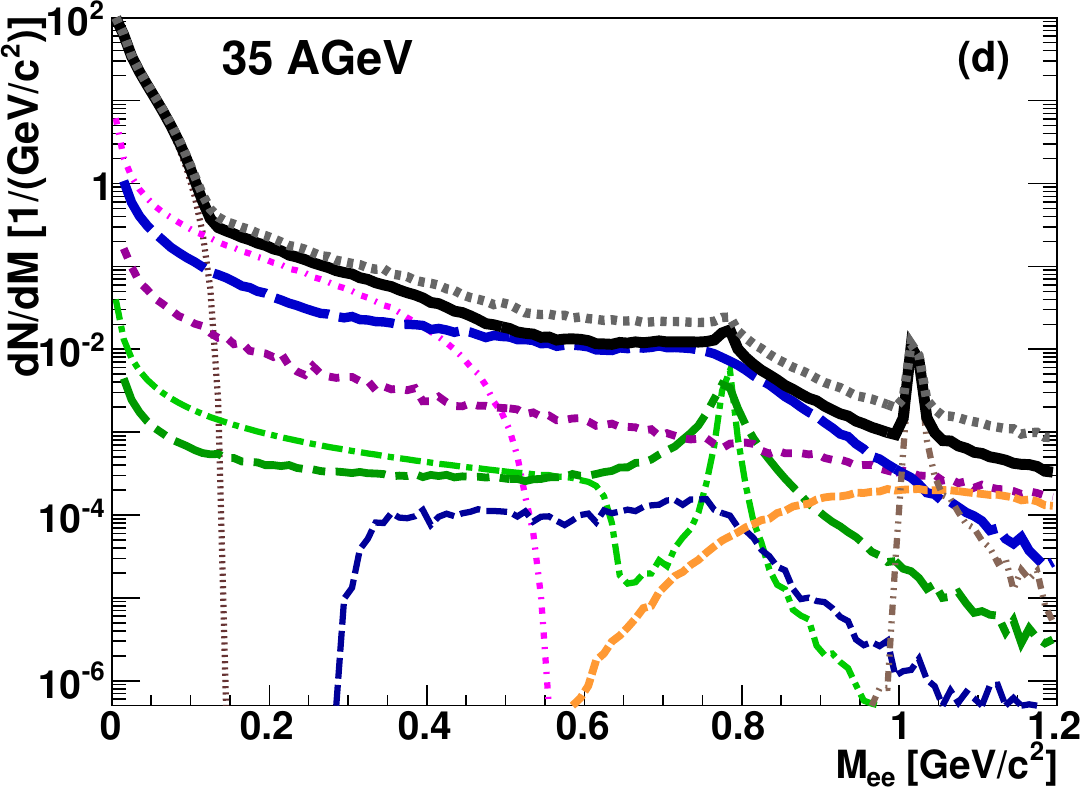} 
\caption{Dilepton invariant mass spectra for Au+Au reactions at different
  energies $E_{\text{lab}}=2$-$35\;A\GeV$ within the centrality class of
  0-10\% most central collisions. The resulting spectra include thermal
  contributions from the coarse-graining of the microscopic simulations
  (CG of UrQMD) and the non-thermal contributions directly extracted
  from the transport calculations (UrQMD). The hadronic thermal
  contributions are only shown for vanishing pion chemical potential,
  while the total yield is plotted for both cases, $\mu_{\pi}=0$ and
  $\mu_{\pi}\neq 0$. Figure taken from {\protect
    \cite{Endres:2015egk}} }
\label{fig.dileps-bes-fair}
\end{figure}
With the motivation to find possible signatures of the various phase
transitions in the phase diagram of strongly interacting matter a
beam-energy scan (BES) program is ongoing at RHIC and is also planned by
the Compressed Baryonic Matter (CBM) experiment at the upcoming Facility
for Antiproton and Ion Research (FAIR). In \cite{Endres:2015egk} we
have thus evaluated the invariant-mass spectra for the four beam
energies, $E_{\text{lab}}=2, \; 8 ,\; 15$ and $35\; A\GeV$ within the
coarse-grained transport approach.

Since particularly at the lower beam energies the pion-chemical
potential becomes quite large in the coarse-graining approach, we study
its influence on the dilepton yield by using lower and upper boundaries
with the following arguments: The lower bound is simply given by
assuming $\mu_{\pi}=0$. For the upper bound it should be noted that
$\mu_{\pi}$ is an effective description for the off-chemical equilibrium
nature of the medium, and thus in the here used Boltzmann approximation,
its influence on the dilepton-production rate is given by a fugacity
factor
\begin{equation}
\label{dil-bes.1}
z_{\pi}^n=\exp \left(\frac{n \mu_{\pi}}{T} \right),
\end{equation}
where $n$ is the difference in the number of pions in the initial and
final state of the corresponding reaction process. For processes
involving the $\rho$ meson, particularly dilepton production in the here
employed VMD model, not only two-pion production
$\pi \pi \rightarrow \rho$ is relevant but, particularly at lower beam
energies, baryonic channels like
$\pi N \rightarrow N^*/\Delta \rightarrow \rho$ become important, for
which $n < 2$. Thus, to estimate an upper limit of the influence of the
pion chemical potential on the dilepton yield, we use $n=2$ in our
calculations.

As can be seen in Fig.\ \ref{fig.dileps-bes-fair}, at all beam energies
in the very-low-mass region, $M_{\ee^+ \ee^-}<0.15 \; \GeV$, the
dilepton yield is dominated by the Dalitz decays of neutral pions,
$\pi^0 \rightarrow \gamma \ee^+ \ee^-$, while beyond that region up to
the vacuum-pole mass of the $\rho$ meson of $770 \; \MeV$ the main
contribution is radiation from thermal sources with medium-modified
$\rho$- and $\omega$-meson spectral functions. Although the absolute
yield of the thermal component increases with $E_{\text{lab}}$, its
relative weight compared to the non-thermal $\eta$-Dalitz component and
thus the enhancement above the hadronic cocktail contribution decreases.

Concerning the onset of deconfinement the calculations show that
temperatures $T \simeq 170 \; \GeV$ are reached in the region of
$E_{\text{lab}}=6$-$8 \;A \GeV$. 

Note that here we use the same cross-over-transition EoS for the
coarse-graining procedure as for the higher beam energies. A possible
deviation of the dilepton spectra from these predictions might thus
indicate a possible change in the nature of the
confinement-deconfinement and/or chiral phase transition.

\begin{figure}[h]
\centerline{\includegraphics[width=0.9\linewidth]{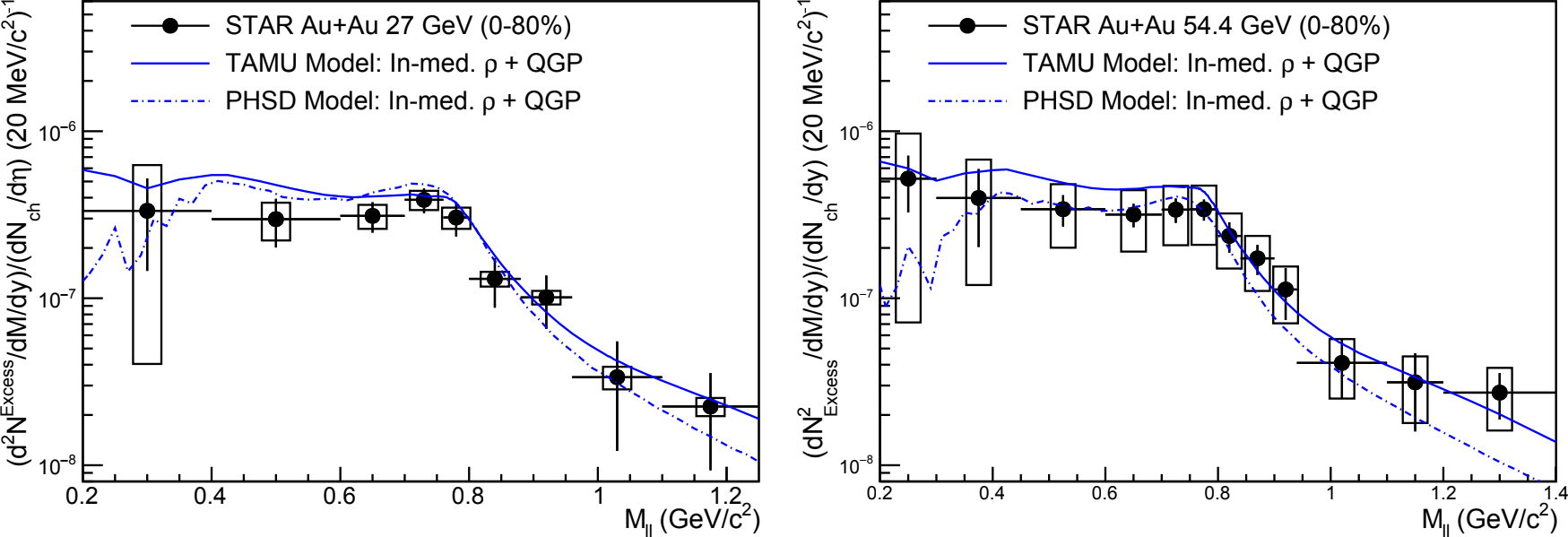}}
\caption{Low mass excess dielectron invariant mass spectrum for Au+Au
  collisions at $\sqrt{s_{NN}} = 27 \,\GeV$ (left) and $54.4 \, \GeV$
  (right), normalized by $\dd N_{\text{ch}}/\dd y$, compared to the
  theoretical calculations from the TAMU
  \cite{vanHees:2006ng,Rapp:2000pe,Rapp:2014hha} and PHSD
  \cite{Cassing:1997jz,Cassing:1999es} models. Vertical bars and boxes
  around data points represent the statistical and systematic
  uncertainties, respectively. Figure taken from \cite{STAR:2024bpc}.}
\label{fig.dileps-besii-star-M-spec}
\end{figure}

Recently a comprehensive analysis of dielectron measurements by the STAR
collaboration within the beam-energy-scan program (BES II) at RHIC has
become available \cite{STAR:2024bpc}. The measured acceptance corrected
access yield are well compatible with models of the type described
above. This is illustrated in Fig.\ \ref{fig.dileps-besii-star-M-spec}.

\subsection{Dileptons at RHIC and LHC energies}

\begin{figure}[t]
\centering
\includegraphics[width=0.45\linewidth]{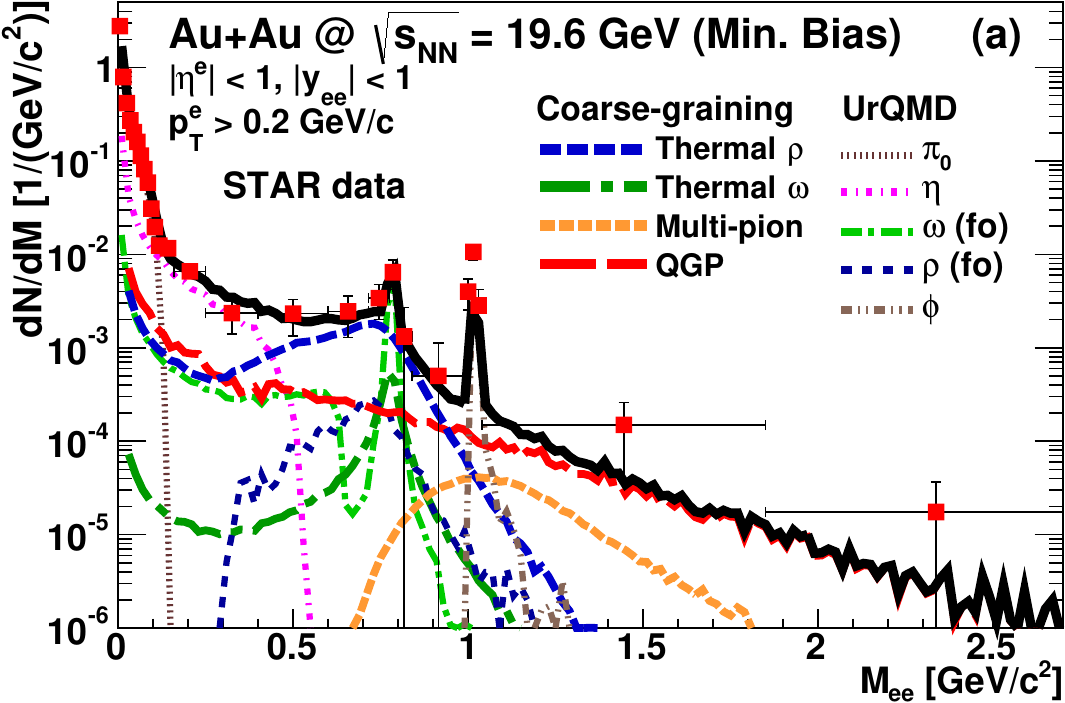} \hspace{7mm}
\includegraphics[width=0.45\linewidth]{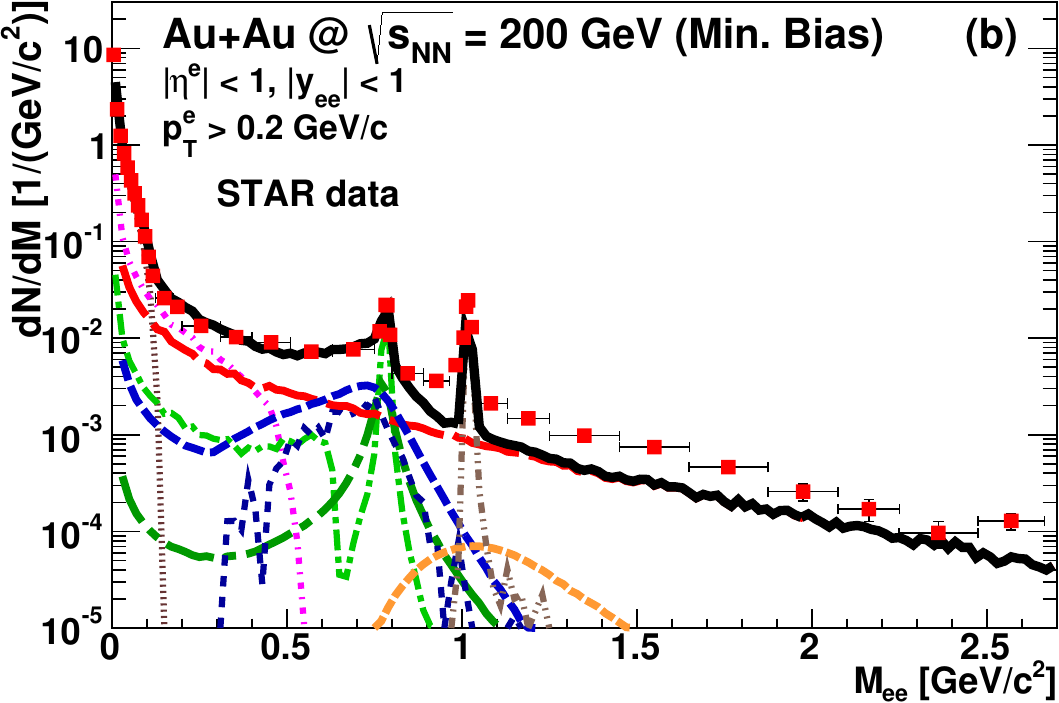}
\\[1mm]
\centering
\includegraphics[width=0.45\linewidth]{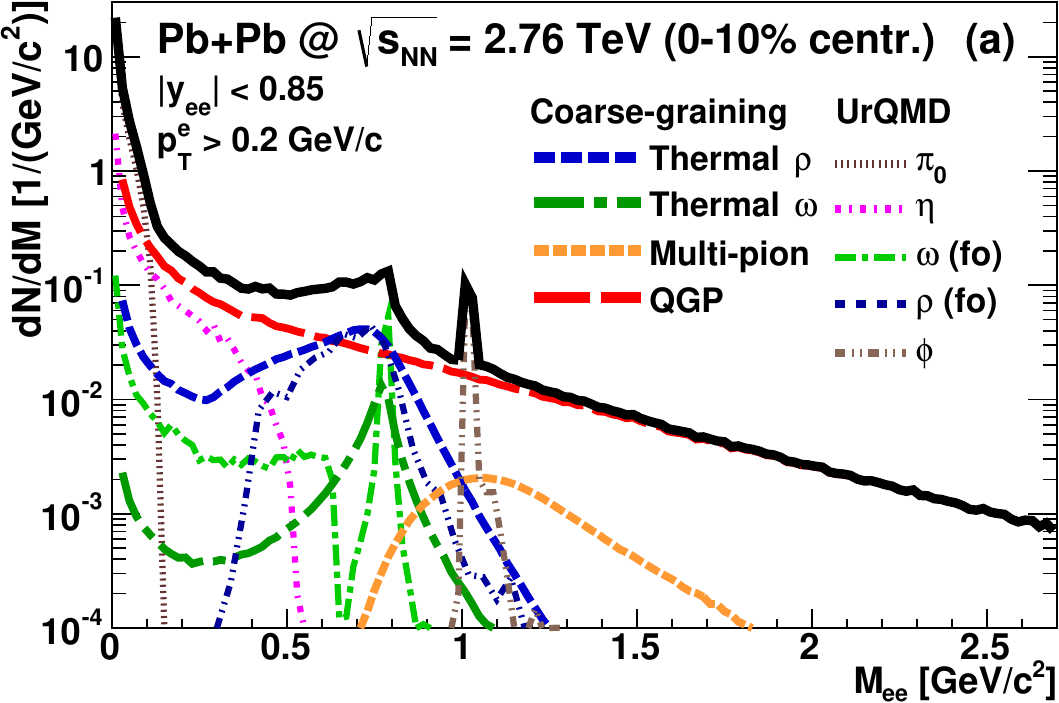} \hspace{7mm}
\includegraphics[width=0.45\linewidth]{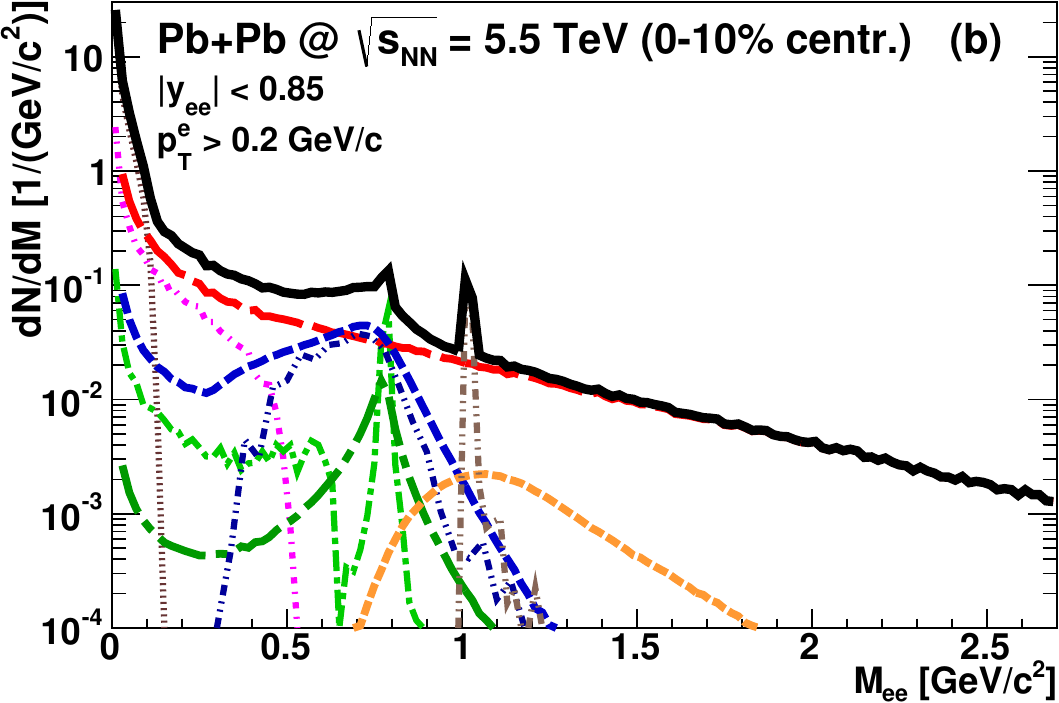}
\caption{Upper panel: Dielectron invariant-mass spectra for
  minimum bias (i.e., 0-80\% most central) Au+Au collisions at
  $\sqrt{s_{NN}}=19.6 \;\GeV$ (a) and $200\; \GeV$ (b). The sum includes
  the thermal hadronic and partonic emission obtained with the
  coarse-graining procedure, and also the hadronic $\pi$-, $\eta$- and
  $\phi$-decay contributions from UrQMD as well as the ``freeze-out''
  contributions (from cold cells) of the $\rho$ and $\omega$ mesons. The
  model results are compared to experimental data obtained by the STAR
  Collaboration \cite{Adamczyk:2015lme}. \textbf{Lower panel:}
  Dielectron invariant-mass spectra for 0-10\% most central Pb+Pb
  collisions at $\sqrt{s_{NN}}=2.76 \; \TeV$ (a) and $5.5 \; \TeV$
  (b). The sum includes the thermal hadronic and partonic emission
  obtained with the coarse-graining approach, and also the hadronic
  $\pi$, $\eta$, and $\phi$ decay contributions from UrQMD as well as
  the ``freeze-out'' contributions (from cold cells) of the $\rho$ and
  $\omega$ mesons. Figures taken from {\protect
    \cite{Endres:2016tkg}}}.
\label{fig:dileps-rhich-lhc}
\end{figure}
In \cite{Endres:2016tkg} we have employed the coarse-grained
transport approach to the evaluation of dilepton spectra at RHIC and LHC
energies. In Fig.\ \ref{fig:dileps-rhich-lhc} the result is shown in
comparison to experimental data on Au-Au collisions at the two beam
energies $\sqrt{s_{NN}}=19.6 \; \GeV$ and $200 \; \GeV$
\cite{Adamczyk:2015lme}. The calculation takes into account the
single-electron rapidity, the dielectron pseudorapidity, and
transverse-momentum electron cuts ($\eta_{\ee}<1$, $y_{\ee}<1$,
$p_t^{\ee}>0.2 \; \GeV$) to account for the STAR acceptance. For the
low-mass region, $M_{\ee^+ \ee^-}< 1 \;\GeV$ the data are well described
within the model. In the region
$0.3 \;\GeV <M_{\ee^+ \ee^-}< 0.7 \;\GeV$ an access above the hadronic
cocktail has been observed, and within our model at the lower beam
energy this region is dominated by thermal contributions from
medium-modified $\rho$ mesons, while at top RHIC energy the emission
from the QGP prevails. In both cases the spectral function of the
in-medium $\rho$ meson shows more similarities with its vacuum shape
than at the lower beam energies discussed in the previous sections. This
is understandable by the fact that here the baryon-chemical potential
$\mu_{\text{B}}$ is smaller than at lower beam energies, and a great
part of the broadening in the peak region as well as the low-mass tail
is mainly due to the baryon interactions of the $\rho$ meson, as already
emphasized before. At intermediate masses $M_{\ee^+ \ee^-}>1 \;\GeV$ our
calculation underestimates the measured yield, which can be explained by
the fact that here the contributions from Drell-Yan processes as well as
decays of correlated $\D \overline{\D}$ mesons have been neglected. For
further details on the comparison of the model on the data, see
\cite{Endres:2016tkg}.

Also in Fig.\ \ref{fig:dileps-rhich-lhc} we present our predictions for
Pb-Pb collisions at center-mass energies, available at the LHC,
$\sqrt{s_{NN}}=2.76\;\TeV$ and $5.5 \; \TeV$. As to be expected also
here the thermal $\rho$ contribution in the low-mass region shows the
vacuum-like peak structure as already seen at the lower RHIC energies;
again this is due to the even smaller baryochemical potential at the
higher beam energies. Compared to RHIC energies the fireball at LHC
energies starts with considerable higher temperatures, leading to larger
lifetimes for both the partonic and hadronic phase of the fireball
evolution, resulting in larger contributions from both the QGP and the
thermal vector mesons in the low-mass region.

More recently a detailed study based on calculations of the dilepton
production in the pre-equilibrium stage of the fireball evolution as
well as thermal partonic dilepton rates at next-to-leading (NLO)
perturbative-QCD order in $\alpha_{\text{s}}$
\cite{Laine:2013vma,Jackson:2019mop} including dilepton spectra,
anisotropic flow coefficients, and polarization observables at LHC
energies has been achieved in
\cite{Wu:2024pba,Wu:2024vyc,Gao:2026vxs}. Here the fireball evolution
has been described in a multi-stage model: the initial state is
described, using the IP-Glasma \cite{Schenke:2012wb,Schenke:2012hg} and
K{\O}MP{\O}ST \cite{Kurkela:2018vqr,Kurkela:2018wud} models as described
in \cite{Gale:2021emg}. The corresponding dilepton rate is estimated by
determining a quasi-hydrodynamic description in terms of a local
temperature and fluid velocity of the off-equilibrium medium by Landau
matching with the HotQCD Equation of State. Since the IP-Glasma and
K{\O}MP{\O}ST assume a purely gluonic initial states and the dileptons
can only occur from the charged quarks, which build up in processes
like $gg \rightarrow \e^+ \e^-$ during the pre-equilibrium fireball
evolution, an effective suppression factor is implemented. After a fixed
formation time $\tau_0^{(\text{hydro})}=0.8\,\fm/c$ is decribed by the
IBE-MUSIC hydrodynamic simulation
\cite{Paquet:2015lta,Schenke:2010nt,Schenke:2010rr} with the same
thermal dilepton rates as used for the pre-equilibrium stage.

The hydrodynamic evolution ends by freezing out if a fluid cell's energy
density drops below $\epsilon_{frz}=0.18 \, \GeV/\fm^3$ using the
Cooper-Frye description taking into account viscous effects
\cite{Cooper:1974mv,Ryu:2015vwa,Schenke:2020mbo} followed by a hadronic
UrQMD afterburner \cite{Shen:2014vra}.

\begin{figure}[H]
\includegraphics[width=0.5\linewidth]{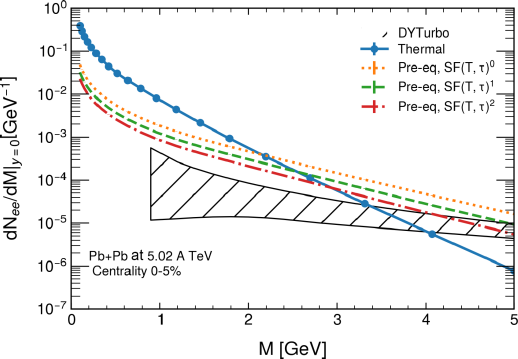} \hfill
\includegraphics[width=0.445\linewidth]{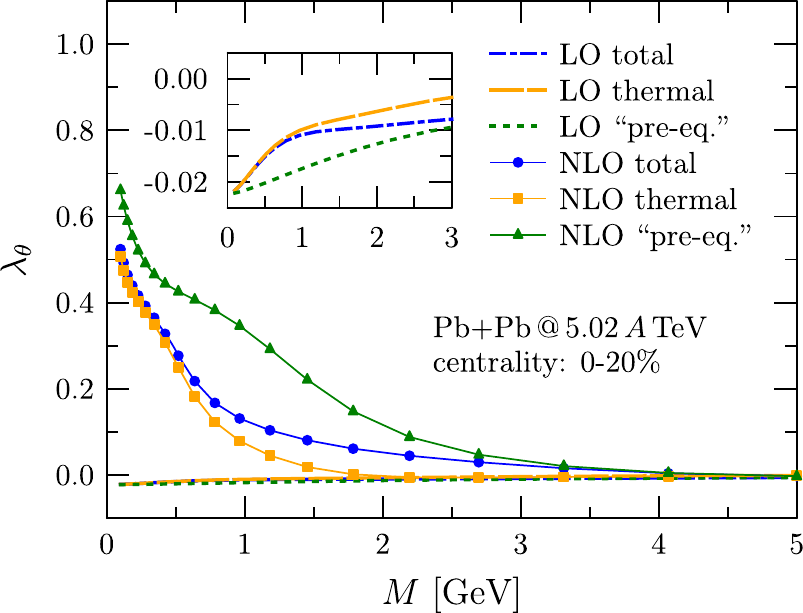}
\caption{Left panel: The thermal dilepton-invariant mass spectrum in
  comparison to the off-equilibrium distribution including the effective
suppression factor (0: none, 1: linear, 2: squared) as well as the
contribution from the Drell-Yan processes employing the DYTURBO package
\cite{Camarda:2019zyx}. Figure taken from \cite{Wu:2024pba}. Right
panel: The polarization coefficient $\lambda_{\theta}$ in the HX
frame. Figure taken from \cite{Wu:2024vyc}.}
\label{fig.nlo-off-eq-dileps}
\end{figure}

As illustrative examples for the results of this comprehensive model we
depict in Fig.\ \ref{fig.nlo-off-eq-dileps} the invariant mass spectrum (left),
which sholws the sensitivity of the dilepton yield to
off-equilibrium (``non-thermel'') contributions, and the polarization
coefficient $\lambda_{\theta}$ in the HX frame, underlining the
sensititivity of the polarization observables to NLO pQCD contributions.

\begin{figure}[h]
\centerline{\includegraphics[width=0.95\linewidth]{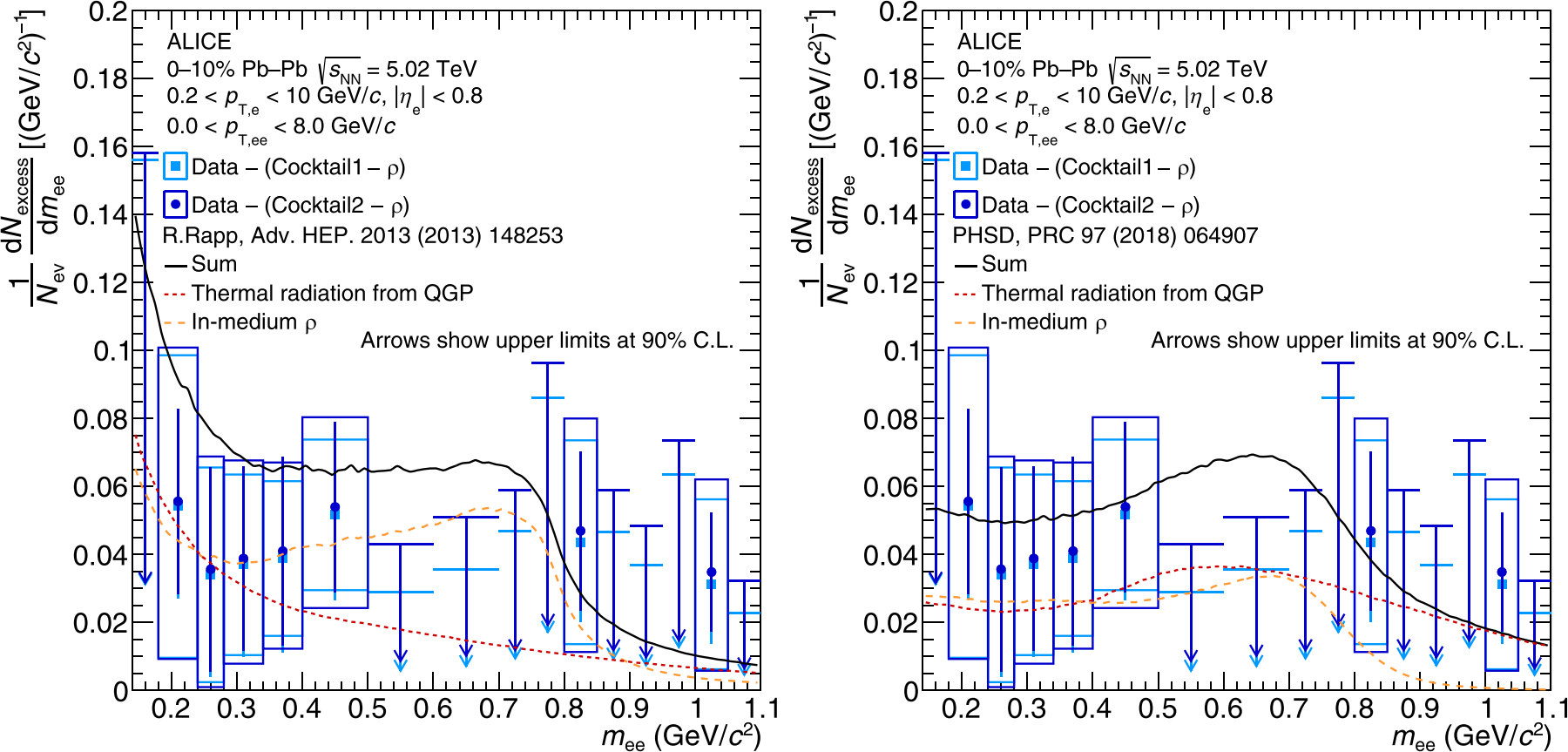}}
\caption{Excess yield of dielectrons in the 10{\%} most central Pb-Pb
  collisions at $\sqrt{s_{NN}}= 5.02 \, \TeV$ with respect to the
  expected $\e^{+} \e^{-}$ contributions from known hadronic sources,
  including (Cocktail2) or not including (Cocktail1) medium effects for
  the heavy-flavor contributions, and compared with predictions from the
  model of Rapp \cite{Rapp:2013nxa} (left) and from the PHSD transport
  approach \cite{Song:2018xca} (right). Figure taken from
  \cite{ALICE:2023jef}.}
\label{fig.lhc-dielectrons-5_02TeV}
\end{figure}

Recently the ALICE collaboration has measured dielectrons at the LHC in
Pb-Pb collisions at $s_{NN}=5.02 \, \TeV$ \cite{ALICE:2023jef}. As
illustrated in Fig.\ \ref{fig.lhc-dielectrons-5_02TeV} the excess yield
is in accordance with the Rapp-Wambach as well as with the PHSD
models. It is notable that for the first time the ``distances of closest
approach'' method has been employed to subtract possible (thermal)
contributions of dileptons from correlated $\D \overline{D}$ decays in
the IMR ($1.2 \, \GeV < M_{\e^+ \e^-} < 2.6 \, \GeV$).

\subsection{Dileptons and the QCD phase diagram}
\label{sect:dileps-qcd-phasediag}

One of the most challenging aims of contemporary heavy-ion-collision
research is the identification of observables indicating changes in the
nature of the confinement-deconfinement or the chiral phase transitions,
e.g., the cross-over transition at low baryochemical potential to a
first-order transition at higher net-baryon densities with a critical
point at the end of the corresponding phase-transition line in the QCD
phase diagram. This has been the motivation for a concise ``beam-energy
scan'' at RHIC and in the future at FAIR.
\begin{figure}[H]
\centering
  \includegraphics[width=0.45\linewidth]{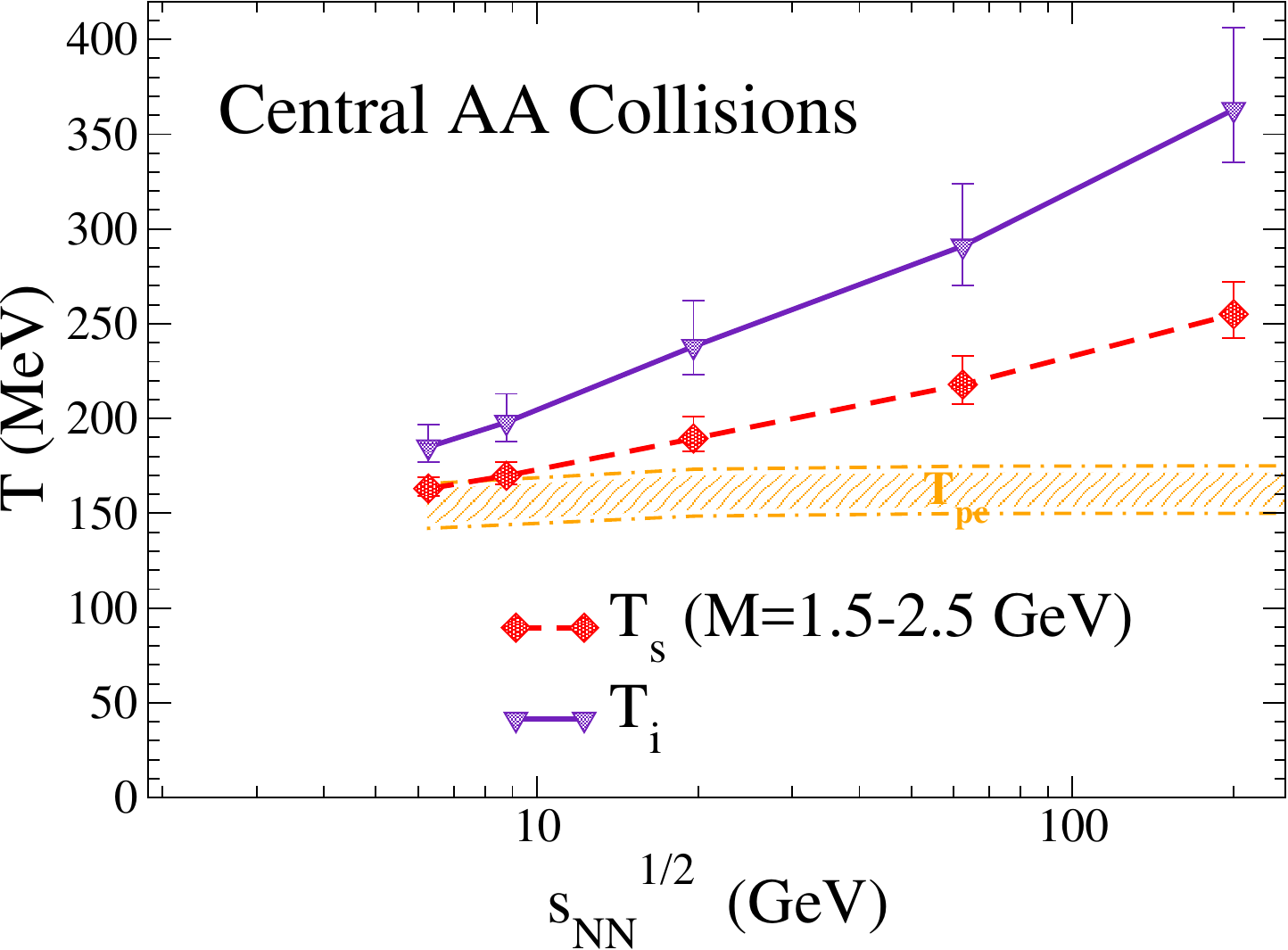} \hspace{3mm}
  \includegraphics[width=0.47\columnwidth]{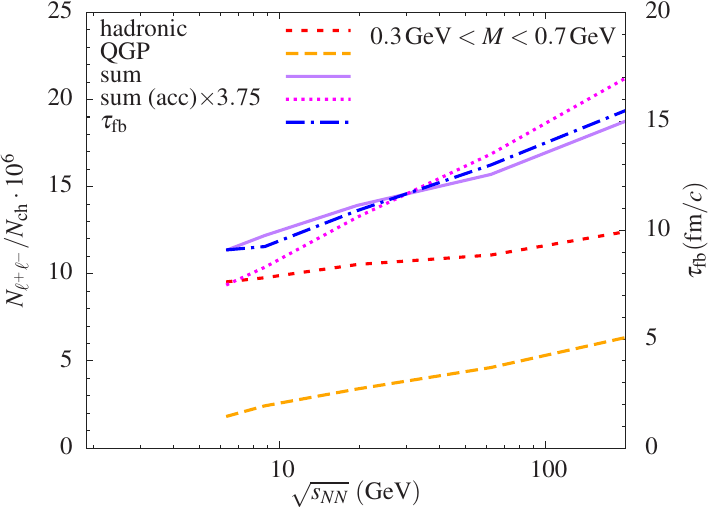}
  \caption{Left panel: Excitation function of the inverse-slope
    parameter, $T_{\mathrm{s}}$, from intermediate-mass dilepton spectra
    ($M=1.5$-$2.5\; \GeV$, diamonds connected with a dashed line) and
    initial temperature $T_{\mathrm{i}}$ (triangles connected with a
    solid line) in central heavy-ion collisions ($A \simeq 200$). The
    error bars on $T_{\mathrm{s}}$ and $T_{\mathrm{i}}$ correspond to a
    variation in the initial longitudinal fireball size, $z_0$, by
    $\pm 30\%$ around the central values. The hatched area schematically
    indicates the pseudo-critical temperature regime at vanishing (and
    small) baryochemical potential as extracted from various quantities
    computed in lQCD \cite{Borsanyi:2010bp}. \textbf{Right
      panel:} Excitation function of low-mass thermal dilepton radiation
    (``excess spectra'') in 0-10\% central AA collisions
    ($A \simeq 200$), integrated over the mass range
    $M = 0.3 \; \GeV$-$0.7 \; \GeV$, for QGP (dashed line) and in-medium
    hadronic (short-dashed line) emission and their sum (solid
    line). The underlying fireball lifetime (dot-dashed line) is given
    by the right vertical scale. Figures taken from {\protect
      \cite{Rapp:2014hha}}}
\label{fig:phase-dil.1}
\end{figure}

\begin{figure}[t]
\centering
\includegraphics[width=0.45\linewidth]{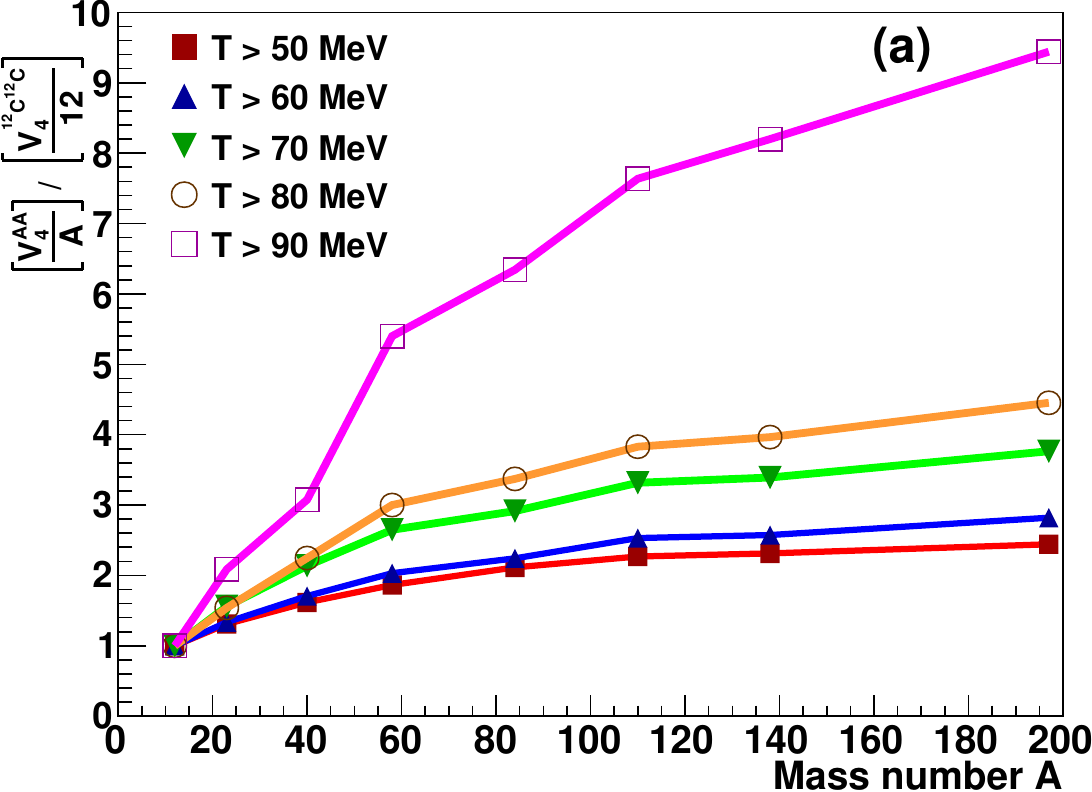} \hspace{3mm}
\includegraphics[width=0.45\linewidth]{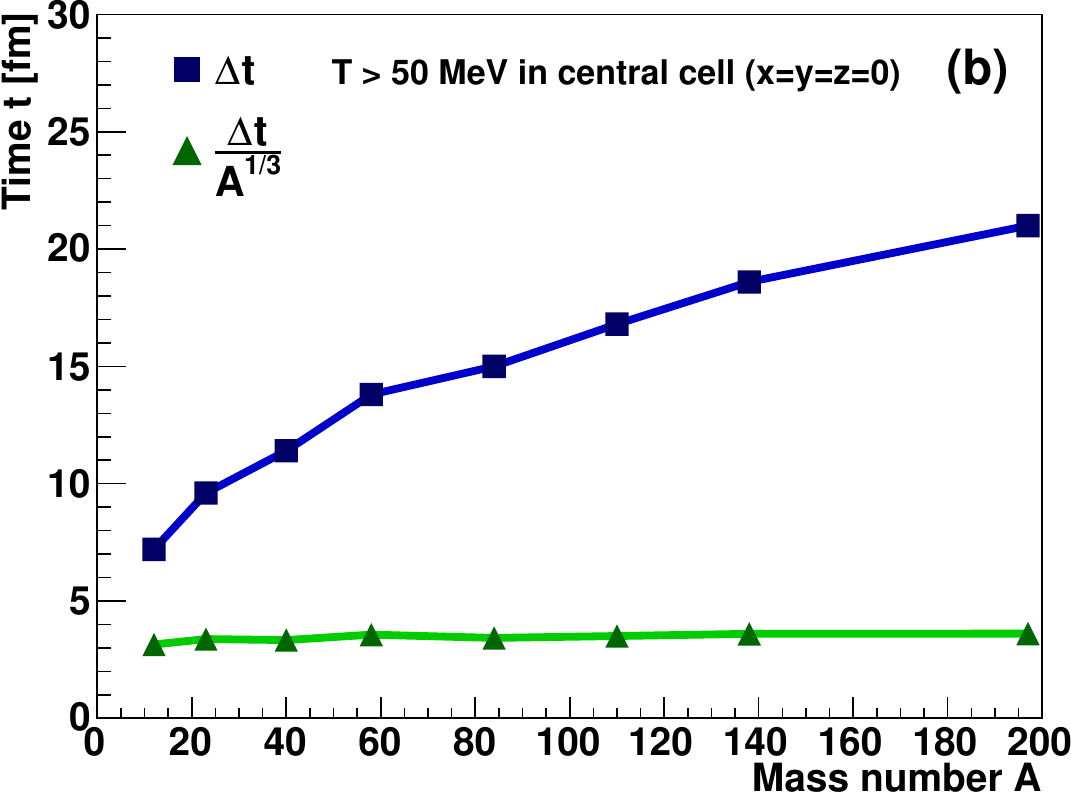} \\[2mm]

\includegraphics[width=0.45\linewidth]{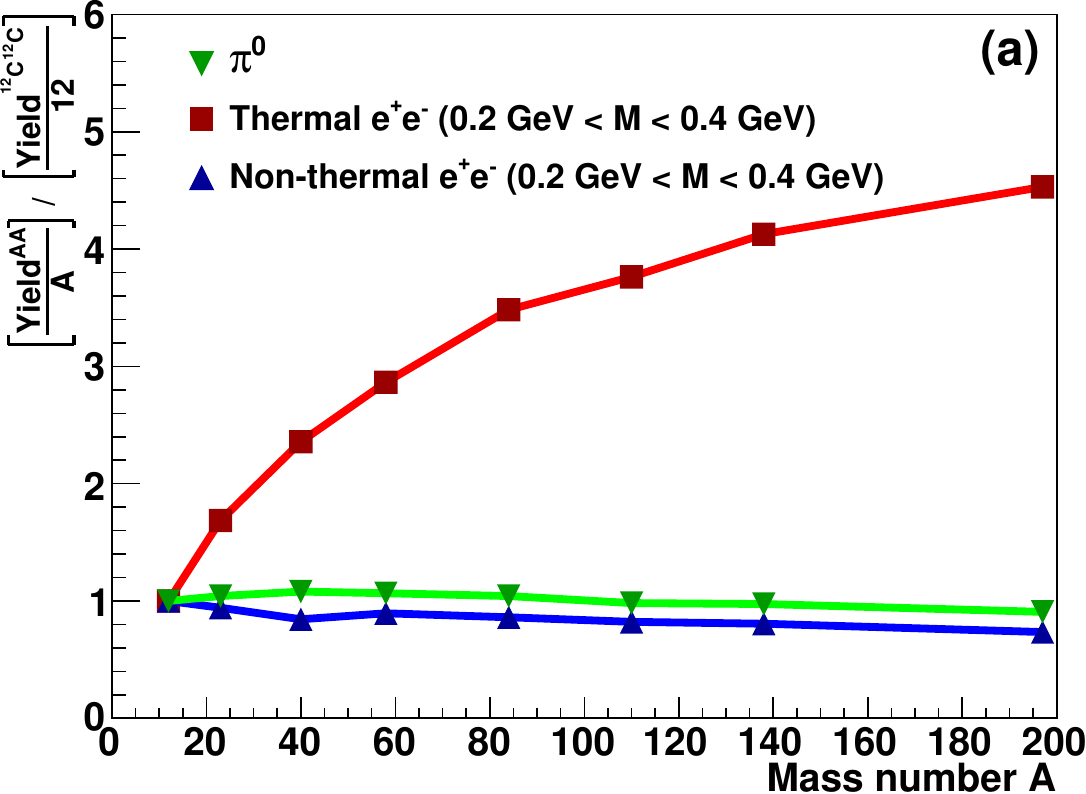} \hspace{3mm}
\includegraphics[width=0.45\linewidth]{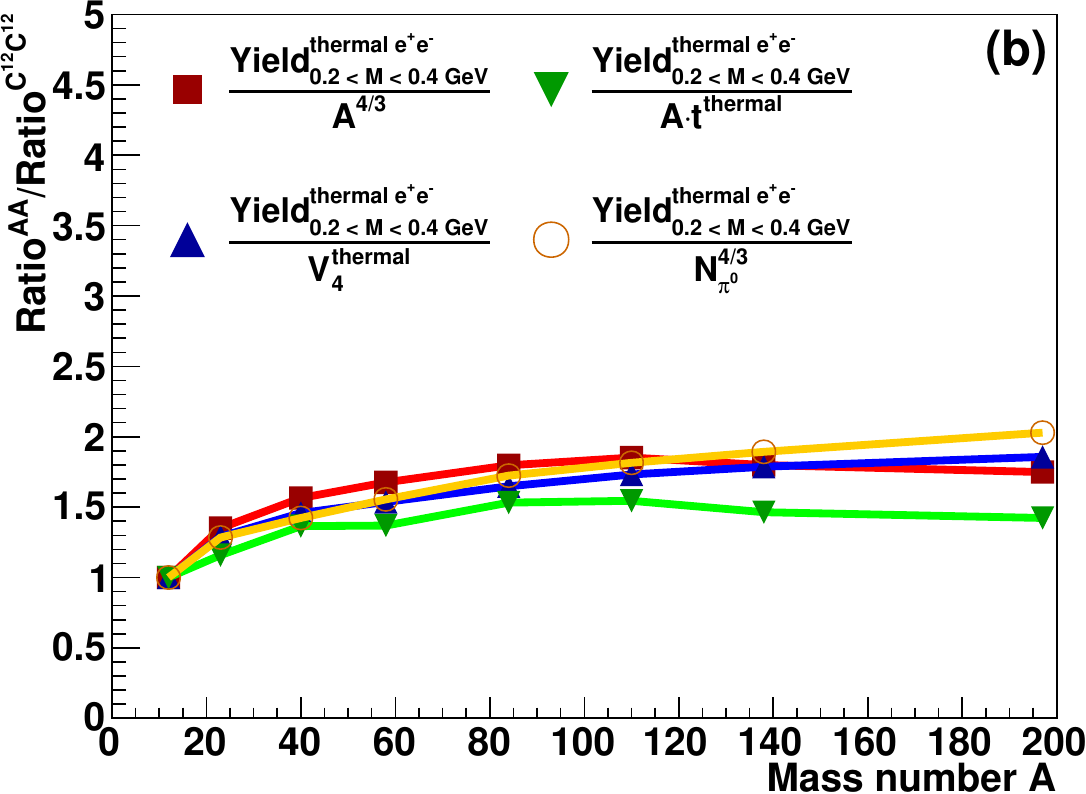}
\caption{Upper panels: (a) Ratio of the thermal four-volume
  $V_{4}$ for different temperatures to the mass number $A$ of the
  colliding nuclei. (b) Time duration over which the central cell of the
  coarse-graining grid (for $x=y=z=0$) emits thermal
  dileptons. \textbf{Lower panels:} (a) Ratio of the thermal (red
  squares) and non-thermal dilepton yield (blue triangles) in the
  invariant-mass range from 0.2 to $0.4\;\GeV/c^{2}$ to the mass number
  $A$ of the colliding nuclei, and the number of $\pi^{0}$ (green
  triangles). The results are normalized to the ratio obtained with
  $\nuclide[12][]{C}+ \nuclide[12][]{C}$ collisions. (b) Ratio of the
  thermal dilepton yield in the invariant mass range from 0.2 to
  $0.4\;\GeV/c^{2}$ scaled by various quantities. All four plots show
  the results for central collisions and a collision energy
  $E_{\mathrm{lab}} = 1.76 \;A \GeV$. Figures taken from {\protect
    \cite{Endres:2015fna}} }
\label{fig:phase-dil.2}
\end{figure}
Since electromagnetic probes are emitted during all stages of the
fireball evolution, leaving the hot and dense medium nearly unaffected
by final-state interactions, they provide information not only about the
in-medium properties of the em.\ current-current correlation function
but also about the space-time evolution of the medium. Together with
future high-precision measurements of dilepton production in heavy-ion
collisions at various beam energies and for different system sizes,
theoretical studies may also shed light on the phase structure of
strongly interacting matter. E.g., the slope of the invariant-mass
spectrum in the mass region $m_{\phi}<M_{\ell^+ \ell^-}<m_{J/\psi}$
leads to a space-time weighted average of the invariant temperature of
the medium (i.e., without blue shifts from radial flow as for slopes of
$q_{\text{t}}$ spectra, cf.\ the discussion in \cite{vanHees:2007th}),
provided the hadronic cocktail, the contribution from decays of
correlated D-$\overline{\D}$ (at higher beam energies also
$\text{B}$-$\overline{\text{B}}$) decay, and Drell-Yan pairs can be
subtracted from the experimental data with sufficient precision.

In \cite{Rapp:2014hha} we have explored the possibilities for such
studies within the above discussed theoretical model for dilepton
production using the thermal-fireball parameterization of the medium,
assuming the cross-over EoS (combining a hadron-resonance gas with a
lQCD EoS as discussed in Sect.\ \ref{sect:bulk-evolution-mods}) for beam
energies corresponding to center-mass energies in the range
$\sqrt{s_{NN}} = 6.3 \;\GeV$-$200 \;\GeV$. We have determined the
average temperature from the slopes of the invariant-mass spectrum of
dileptons by fitting the dilepton spectrum in the range
$M_{\ell^+ \ell^-}=1.5$-$2.5 \;\GeV$ to
\begin{equation}
\label{phase-dil.1}
\frac{\dd R_{\ell^+ \ell^-}}{\dd M_{\ell^+ \ell^-}} \propto (M T)^{3/2} \exp \left
  (-\frac{M_{\ell^+ \ell^-}}{T} \right).
\end{equation}
Since in this mass range $M_{\ell^+\ell^-} \ll T$ the average is
weighted towards the hot and early phases of the fireball evolution. As
can be seen in the left panel of Fig.\ \ref{fig:phase-dil.1} the
resulting ``slope temperatures'' are smoothly increasing with beam
energy and ranging from $T_{\text{s}} \simeq 160 \; \MeV$ at
$\sqrt{s_{NN}}=6\; \GeV$ to $260 \; \MeV$ at
$\sqrt{s_{NN}}=200 \; \GeV$. The values at the higher beam energies are
clearly above the pseudo-critical temperature of
$T_{\text{c}} \simeq 160 \; \MeV$ but considerably lower than the
initial fireball temperatures. This difference becomes smaller at the
lower beam energies which is due to the (pseudo-)latent heat in the
transition. This indicates that the beam-energy range around
$10 \; \GeV$ is promising to map out the phase-transition region, maybe
indicating the onset of a first-order transition by developing a plateau
of the slope curve, resembling a ``caloric curve''.

Further, for a given model for the dilepton-production rates, together
with the determination of the fireball parameters from the hadronic
observables (cf.\ Sect.\ \ref{sect:bulk-evolution-mods}) the total yield
of dileptons is a quite precise measure for the fireball lifetime. This
is shown in Fig. \ref{fig:phase-dil.1}: The yield is determined by
integrating the invariant-mass spectra over the range
$M_{\ell^+ \ell^-}=0.3$-$0.7\;\GeV$, which is just below the
$\rho$- and $\omega$-vacuum mass, so that it consists of contributions
from both partonic and hadronic sources and is quite representative for
the dilepton enhancement due to medium effects, dominated by
interactions of the vector mesons with baryons. As can be seen, the
yield follows closely the proper lifetime $\tau_{\text{fb}}$ of the
fireball. It is important to note that this correlation is disturbed by
either changing the invariant-mass range or by using yields within the
typical single-electron cuts describing the detector acceptance (e.g.,
$p_{\text{t}}>0.2 \; \GeV$, $y<0.9$ for the STAR detector). Thus for
such studies the availability of fully acceptance-corrected dilepton
excess spectra is mandatory.

In \cite{Endres:2015fna} we have made similar studies concerning the
system-size dependence of the bulk-medium dynamics at GSI-SIS energy,
$E_{\text{lab}}=1.76 \; A\GeV$, within the coarse-grained transport
approach (cf.\ Sect. \ref{sect:dileps-SIS}). At these low energies one
expects that the lifetime of the medium is defined by the time the
colliding nuclei overlap, forming a highly excited hadronic medium
dominated by baryons. This can be confirmed by investigating the bulk
properties and associated with dilepton observables within the
coarse-grained transport approach. In the upper panels of
Fig.\ \ref{fig:phase-dil.2} the ``thermal four-volume'' is plotted, i.e.,
the sum of all spacetime cells $\Delta V \Delta t$ for which the
temperature, obtained from the coarse-graining procedure (as described
in Sect.\ \ref{sect:bulk-evolution-mods}), is above various given values
as indicated in plot (a). Since the total volume of each of the nuclei
is $V_{\text{nucl}} \propto A$ and the lifetime of the thermal medium is
expected to be determined by the time the nuclei overlap during the
collision one concludes that
$\Delta t \propto r_{\text{Nucl}} \propto A^{1/3}$, and this is indeed
confirmed in plot (b) by the fact that
$\Delta t/A^{1/3} \simeq \text{const}$.

Also in Fig. \ref{sect:bulk-evolution-mods} the system-size dependence
of the $\e^+ \e^-$ yields (in the mass window
$M_{\e^+ \e^-}=0.2 \; \GeV$-$0.4 \; \GeV$) for emission from the medium
(``thermal dileptons'') and from $\rho$ decays after thermal freeze-out
(``non-thermal dileptons'') as well as of neutral pions is shown. Since
the ``thermal dileptons'' are emitted during the entire fireball
evolution, one expects a scaling with the four-volume. On the other
hand, the ``non-thermal dilepton'' as well as the pion yield is
determined by the situation at thermal freezeout and thus scales with
the three-volume of the corresponding cells (``freeze-out
hypersurface'') with coarse-graining temperatures below $50
\;\MeV$. This is approximately confirmed by plot (b): Scaling the yield
with $A^{4/3}$, $A \cdot t^{\text{thermal}}$, $V_4^{\text{thermal}}$, or
$N_{\pi_0}^{4/3}$ leads to a roughly flat system-size dependence.

As we have demonstrated in all these studies, dilepton and photon
production in heavy-ion collisions can be well described with partonic
and effective hadronic models for the in-medium em.\ current-current
correlation function. Thus, together with adequate descriptions of the
bulk-medium evolution, these techniques provide a promising tool to
understand probable signals of the details of the QCD phase diagram as
soon as high-precision data on dileptons and photons in heavy-ion
collisions at various beam energies become available in the
future. Particularly, when deviations from the here provided results are
observed, sensitivity to Equations of State with different
phase-transition properties may be reached.

\begin{figure}[h]
\begin{minipage}{0.58 \linewidth}
\includegraphics[width=\linewidth]{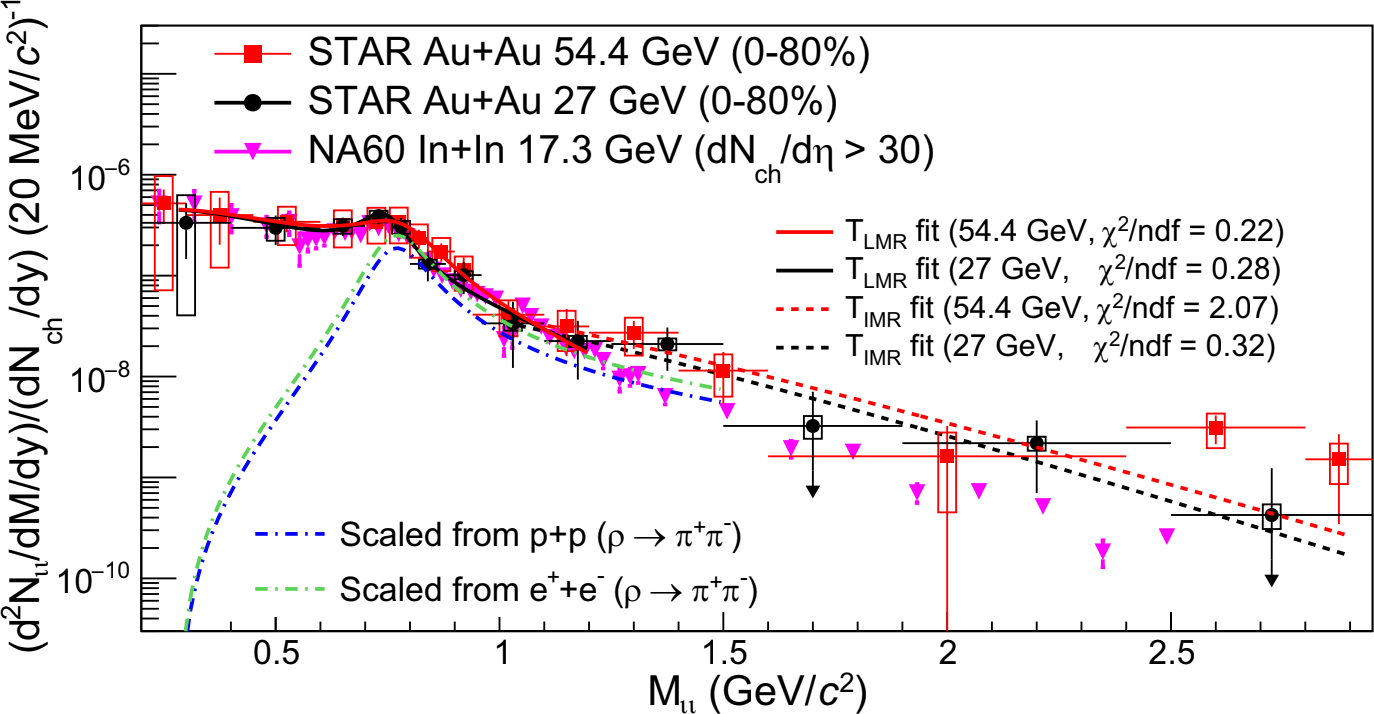}
\end{minipage} \hfill
\begin{minipage}{0.375 \linewidth}
\includegraphics[width=\linewidth]{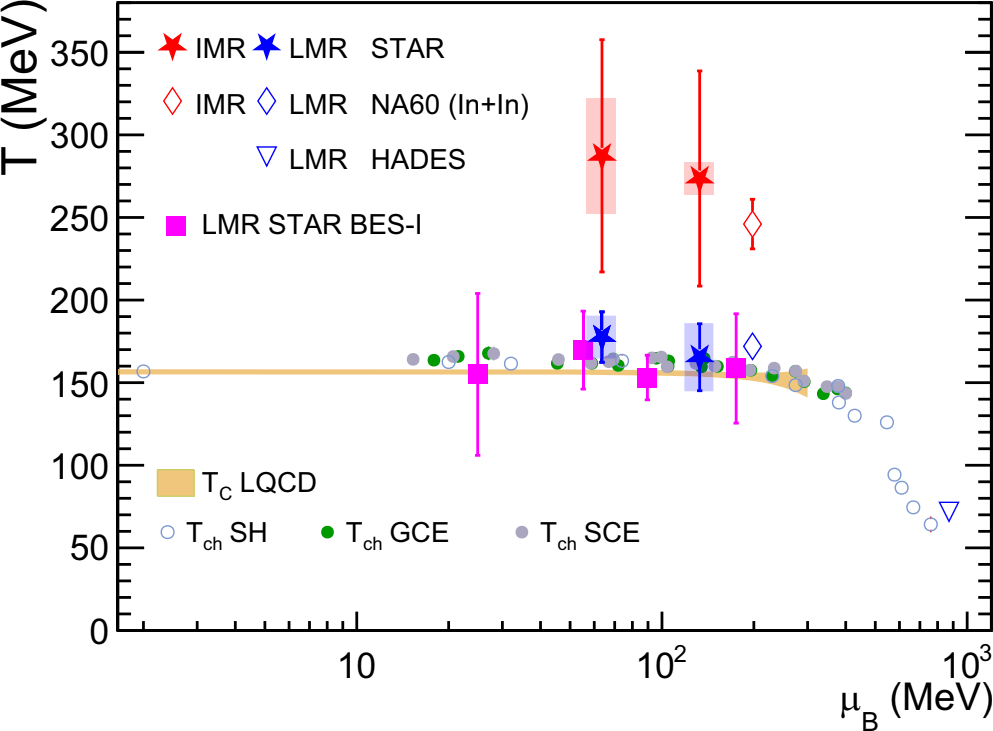}
\end{minipage}
\caption{Left: Thermal dielectron mass spectra from
  $\sqrt{s_{NN}}=54.4 \,\GeV$ (red squares) and $27 \, \GeV$ (black
  dots) compared to the NA60 thermal dimuon data (magenta inverted
  triangles). $ll$ denotes the dielectron or dimuon pairs. Dashed lines
  show the fitting curves for the corresponding temperature
  extractions. Dot-dashed lines display the expected vacuum-$\rho$
  spectra based on the p+p \cite{AguilarBenitez:1991yy} and $\e^+ \e^-$
  \cite{Derrick:1985jx} collision data.  Vertical bars and boxes around
  data points represent the statistical and systematic uncertainties,
  respectively. Downward arrows indicate statistical uncertainties
  exceeding 100{\%}. Right: Temperatures vs. baryon chemical potential. Temperatures extracted
from in-medium $\rho^0$
; the region of the later QGP stage (blue stars), and the region of
the earlier QGP stage (red stars) from STAR data are compared to the temperatures
extracted from NA60 data \cite{Arnaldi:2008er} (diamonds) and HADES data
\cite{HADES:2019auv} (inverted triangle).
Chemical freeze-out temperatures extracted from the statistical thermal models
(SH, GCE, SCE) \cite{Andronic:2017pug,STAR:2017sal} are shown as open
and filled circles. The QCD critical temperature $T_c$ at finite
$\mu_{\text{B}}$ predicted by LQCD calculations \cite{HotQCD:2018pds} is shown as a yellow band. All
temperatures are plotted at the $\mu_{\text{B}}$ determined at chemical freeze-out. Vertical bars
and boxes around the data points represent the statistical and
systematic uncertainties, respectively. Figure taken from \cite{STAR:2024bpc}.}
\label{fig.temp-fit-temps-bes-ii-STAR.png}
\end{figure}

Recently the STAR collaboration has extracted the space-time averaged
temperatures from the measured acceptance corrected dielectron excess
spectra within the beam-energy-scan program (BES II)
\cite{STAR:2024bpc}. Interestingly they fitted to both, low-mass (LMR)
and intermediate-mass (IMR) regions. This is achieved by fitting the
$\e^+ \e^-$-invariant-mass spectra in the regions
$0.4 \;\GeV < M_{\e^+ \e^-}<1.2 \; \GeV$ for the LMR and
$1.0 \; \GeV < M_{\e^+ \e^-}<2.9 \; \GeV$ for the IMR. In the LMR the
fit is to an ansatz function of the shape
\begin{equation}
\frac{\dd N_{\e^+ \e^-}}{\dd N} \propto \Gamma_{\e^+ \e^-}
f_{\text{BW}}(M) M^{3/2} \exp(-M/T)
\end{equation}
with the relativistic Breit-Wigner distribution,
\begin{equation}
f_{\text{BW}}(M) = \frac{M M_0}{(M_0^2-M^2)^2+M_0^2 \Gamma^2(M)}, \quad
\Gamma(M)=\Gamma_0 \frac{M_0}{M} \left (\frac{M^2-4m_{\pi}^2}{M_0^2-4 m_{\pi}^2} \right )^{3/2}
\end{equation}
and the dilepton phase-space factor,
\begin{equation}
\Gamma_{\e^+ \e^-} \propto \left (1+\frac{2m_{\e}^2}{M^2} \right)
\sqrt{1-\frac{4m_{\e}^2}{M^2}}.
\end{equation}
For the IMR the spectral function is assumed to be flat, i.e., fitting
the excess yiels to a function $\propto M^{3/2} \exp(-M/T)$. In the IMR
both the correlated-$\D \overline{\D}-decay$ and the Drell"=Yan
background are subtracted using corresponding pp cross sections
(``hadronic cocktail''). The temperature fits as well as the extracted
temperatures in both mass regions are shown in Fig.\
\ref{fig.temp-fit-temps-bes-ii-STAR.png}. The results are in accordance
with the expectation that the space-time weighted average is biased
towards lower temperatures and larger fireball volumes in the LMR vs.\
higher temperatures and lower fireball volumes in the IMR.

From the theoretical side, e.g., the exploitation of
functional-renormalization-group methods that allow a consistent
description of both the EoS and the in-medium spectral functions of the
light vector mesons, can be a promising way for such further studies
\cite{Tripolt:2016cey,Jung:2016yxl}.

\section{Conclusions and Outlook}
\label{sect:conclusions}

As has been summarized in this review the dilepton production in
heavy-ion collisions is theoretically quite well understood in terms of
models using hard-thermal-loop or input from lattice-QCD calculations of
$q\bar{q}$ annihilation in the partonic and effective hadronic models in
the hadronic stages of the fireball evolution to describe the in-medium
electromagnetic current-current-correlation function and the (thermal)
dilepton-production rates from the strongly interacting medium created
in heavy-ion collisions over all beam-energy ranges. There are strong
indications that chiral symmetry is realized via the
mirror-assignment/chiral doubler representation in the hadronic models,
where the spontaneous breaking of chiral symmetry due to the formation
of a quark condensate $\erw{\overline{q} q} \neq 0$ at lower
temperatures is responsible only for the non-degeneracy of the mass
spectra of chiral partners. It has also been demonstrated that accurate measurements
of the invariant-mass spectrum of dileptons in the intermediate-mass
region can be used to determine a space-time-evolution weighted average
of the fireball temperature, provided non-thermal contributions like the
dileptons from hard Drell-Yan processes and the decay of correlated
open-heavy flavor (D and B) mesons can be reliably subtracted. The
models also describe first measurements of polarization observables, and
with realistic bulk-evolution models of the fireball it is possible to
extract the life time of the strongly interacting medium
\cite{Rapp:2014hha}.

This makes the electromagnetic probes a promising tool for constraining
the rich phase diagram of strongly interacting (QCD) matter via
high-precision measurements of the excitation function of dilepton
emission: e.g., a first-order chiral phase transition should manifest
itself in a plateau of the fireball temperatures (``latent heat'') and if
the ``thermal trajectory'' of the fireball comes close to a critical
point this might be reflected in a prolonged fireball lifetime,
indicating ``critical slowing-down''.



\end{document}